\documentclass[11pt]{article}
\usepackage[a4paper,margin=2.5cm]{geometry}
\usepackage{amsmath,amssymb,amsthm}
\usepackage{bbold}      
\usepackage{xcolor}
\usepackage{listings}
\usepackage{booktabs,array}
\usepackage{url}
\usepackage[authoryear,round]{natbib}
\usepackage[hidelinks]{hyperref}
\usepackage{thmtools}   

\usepackage{bbold}      
\usepackage{xcolor}
\usepackage{listings}
\usepackage{array}
\usepackage{thmtools}   

\newif\iftechapp \techapptrue
\newif\ifarxiv \arxivfalse

\makeatletter
\lst@Key{elixirtokens}{false}[t]{\lstKV@SetIf{#1}\lst@ifelixirtokens}
\lst@AddToHookExe{SetLanguage}{\let\lst@ifelixirtokens\iffalse}
\lst@AddToHook{Output}{%
  \lst@ifelixirtokens
    \ifx\lst@thestyle\lst@identifierstyle 
      \expandafter\elx@classify\the\lst@token\elx@end:\elx@end\elx@stop
    \fi
  \fi}
\def\elx@end{\elx@end}
\def\elx@classify#1#2:\elx@end#3\elx@stop{%
  \ifx:#1\let\lst@thestyle\elx@atomstyle
  \else\if\relax\detokenize{#3}\relax
    \ifcat\noexpand#1.%
      \ifnum`#1>64 \ifnum`#1<91 \let\lst@thestyle\elx@modulestyle\fi\fi
    \fi
  \else\let\lst@thestyle\elx@atomstyle
  \fi\fi}
\def\elx@atomstyle{\color{atc}}
\def\elx@modulestyle{\color{mdc}}
\makeatother
\lstdefinelanguage{ElixirNew}{%
  morekeywords={after,and,case,catch,cond,def,defmacro,defmodule,defp,defstruct,do,do:,
    else,else:,end,fn,for,if,import,in,not,or,raise,receive,require,rescue,try,unless,
    use,when},
  morekeywords=[2]{nil,true,false},
  alsoletter={:}, elixirtokens,
  sensitive=true, comment=[l]{\#}, morestring=[b]", escapeinside={(*@}{@*)}}
\lstdefinelanguage{ErlangNew}{%
  morekeywords={after,and,andalso,band,begin,bnot,bor,bsl,bsr,bxor,case,catch,cond,
    div,end,fun,if,let,not,of,or,orelse,receive,rem,try,when,xor},
  morekeywords=[2]{true,false},
  otherkeywords={-spec,-type,-opaque,-callback,-module,-export,-behaviour,-record},
  sensitive=true, morecomment=[l]\%, morestring=[b]"}

\ExplSyntaxOn
\clist_map_inline:nn
  { after,and,case,catch,cond,def,defmacro,defmodule,defp,defstruct,do,else,end,fn,
    for,if,import,in,not,or,raise,receive,require,rescue,try,unless,use,when }
  { \tl_const:cn { c_code_kwc_#1_tl } { kwc } }
\clist_map_inline:nn { nil,true,false,ok,undefined }
  { \tl_const:cn { c_code_atc_#1_tl } { atc } }
\clist_map_inline:nn { gen,gen_server,gen_statem }
  { \tl_const:cn { c_code_mdc_#1_tl } { mdc } }
\tl_new:N \l_code_word_tl
\tl_new:N \l_code_quote_tl
\tl_new:N \l_code_tok_tl
\tl_new:N \l_code_math_tl
\bool_new:N \l_code_atom_bool
\bool_new:N \l_code_colon_bool
\bool_new:N \l_code_math_bool
\bool_new:N \l_code_quote_bool
\bool_new:N \l_code_dollar_bool
\bool_new:N \l_code_attr_bool
\cs_new_protected:Npn \code_emit:nn #1#2 { \textcolor{#1}{#2} }
\cs_new_protected:Npn \code_flush:
  {
    \tl_if_empty:NF \l_code_word_tl
      {
        \bool_if:NTF \l_code_atom_bool
          { \code_emit:nn { atc } { : \l_code_word_tl } }
          {
            \bool_if:NTF \l_code_dollar_bool
              { \code_emit:nn { newc } { \l_code_word_tl } }
              {
            \bool_if:NTF \l_code_attr_bool
              { \code_emit:nn { kwc } { \l_code_word_tl } }
              {
            \tl_set_eq:NN \l_tmpa_tl \l_code_word_tl
            \tl_replace_all:Nnn \l_tmpa_tl { \_ } { _ }
            \cs_if_exist:cTF { c_code_kwc_ \l_tmpa_tl _tl }
              { \code_emit:nn { kwc } { \l_code_word_tl } }
              {
                \cs_if_exist:cTF { c_code_atc_ \l_tmpa_tl _tl }
                  { \code_emit:nn { atc } { \l_code_word_tl } }
                  {
                    \cs_if_exist:cTF { c_code_mdc_ \l_tmpa_tl _tl }
                      { \code_emit:nn { mdc } { \l_code_word_tl } }
                      {
                        \tl_set:Nx \l_tmpb_tl { \str_head:N \l_code_word_tl }
                        \exp_args:NnV \str_if_in:nnTF { ABCDEFGHIJKLMNOPQRSTUVWXYZ }
                          \l_tmpb_tl
                          { \code_emit:nn { mdc } { \l_code_word_tl } }
                          { \l_code_word_tl }
                      }
                  }
              }
              }
              }
          }
        \tl_clear:N \l_code_word_tl
        \bool_set_false:N \l_code_atom_bool
      }
    \bool_set_false:N \l_code_dollar_bool
    \bool_set_false:N \l_code_attr_bool
  }
\cs_new_protected:Npn \code_main:nnn #1#2#3
  {
    \bool_if:NTF \l_code_math_bool
      {
        \str_if_eq:nnTF { #3 } { 3 }
          {
            \bool_set_false:N \l_code_math_bool
            $ \l_code_math_tl $
          }
          { \tl_put_right:Nn \l_code_math_tl { #1 } }
      }
      {
        \bool_if:NTF \l_code_quote_bool
          {
            \int_compare:nTF { #2 = 34 }
              {
                \code_emit:nn { atc } { :"\l_code_quote_tl" }
                \bool_set_false:N \l_code_quote_bool
              }
              { \tl_put_right:Nn \l_code_quote_tl { #1 } }
          }
          { \code_plain:nnn { #1 } { #2 } { #3 } }
      }
  }
\prg_new_conditional:Npnn \code_letter:nnn #1#2#3 { TF }
  { \bool_lazy_or:nnTF { \str_if_eq_p:nn { #3 } { B } }
      { \str_if_eq_p:nn { #1 } { \_ } } { \prg_return_true: } { \prg_return_false: } }
\cs_new_protected:Npn \code_plain:nnn #1#2#3
  {
    \bool_if:NT \l_code_colon_bool
      {
        \bool_set_false:N \l_code_colon_bool
        \code_letter:nnnTF { #1 } { #2 } { #3 }
          {
            \tl_if_empty:NTF \l_code_word_tl
              { \bool_set_true:N \l_code_atom_bool }
              { \code_flush: : }
          }
          {
            \tl_if_empty:NTF \l_code_word_tl
              {
                \int_compare:nTF { #2 = 34 }
                  { \bool_set_true:N \l_code_quote_bool \tl_clear:N \l_code_quote_tl }
                  { : }
              }
              { \code_emit:nn { atc } { \l_code_word_tl : } \tl_clear:N \l_code_word_tl }
          }
      }
    \bool_if:NF \l_code_quote_bool
      {
        \code_letter:nnnTF { #1 } { #2 } { #3 }
          { \tl_put_right:Nn \l_code_word_tl { #1 } }
          {
            \bool_lazy_and:nnTF { \str_if_eq_p:nn { #3 } { C } }
              { \int_compare_p:n { 48 <= #2 <= 57 } }
              {
                \tl_if_empty:NTF \l_code_word_tl
                  { #1 }
                  { \tl_put_right:Nn \l_code_word_tl { #1 } }
              }
              {
                \int_compare:nTF { #2 = 58 }
                  {
                    \bool_if:NT \l_code_atom_bool { \code_flush: }
                    \bool_set_true:N \l_code_colon_bool
                  }
                  {
                    \code_flush:
                    \str_if_eq:nnTF { #3 } { 3 }
                      {
                        \bool_set_true:N \l_code_math_bool
                        \tl_clear:N \l_code_math_tl
                      }
                      {
                        \tl_if_eq:nnTF { #1 } { \$ }
                          {
                            \code_emit:nn { newc } { \$ }
                            \bool_set_true:N \l_code_dollar_bool
                          }
                          {
                            \int_compare:nTF { #2 = 64 }
                              {
                                \code_emit:nn { kwc } { @ }
                                \bool_set_true:N \l_code_attr_bool
                              }
                              { #1 }
                          }
                      }
                  }
              }
          }
      }
  }
\cs_generate_variant:Nn \code_main:nnn { V }
\cs_new_protected:Npn \code_item:nnn #1#2#3
  {
    \tl_set:Nx \l_code_tok_tl { #1 }
    \code_main:Vnn \l_code_tok_tl { #2 } { #3 }
  }
\cs_new_protected:Npn \code_scan:n #1
  {
    \group_begin:
    \tl_clear:N \l_code_word_tl
    \bool_set_false:N \l_code_atom_bool
    \bool_set_false:N \l_code_colon_bool
    \bool_set_false:N \l_code_math_bool
    \bool_set_false:N \l_code_quote_bool
    \bool_set_false:N \l_code_dollar_bool
    \bool_set_false:N \l_code_attr_bool
    \tl_analysis_map_inline:nn { #1 } { \code_item:nnn { ##1 } { ##2 } { ##3 } }
    \bool_if:NT \l_code_colon_bool
      {
        \bool_set_false:N \l_code_colon_bool
        \tl_if_empty:NTF \l_code_word_tl
          { : }
          { \code_emit:nn { atc } { \l_code_word_tl : } \tl_clear:N \l_code_word_tl }
      }
    \code_flush:
    \group_end:
  }
\cs_set_eq:Nc \code_orig_texttt:n { texttt~ }
\cs_new_protected:Npn \castagnacodefont
  { \DeclareRobustCommand \texttt [1] { \code_orig_texttt:n { \code_scan:n { ##1 } } } }
\ExplSyntaxOff

\AtBeginDocument{\xdef\castagnatocdepth{\the\value{tocdepth}}}

\makeatletter
\@ifundefined{claim}{}{}
\@ifundefined{lemma}{\newtheorem{lemma}{Lemma}}{}
\@ifundefined{theorem}{\newtheorem{theorem}{Theorem}}{}
\@ifundefined{proposition}{\newtheorem{proposition}[theorem]{Proposition}}{}
\@ifundefined{corollary}{\newtheorem{corollary}[theorem]{Corollary}}{}

\@ifpackageloaded{biblatex}{}{\makeatother }

\DeclareSourcemap{
  \maps[datatype=bibtex]{
    \map{
      \perdatasource{cdr80.bib}
      \step[fieldset=options,
            fieldvalue={maxcitenames=2,mincitenames=1,maxbibnames=99}]
    }
  }
}

\newbool{castagna:parens}

\newbibmacro*{castagna:init}{%
  \ifnumless{\value{multicitecount}}{2}
    {\global\boolfalse{castagna:parens}%
     \global\undef\castagna@lasthash
     \global\undef\castagna@lastyear}
    {\iffieldundef{prenote}
       {}
       {\global\undef\castagna@lasthash
        \global\undef\castagna@lastyear}}}

\newbibmacro*{castagna:reinit}{%
  \global\undef\castagna@lasthash
  \global\undef\castagna@lastyear}

\newbibmacro*{castagna:label}{%
  \iffieldundef{label}
    {\printtext[bibhyperref]{\printfield[citetitle]{title}}}
    {\printtext[bibhyperref]{\printfield{label}}}}

\newbibmacro*{castagna:date}{%
  \iffieldundef{labelyear}
    {}
    {\printtext[bibhyperref]{\printlabeldateextra}}}

\newbibmacro*{castagna:extradate}{%
  \iffieldundef{extradate}
    {}
    {\printtext[bibhyperref]{\printfield{extradate}}}}

\newbibmacro*{castagna:cite}{%
  \ifthenelse{\ifnameundef{labelname}\OR\iffieldundef{labelyear}}
    {\usebibmacro{castagna:label}%
     \setunit{\addcomma\space}%
     \usebibmacro{castagna:date}%
     \usebibmacro{castagna:reinit}}
    {\iffieldequals{namehash}{\castagna@lasthash}
       {\ifthenelse{\iffieldequals{labelyear}{\castagna@lastyear}\AND
                    \(\value{multicitecount}=0\OR\iffieldundef{postnote}\)}
          {\setunit{\addcomma}%
           \usebibmacro{castagna:extradate}}
          {\setunit{\compcitedelim}%
           \usebibmacro{castagna:date}%
           \savefield{labelyear}{\castagna@lastyear}}}
       {\printnames{labelname}%
        \setunit{\addcomma\space}%
        \usebibmacro{castagna:date}%
        \savefield{namehash}{\castagna@lasthash}%
        \savefield{labelyear}{\castagna@lastyear}}}%
  \setunit{\addsemicolon\space}}

\newbibmacro*{castagna:textcite}{%
  \iffieldequals{namehash}{\castagna@lasthash}
    {\ifthenelse{\iffieldequals{labelyear}{\castagna@lastyear}\AND
                 \(\value{multicitecount}=0\OR\iffieldundef{postnote}\)}
       {\setunit{\addcomma}%
        \usebibmacro{castagna:extradate}}
       {\setunit{\compcitedelim}%
        \usebibmacro{castagna:date}%
        \savefield{labelyear}{\castagna@lastyear}}}
    {\ifnameundef{labelname}
       {\usebibmacro{castagna:label}%
        \setunit{%
          \global\booltrue{castagna:parens}%
          \addspace\bibopenparen}%
        \ifnumequal{\value{citecount}}{1}
          {\usebibmacro{prenote}}
          {}%
        \usebibmacro{castagna:date}}
       {\printnames{labelname}%
        \setunit{%
          \global\booltrue{castagna:parens}%
          \addspace\bibopenparen}%
        \ifnumequal{\value{citecount}}{1}
          {\usebibmacro{prenote}}
          {}%
        \iffieldundef{labelyear}
          {\usebibmacro{castagna:label}}
          {\usebibmacro{castagna:date}}%
        \savefield{labelyear}{\castagna@lastyear}}%
     \stepcounter{textcitecount}%
     \savefield{namehash}{\castagna@lasthash}}%
  \setunit{%
    \ifbool{castagna:parens}
      {\bibcloseparen\global\boolfalse{castagna:parens}}
      {}%
    \addsemicolon\space}}

\newbibmacro*{castagna:textcite:postnote}{%
  \usebibmacro{postnote}%
  \ifthenelse{\value{multicitecount}=\value{multicitetotal}}
    {\setunit{}%
     \printtext{%
       \ifbool{castagna:parens}
         {\bibcloseparen\global\boolfalse{castagna:parens}}
         {}}}
    {\setunit{%
       \ifbool{castagna:parens}
         {\bibcloseparen\global\boolfalse{castagna:parens}}
         {}%
       \addsemicolon\space}}}

\DeclareCiteCommand{\castagnacitep}[\mkbibparens]
  {\usebibmacro{castagna:init}%
   \usebibmacro{prenote}}
  {\usebibmacro{citeindex}%
   \usebibmacro{castagna:cite}}
  {}
  {\usebibmacro{postnote}}

\DeclareCiteCommand{\castagna@textcite}
  {\usebibmacro{castagna:init}}
  {\usebibmacro{citeindex}%
   \usebibmacro{castagna:textcite}}
  {}
  {\usebibmacro{castagna:textcite:postnote}}

\DeclareCiteCommand{\castagnacitet}[\castagna@textcite@init\castagna@textcite]
  {\gdef\castagna@savedkeys{}%
   \citetrackerfalse%
   \pagetrackerfalse%
   \DeferNextCitekeyHook%
   \usebibmacro{castagna:init}}
  {\ifthenelse{\iffirstcitekey\AND\value{multicitetotal}>0}
     {\protected@xappto\castagna@savedcites{()(\thefield{multipostnote})}%
      \global\clearfield{multipostnote}}
     {}%
   \xappto\castagna@savedkeys{\thefield{entrykey},}%
   \iffieldequals{namehash}{\castagna@lasthash}
     {}
     {\stepcounter{textcitetotal}%
      \savefield{namehash}{\castagna@lasthash}}}
  {}
  {\protected@xappto\castagna@savedcites{%
     [\thefield{prenote}][\thefield{postnote}]{\castagna@savedkeys}}}

\DeclareDelimcontextAlias{castagna@textcite}{textcite}

\newrobustcmd{\castagna@textcite@init}[2]{%
  \setcounter{textcitetotal}{0}%
  \setcounter{textcitecount}{0}%
  \def\castagna@savedcites{#1}#2%
  \ifnumgreater{\value{textcitetotal}}{0}
    {\castagna@savedcites}
    {}}

\defbibenvironment{castagnabib}
  {\list{}
     {\setlength{\leftmargin}{1.5em}%
      \setlength{\itemindent}{-\leftmargin}%
      \setlength{\itemsep}{\bibitemsep}%
      \setlength{\parsep}{\bibparsep}}}
  {\endlist}
  {\item}
\makeatother

\arxivtrue
\techapptrue

\providecommand{\setcommand}{}
\renewcommand{\setcommand}[1]{\providecommand{#1}{}\renewcommand{#1}}

\setcommand{\techref}[1]{\iftechapp\ (see Appendix~\ref{#1})\fi}
\setcommand{\techthm}[2]{\iftechapp\ (see #1~\ref{#2} in Appendix~\ref{app:tech})\fi}

\setcommand{\parref}[1]{the paragraph ``\nameref{#1}'' of Appendix~\ref{#1}}
\setcommand{\techparref}[1]{\iftechapp\ (see the paragraph ``\nameref{#1}'' of Appendix~\ref{#1})\fi}

\setcommand{\noappnote}{\begingroup%
\footnotetext{\hspace*{-6mm}A version of this essay with an appendix containing the full technical details
  is available online.}%
\endgroup}

\setcommand{\beppe}[1]{\textcolor{red}{[Beppe: #1]}}
\setcommand{\tocheck}[1]{\textcolor{orange}{[check: #1]}}

\definecolor{newc}{HTML}{820073}
\definecolor{kwc}{HTML}{1E754F}   
\definecolor{atc}{HTML}{1F4E9C}   
\definecolor{mdc}{HTML}{8A4B0F}   
\definecolor{stc}{HTML}{A3311E}   
\definecolor{cmc}{HTML}{7A8A7A}   
\setcommand{\N}[1]{\textcolor{newc}{#1}}
\setcommand{\inst}[2]{\textnormal{\ttfamily\textcolor{mdc}{#1}\N{[}#2\N{]}}}
\setcommand{\prm}[1]{\N{#1:}}
\setcommand{\named}{\mathrel{\N{\textnormal{\ttfamily ::}}}}

\setcommand{\pid}{\mathsf{pid}}
\setcommand{\srv}{\mathsf{server}}
\setcommand{\so}{\mathsf{server\_of}}
\setcommand{\from}{\mathsf{from}}
\setcommand{\rtag}{\mathsf{reply\_tag}}
\setcommand{\reqid}{\mathsf{request\_id}}
\setcommand{\unit}{\mathsf{unit}}
\setcommand{\ch}{\mathsf{ch}^{-}}
\setcommand{\chl}{\mathsf{ch}^{\lambda}}
\setcommand{\MT}{\mathrm{MT}}
\setcommand{\LC}{\mathrm{LC}}
\setcommand{\at}[1]{\textup{\texttt{:#1}}}
\setcommand{\Tempty}{\mathbb{0}}
\setcommand{\Tany}{\mathbb{1}}
\setcommand{\Int}{\mathtt{int}}
\setcommand{\Bool}{\mathtt{bool}}
\setcommand{\Nil}{\mathtt{nil}}

\setcommand{\state}{\mathit{state}}
\setcommand{\Hpid}{H^{\pid()}_m}
\setcommand{\Hr}{H^{r}_m}
\setcommand{\Tm}{T_m}

\castagnacodefont   

\title{From Strong Normalization to WhatsApp:\\ the Journey of an Axiom}
\author{Giuseppe Castagna\\[1mm]
  \small CNRS, IRIF, Universit\'e Paris Cit\'e}
\date{\small\emph{To Mario Coppo, Mariangiola Dezani-Ciancaglini, and Simona Ronchi Della
    Rocca,\\ for their eightieth birthdays}}

\begin{document}
\maketitle
\iftechapp\else\noappnote\fi

\newcommand{\journeytable}{%
\begin{table}[t]
  \centering
  \ifarxiv \small \def\colauthors{4.8cm}\def\colcontrib{5.9cm}%
  \else \footnotesize \def\colauthors{.31\textwidth}\def\colcontrib{.38\textwidth}\fi
  \begin{tabular}{@{}l>{\raggedright\arraybackslash}p{\colauthors}>{\raggedright\arraybackslash}p{\colcontrib}l@{}}
    \toprule
    Year & Authors & Contribution & Here\\
    \midrule
    1978--80 & Coppo, Dezani-Ciancaglini & intersection types, strong normalization
      & \S\ref{sec:intersection}, \S\ref{sec:elixir}\\[1.5mm]
    1980 & Coppo, Dezani-Ciancaglini, Venneri & principal type schemes, expansions
      & \S\ref{sec:intersection}, \S\ref{sec:poly}\\[1.5mm]
    1983 & Barendregt, Coppo, Dezani-Ciancaglini & filter model, \eqref{eq:bcd}
      & \S\ref{sec:intersection}--\S\ref{sec:server-of}\\[1.5mm]
    1984, 1988 & Ronchi Della Rocca (and Venneri) & principal types for the filter theory,
      unification & \S\ref{sec:intersection}, \S\ref{sec:poly}\\[1.5mm]
    1992 & Milner & functions as processes & \S\ref{sec:servers}, \S\ref{sec:lost}\\[1.5mm]
    1995 & Barbanera, Dezani-Ciancaglini, de'Liguoro & union types
      & \S\ref{sec:union}, \S\ref{sec:server-of}\\[1.5mm]
    2002, 2008 & Frisch, Castagna, Benzaken & semantic subtyping
      & \S\ref{sec:union}, \S\ref{sec:server-of}\\[1.5mm]
    2002 & Dezani-Ciancaglini, Frisch, Giovannetti, Motohama & relevance of semantic
      subtyping & \S\ref{sec:union}\\[1.5mm]
    2002, 2007 & Ronchi Della Rocca (and Liquori) & explicitly typed intersections
      & \S\ref{sec:poly}\\[1.5mm]
    2006 & Castagna, Dezani-Ciancaglini, Varacca & functions as processes: the encoding
      loses \eqref{eq:bcd} & \S\ref{sec:lost}, \S\ref{sec:server-of}\\[1.5mm]
    2008 & Castagna, De Nicola, Varacca & semantic subtyping for channels, hence types
      for processes & \S\ref{sec:servers}, \S\ref{sec:lost}\\[1.5mm]
    2014 & Castagna et al. & polymorphic functions with set-theoretic types
      & \S\ref{sec:poly}\\[1.5mm]
    2024--26 & Castagna, Duboc, Valim & the type system of Elixir & \S\ref{sec:elixir}\\[1.5mm]
    2026 & this essay & typing \texttt{call} from \texttt{handle\_call} & \S\ref{sec:server-of}\\
    \bottomrule
  \end{tabular}
  \caption{The journey of the axiom \eqref{eq:bcd}.}\label{tab:journey}
\end{table}%
}

\begin{abstract}
  Twenty years ago, Mariangiola Dezani-Ciancaglini, Daniele Varacca and I showed that
  Milner's classic encoding of functions into processes does not respect one of the laws of intersection
  types, the one stating that $(\sigma\to\tau)\land(\sigma\to\rho)$ is a subtype of
  $\sigma\to(\tau\land\rho)$, which Barendregt, Coppo and Dezani-Ciancaglini had stated in
  1983 for their filter $\lambda$-model. In this essay I show that this observation is the
  key to closing what a colleague at Meta calls ``the main typing gap in Erlang''---the language in which the servers of WhatsApp are written: the type of
  the function that implements a server is not related to the type of the server, so that
  a client may expect from the server a reply that the server will never send.

  Actually, closing this gap is just an excuse to follow the journey of the law above, and
  of some of its relatives, from the study of strong normalization in the $\lambda$-calculus
  to the type systems of Elixir and Erlang: a journey that shows how much the practical
  typing of these languages owes to the work of Mario Coppo, Mariangiola
  Dezani-Ciancaglini and Simona Ronchi Della Rocca, to whom this essay is dedicated for
  their eightieth birthdays.
\end{abstract}

\section{Introduction}\label{sec:intro}

In October 2026 I gave a talk at an Elixir/Erlang meetup at Meta in Stockholm about the types that we are
designing for the processes and the modules of Elixir. One of my examples was a server
written with the standard library of Elixir, and a couple of people in the audience pointed out
that in real servers the type of the reply usually depends on the type of the request. The
natural answer is to give an intersection type to the function that computes the
replies, and the question becomes how to make the clients of the server aware of this
type. A couple of days later Daniel Gorín, of Meta, sent me an email that
described a small Erlang server written in the way most Erlang code is written, and that
concluded:\footnote{~Quoted with permission.}
\begin{quote}
  ``This one is IMO the main typing gap in Erlang atm: eqwalizer (or any other
  type-checker) can't ensure the types of api and implementation are in sync. I consider
  this one the main gap, as most actual Erlang code is written this way, by implementing
  behaviours like \texttt{gen\_server} and not that much sending/receiving messages
  directly, which are more like low-level primitives.''
\end{quote}
Erlang is the language in which the servers of Meta's WhatsApp are written, and eqWAlizer, the type
checker that the email mentions, has been developed at Meta to type the code of WhatsApp \citep{eqWAlizer}: so when a gap is
called ``the main typing gap in Erlang'' by somebody who works with this code, it is worth
taking it seriously.

The problem, which I explain in detail in Section~\ref{sec:servers}, is the following. In
Erlang and in Elixir a server is a process that answers requests. The programmer does not
write the process: a library supplies it, provided that the programmer writes a function,
\texttt{handle\_call} (a ``callback'' for this library), that computes the reply to each request. Clients use another function
of the library, \texttt{call}, that sends a request to the server and waits for the reply.
The reply usually depends on the request, so the type of \texttt{handle\_call} is
overloaded. Thanks to the work of Mario Coppo, Mariangiola Dezani-Ciancaglini, and Simona Ronchi Della Rocca,
and also of John Reynolds
(not exactly minor figures, all four), this overloading can be expressed as an
intersection of arrows, one for each kind of request; Erlang programmers already write
such types, as we will see. But nothing relates this type to the type of \texttt{call},
whose result has the type of all values (the top type in a subtyping hierarchy): a client may expect an integer where the server
can reply with an undefined value, and no checker will notice.

The solution, which I describe in Sections~\ref{sec:lost}--\ref{sec:poly}, is made of old
pieces. When I tried to understand what the type of a server that answers requests of
different kinds should be (what I ended up calling an ``overloaded process''), I realized
that the answer was in a paper that Mariangiola Dezani-Ciancaglini, Daniele Varacca and I
wrote twenty years ago \citep{CDV06},\footnote{~The careful reader will notice that five of
  the papers cited in this essay are by a C, a D and a V: two by Coppo, Dezani-Ciancaglini
  and Venneri, and one each by Castagna, Dezani-Ciancaglini and Varacca; Castagna, De Nicola
  and Varacca; and Castagna, Duboc and Valim. I swear that this is a coincidence. To avoid any confusion, I
  cite all bibliographic references by name and year.} and that it rests on a law that Barendregt, Coppo and
Dezani-Ciancaglini stated in 1983 for their filter $\lambda$-model \citep{BCD83}:
\begin{equation}\label{eq:bcd}
  (\sigma\to\tau)\land(\sigma\to\rho)\ \le\ \sigma\to(\tau\land\rho)
  \tag{$\to\land$}
\end{equation}
where $\sigma$, $\tau$ and $\rho$ denote types, $\sigma\land\tau$ is the intersection of
$\sigma$ and $\tau$, and $\le$ is subtyping. In words: a function that maps $\sigma$ into
$\tau$ and maps $\sigma$ into $\rho$ maps $\sigma$ into values that are both in $\tau$ and
in $\rho$.\footnote{~The converse inequality holds by covariance of the codomain, so the two
  types are equivalent: the law is an equation, and this is how it will appear in the $\lambda$-calculus with semantic subtyping, that is, CDuce \citep{BCF03}.}
As we showed in \citet{CDV06}, this law is lost when functions are encoded as
processes; the solution that I propose here consists in applying the law before the encoding.

This is the reason why I chose this problem to celebrate the eightieth birthdays of Mario
Coppo, Mariangiola Dezani-Ciancaglini and Simona Ronchi Della Rocca: it is the perfect
excuse to tell how a law born in the study of the models of $\lambda$-calculus may end up typing the
servers of WhatsApp, and to show how much the practical typing of Erlang and Elixir owes to
their work.

The essay is informal. It relies on work in progress on the typing of processes
\citep{YHM26} and of modules \citep{Bretagne26,BC26} in Elixir, and I state results as
claims with sketches of proofs.\iftechapp{} The reader who wants the details will find them
in Appendix~\ref{app:tech}, a technical companion to Sections~\ref{sec:servers}--\ref{sec:poly}
drawn from the note from which this essay originated.\fi

\ifarxiv\else
\journeytable
\fi

\paragraph{Plan of the article.} The essay is organized as a journey, and its first stops
are historical. Section~\ref{sec:intersection} starts in Turin in 1978, with the
intersection types that Coppo and Dezani-Ciancaglini introduced to characterize the
strongly normalizing $\lambda$-terms, and stops at the filter model, where our law first
appears. Section~\ref{sec:union} adds unions and negations, and shows how semantic
subtyping turned the laws of intersection and union types into properties of sets of
values. Section~\ref{sec:elixir} arrives in Elixir and Erlang, where functions defined by
several clauses are typed by intersections. The second part of the essay is about servers.
Section~\ref{sec:servers} is a primer for the $\lambda$-calculist who never wrote a server
in Erlang or Elixir: it explains what processes and servers are, how they can be typed, and
what the gap pointed out by Daniel is. Section~\ref{sec:lost} shows that the natural way of
typing a server, that is, translating the type of \texttt{handle\_call} as one translates
functions into processes, loses the law, and recalls how we diagnosed this loss twenty
years ago. Section~\ref{sec:server-of} defines the operator that restores the law, and
Section~\ref{sec:poly} shows that this operator is not ad hoc, since it is what
polymorphism yields, by way of the expansions of Coppo, Dezani-Ciancaglini and Venneri and
of the explicitly typed intersections of Ronchi Della Rocca. Section~\ref{sec:whatsapp}
goes back to Erlang and lists what is still missing. The essay has thus two keys of
reading: the reader interested only in the
history can stop at the end of Section~\ref{sec:elixir}; the reader interested only in
servers can start from Section~\ref{sec:elixir}, going back to Section~\ref{sec:union} for the
notation of set-theoretic types. Table~\ref{tab:journey} lists the stops
of the journey and the sections where each one is discussed.

\ifarxiv
\journeytable
\fi

\section{Several functional characters}\label{sec:intersection}

Let me start at the beginning (how original, I know). Curry's theory of functionality assigns simple types to
$\lambda$-terms: if $M$ has type $\sigma\to\tau$ and $N$ has type $\sigma$, then $MN$ has
type $\tau$. Types are preserved by reduction but not by expansion, and many terms in normal
form, such as $\lambda x.xx$, have no type. \citet{CD78,CD80} extended the theory with an
intersection $\sigma\land\tau$, the type of the terms that have both type $\sigma$ and type
$\tau$, so that, for instance, $\lambda x.xx$ has type $(\sigma\land(\sigma\to\tau))\to\tau$.
A type describes a functional behavior of a term, one of its ``functional characters''
(Curry's words, 
which \citet{CDV81} kept in the title of their paper on solvable terms)
and the novelty is that a term may have several of them at once, as $x$ has in $\lambda x.xx$.
In the resulting system the typable terms are exactly the strongly normalizing ones
\citep{CD80,Pottinger80}.

Then came the model. \citet{BCD83} ordered the types, which by then included a universal
type $\omega$ that every term has, by a preorder $\le$ (a subtyping
relation, in today's terms) for which $\omega$ is the largest type, $\land$ is the greatest
lower bound, the arrow is contravariant in its domain and covariant in its codomain,
$\omega\le\omega\to\omega$, and \eqref{eq:bcd} holds. A \emph{filter} is a set of types that
contains $\omega$ and is closed upwards with respect to $\le$ and under $\land$. The filter
model interprets a term as the filter of its types; an application $MN$ is then interpreted
as the application $F\cdot G$ of the filter $F$ of $M$ to the filter $G$ of $N$, defined as
\[
  F\cdot G=\{\tau \mid \sigma\to\tau\in F \text{ for some } \sigma\in G\}
\]
This is where \eqref{eq:bcd} enters the scene: it is what makes $F\cdot G$ closed under
intersection (that $F\cdot G$ contains $\omega$ and is closed upwards follows from
$\omega\le\omega\to\omega$ and from covariance). To see why, take $\tau_1$ and $\tau_2$ in
$F\cdot G$; if they are there it is because $\sigma_1\to\tau_1$ and
$\sigma_2\to\tau_2$ are in $F$ for some $\sigma_1,\sigma_2\in G$. Then $\sigma_1\land\sigma_2$
is in $G$, contravariance puts $\sigma_1\land\sigma_2\to\tau_1$ and
$\sigma_1\land\sigma_2\to\tau_2$ in $F$, hence their intersection, and \eqref{eq:bcd} puts
$\sigma_1\land\sigma_2\to\tau_1\land\tau_2$ in $F$, so that $\tau_1\land\tau_2$ is in $F\cdot
G$. With this model, Barendregt, Coppo and Dezani-Ciancaglini proved that type assignment is
complete for the interpretation of types as sets of elements of a $\lambda$-model, where
$\sigma\land\tau$ denotes the intersection of the sets and $\sigma\to\tau$ the set of the
elements that map every element of $\sigma$ into $\tau$; from this result they deduced the
completeness of simple type assignment, which solved an open problem of \citet{Scott75}.
In this interpretation \eqref{eq:bcd} is an equality between sets,
which is a first hint of what will come: once types are sets, the laws of intersection
types become properties of sets.

The same group studied type inference, and this is where Simona Ronchi Della Rocca enters
our story. A term has infinitely many types, and a principal type is one from which all the
others can be obtained. For intersection types substitutions are not enough: one also needs
\emph{expansions}, which duplicate a part of a type into an intersection of copies.
\citet{CDV80} introduced principal type schemes with expansions, \citet{RDRV84} extended
them to the type theory of Barendregt, Coppo and Dezani-Ciancaglini, and \citet{RDR88}
studied unification for intersection types, from which she obtained a semi-algorithm for
type inference. We will meet expansions again in
Section~\ref{sec:poly}, where they explain how polymorphism produces the type of a server.

Two ideas of this period will travel all the way to the servers of WhatsApp: types denote
sets, with intersection as set intersection; and an intersection of arrows describes a
function with several behaviors (an overloaded function in modern lingo), which \eqref{eq:bcd} combines when they concern the same
arguments.

\section{Unions, negations, and types as sets of values}\label{sec:union}

Unions came next. \citet{MPS86} interpreted types as ideals of a domain and gave a rule to
eliminate unions,\footnote{~As we showed in \citet{CLNL22,CLN24}, this rule, known as
  \emph{union elimination}, is the essence of \emph{occurrence typing}, the technique,
  first defined by \citet{THF08}, that assigns different types to the same expression in
  the different branches of a type test. But this is another story.} and \citet{BDdL95} gave intersection and union types for the
$\lambda$-calculus a syntax and a semantics. The union $\sigma\lor\tau$ is the type of the
terms that have type $\sigma$ or type $\tau$, and their type theory contains, besides the
distributivity of $\land$ over $\lor$, the law
\begin{equation}\label{eq:bddl}
  (\sigma\to\rho)\land(\tau\to\rho)\ \le\ (\sigma\lor\tau)\to\rho
  \tag{$\lor\to$}
\end{equation}
an axiom that states that a function that maps $\sigma$ into $\rho$ and maps
$\tau$ into $\rho$ maps $\sigma\lor\tau$ into $\rho$. As for \eqref{eq:bcd}, the converse
inequality holds, here by contravariance, so the law is an equation.

Intersection and union types then entered programming languages: intersection types with
Forsythe \citep{Forsythe} and the thesis of \citet{Pierce92}, union types with
\citet{Pierce91}. But there is a catch. These languages, like the system of Barbanera,
Dezani-Ciancaglini and de'Liguoro, have no construct that tests the type of a value at run
time, and without one a function cannot select its behavior according to the type of its
argument: a type such as
$(\sigma_1\to\tau_1)\land(\sigma_2\to\tau_2)$ does not describe two behaviors, it refines the
description of a single one, a form of ad hoc polymorphism known as coherent overloading
\citep{Pierce92,Forsythe}. To behave differently on arguments of different types, a function must be able to test
the type of its argument \citep{CGL92,CGL95,FSOP}, and this is what the functions of CDuce (discussed in the next paragraph), as well as those of Erlang and Elixir, can do.

The last step was to take sets as primary, which is what semantic subtyping does.
\citet{HP03} defined the subtyping relation of XDuce as the inclusion of sets of values, for
first-order types that describe XML documents. \citet{FCB02,FCB08} extended the approach to
functions and to negation, for the language CDuce, whose functional core is a $\lambda$-calculus with run-time type tests and set-theoretic types. A type $t$ denotes a set of values; union,
intersection and negation $\neg t$ denote set-theoretic union, intersection and complement; $\Tempty$ is the
empty type, $\Tany$ the type of all values, $t\setminus s$ stands for $t\land\neg s$; and $t\le
s$ holds if and only if the set denoted by $t$ is included in the set denoted by $s$.
To break the circularity between the subtyping of arrows and the typing of functions, types are
interpreted in a domain that does not depend on the terms of the language. For cardinality
reasons, this domain cannot contain all the relations on itself; but, as in the graph model
of \citet{Scott76}, finite approximations suffice: arrow types are interpreted as sets of finite
relations, which induce the same subtyping relation as sets of arbitrary relations would
\citep{FCB08}. As a consequence, the laws
\eqref{eq:bcd} and \eqref{eq:bddl} are no longer axioms but properties of the interpretation.
And although defined semantically, subtyping is decidable \citep{FCB08}.

In a sense, semantic subtyping reverses the construction of Barendregt, Coppo and
Dezani-Ciancaglini: there, a model of the $\lambda$-calculus is built from a subtyping
relation; here, a subtyping relation is built from a model of the types. The two approaches
meet, and once more Mariangiola is at the meeting point: with Frisch, Giovannetti and
Motohama she showed that the subtyping relation of CDuce restricted to arrows, unions and
intersections coincides with the relevant entailment of the logic $\mathbf{B}_+$
\citep{DFGM03}.

Finally, \citet{CL17}, \citet{CLPS19} and \citet{Lanvin21} extended set-theoretic types to gradual typing,
with a type for the values whose type is checked at run time. This is the theory on which the
type system of Elixir is built, and part of the theory of \citet{Lanvin21} is also used
in eqWAlizer (which is not surprising, since Victor Lanvin is currently at Meta in charge of
eqWAlizer).

\section{Intersection types in Erlang and Elixir}\label{sec:elixir}

So much for the theory: let us now see where these types landed. Erlang was designed at
Ericsson in the 1980s for telecommunication systems, and its programs run on a virtual
machine called the BEAM. Elixir, created by José Valim, runs on the same virtual machine
and uses the same libraries. Both are dynamically typed
functional languages whose values include integers, atoms,\footnote{~Atoms are constants
  that denote themselves, written \texttt{:ok} in Elixir and \texttt{ok} in Erlang. The
  Booleans are the atoms \texttt{true} and \texttt{false}, and \texttt{nil} (Erlang's
  \texttt{undefined}) is an atom too.} tuples \texttt{\{$v_1$,\ldots,$v_n$\}}, lists, maps
(such as \texttt{\%\{current:\,42\}} that associates the key \texttt{current} with the value 42), functions, and process identifiers. Functions are
defined by clauses, each with a pattern and possibly a guard, which are tried in order:
\begin{lstlisting}
def negate(x) when is_integer(x), do: -x
def negate(x) when is_boolean(x), do: not x
\end{lstlisting}
An application \texttt{negate($e$)} executes the first clause whose pattern (here the simple capture variable \texttt{x}) and guard accept the result of the evaluation of $e$, and fails if there is none. A guard is a 
test on the variables of the pattern, written after \texttt{when}: here \texttt{is\_integer(x)}
and \texttt{is\_boolean(x)} test the type of \texttt{x} at run time, which is precisely the
test that, as we saw in Section~\ref{sec:union}, a function needs in order to behave
differently on arguments of different types. A definition by several clauses is thus a type-case
\citep{CD26}, and the natural type of \texttt{negate} is an intersection:\footnote{~In the
  syntax of Elixir this type is written
  \texttt{(integer() -> integer()) and (boolean() -> boolean())}.}
\[
  (\texttt{integer()}\to\texttt{integer()})\land(\texttt{boolean()}\to\texttt{boolean()})
\]
where \texttt{integer()} and \texttt{boolean()} are the types of integers and of Booleans in Elixir.
From now on I abbreviate them as $\Int$ and $\Bool$, I write types with the connectives of
Section~\ref{sec:union}, and I use singleton types for atoms: the type \at{ok} contains only
the atom \texttt{:ok}.

The type system of Elixir \citep{Design24,CD26,Duboc26} is a gradual system with set-theoretic
types, based on the theory of Section~\ref{sec:union}. It has been included in the compiler
since version~1.17 (June 2024); its latest version (v1.20 of June 2026) covers all constructs of the language and infers intersection types for functions defined by
several clauses \citep{Elixir117,Elixir120}. For \texttt{negate} it infers the type
above, and thus it deduces that \texttt{negate($e$)} is an integer whenever $e$ is, and likewise for Booleans
\citep{CD26}.

Erlang programmers, for their part, have been writing intersection types for years, whether
they call them so or not. A specification (a \emph{spec}, in Erlang jargon) of an Erlang
function may have several clauses, one for each clause of the function, and programmers call
it an \emph{overloaded spec}. In his email, Daniel specified the callback of the server of
Section~\ref{sec:servers} as follows:
\begin{lstlisting}[language=ErlangNew,emph={get_int,set_int,reply,ok},emph={[2]gen_server}]
-spec handle_call
    (get_int, gen_server:from(), state()) -> {reply, integer(), state()};
    ({set_int, integer()}, gen_server:from(), state()) -> {reply, ok, state()}.
\end{lstlisting}
This is an intersection of two arrows in Erlang syntax: what Coppo, Dezani-Ciancaglini and
Venneri called the several functional characters of a term is, at WhatsApp, one spec clause
for each function clause. As Daniel's email notes, one can write such a spec today for eqWAlizer.
Dialyzer \citep{LS06}, the traditional analyzer for Erlang, uses specs in an optimistic way
and reports only code that is sure to fail; eqWAlizer is a type checker that uses part of
the theory of gradual set-theoretic types of \citet{Lanvin21} (Section~\ref{sec:union}).

\section{Servers, for \texorpdfstring{$\lambda$}{lambda}-calculists}\label{sec:servers}

If you know the $\pi$-calculus, then you already know most of what you need to understand
the servers of Erlang and Elixir. Let me fill in the details.

\paragraph{Processes.} A process of the BEAM (the virtual machine on which Erlang and Elixir
run) is a lightweight thread with a mailbox, identified by a
process identifier, or \emph{pid}. The expression \texttt{spawn($f$)} creates a process that
runs the function $f$ (a zero-ary function) and returns its pid; \texttt{send($p$, $e$)} puts the result of the evaluation of the expression
$e$ in the mailbox of the process identified by $p$ and continues, since sends are
asynchronous; \texttt{receive} expressions such as
\begin{lstlisting}
receive do
  {:add, k} when is_integer(k) -> k + 42
  :stop                        -> 42
end
\end{lstlisting}
take from the mailbox of the running process the first message that matches one of the
patterns and satisfies its guard, as for the clauses of a function (here, a pair formed by the
atom \texttt{:add} and an integer \texttt{k}, or the atom \texttt{:stop} alone), waiting if there is none, and execute the corresponding branch. A process can read
only its own mailbox. In $\pi$-calculus terms, pids are channel names on which everybody can
write but only one process, their owner, can read: the BEAM implements a \emph{local}
$\pi$-calculus \citep{MS04}.

\paragraph{Pid types.} To type processes I use the types of \citet{YHM26}: $\pid(t)$ is the
type of the pids of the processes that accept the messages of type $t$. It is the type
$\ch(t)$ of the channels on which values of type $t$ can be written, which \citet{CDV08}
introduced in their semantic subtyping for the $\pi$-calculus. Pid types satisfy the
following laws \citep{YHM26}:
\begin{align}
  \pid(t)\le\pid(s) &\iff s\le t \label{eq:contra}\\
  \pid(t_1)\land\pid(t_2) &= \pid(t_1\lor t_2) \label{eq:pidand}\\
  \textstyle\pid(t)\le\bigvee_{n\in N}\pid(s_n) &\iff s_n\le t \text{ for some } n\in N
  \label{eq:union}
\end{align}

The first law is contravariance: a pid that accepts all the messages in $t$ can be used where
only messages in some $s\le t$ are sent. In particular $\pid(\Tempty)$, which I write $\pid()$ as in Elixir,
contains every pid, and no message can be sent safely to a pid of this type.

The second law says that a pid that accepts the messages in $t_1$ and those in $t_2$ accepts
the messages in $t_1\lor t_2$: an intersection of pid types is the type of a process that
handles several kinds of messages. It allows a process to share only a partial view of its mailbox with each of its clients, a view being a factor of the intersection. 

The third law, which I call the \emph{union law}, says that a pid type is included in a union
of pid types only if it is included in one of them. Intuitively, $\pid(t)$ contains the pid of
a process that accepts exactly the messages in $t$, and this pid is in $\pid(s_n)$ only if
$s_n\le t$. It is the law that determines the type of a reply in Sections~\ref{sec:lost}
and~\ref{sec:server-of}.

\paragraph{Behaviours and GenServer.} Actually, Erlang programs rarely use \texttt{send} and
\texttt{receive} directly. They use \emph{behaviours} (sic: Erlang uses the British spelling), generic components of the OTP library
(the Open Telecom Platform) that are parameterized by a \emph{callback module} written by the
programmer. The most common one is the generic server, \texttt{gen\_server} in Erlang and
\texttt{GenServer} in Elixir \citep{GenServerErl,GenServerDoc}. The library provides the server process;
the programmer provides a function \texttt{init}, which computes the initial state of the
server from an argument, and a function \texttt{handle\_call}, which, given a request, the
return address of the client and the current state, returns the reply and the new state.
Stripped of tags, monitors, timeouts and of the other callbacks, the library is the following
code, where \texttt{m} is the callback module:
\begin{lstlisting}
def start_link(m, v) do
  {:ok, s} = m.init(v)
  {:ok, spawn(fn -> loop(&m.handle_call/3, s) end)}
end

def loop(h, s) do
  receive do
    {c, w} ->                    # a request c and a return address w
      {:reply, r, s2} = h.(c, w, s)
      send(w, r)
      loop(h, s2)
  end
end

def call(p, c) do
  w = alias()                    # a fresh return address
  send(p, {c, w})
  receive_on(w)                  # wait for the reply sent to w
end
\end{lstlisting}
The function \texttt{start\_link} computes the initial state with \texttt{init}, spawns the
server process, which runs \texttt{loop} on the function \texttt{handle\_call} of the module,
and returns its pid. OTP does the same: when a server starts, it extracts the function
\texttt{handle\_call} from the module, and its loop applies this function\techref{ta-sec:otp}.
For the reader who does not speak Elixir: \texttt{=} is pattern matching, so that
\texttt{\{:ok, s\} = m.init(v)} binds \texttt{s}, and fails if \texttt{m.init(v)} does not
return a pair tagged by \texttt{:ok}; a module is a value, so that \texttt{m.init(v)} calls the
function \texttt{init} of the module bound to \texttt{m}; the capture operator \texttt{\&}
turns a function of a module into a value, so that \texttt{\&m.handle\_call/3} denotes the
function \texttt{handle\_call} of arity 3 of \texttt{m}; \texttt{fn -> $e$ end} is the
abstraction $\lambda().e$, the zero-ary function that \texttt{spawn} expects; and
\texttt{h.(c, w, s)} applies the function \texttt{h}, the dot marking the application of a
function value (rather than a named function).

A client obtains a reply by
evaluating \texttt{call(p, c)}, where \texttt{p} is the pid of the server and \texttt{c} the call's
request. This function runs in the client process: it creates a fresh return address
\texttt{w}, sends the pair \texttt{\{c, w\}} to the server, and blocks until a message arrives
at \texttt{w}; this message, the reply, is the result of \texttt{call}. On the other side,
\texttt{loop} receives the pair sent by the client \texttt{call}, computes the reply \texttt{r} and the new state \texttt{s2}
with \texttt{h} ---i.e., with \texttt{handle\_call}---, sends \texttt{r} to \texttt{w}, and starts again with
\texttt{s2}. Here \texttt{alias()} stands for \texttt{Process.alias()}, which returns a new
address whose messages arrive in the mailbox of the process that created it, and
\texttt{receive\_on} (which is not an Elixir primitive) stands for what the library does:
\texttt{receive\_on(w)} takes from this mailbox the first message sent to \texttt{w}, and returns it. The address is fresh so that the client
cannot take for the reply another message of its mailbox, for instance the late reply to an
earlier call.\footnote{~OTP implements these addresses with references and
  process aliases: the request is the triple \texttt{\{:"\$gen\_call", from, c\}}, where
  \texttt{from} is the pair formed by the pid of the client and a fresh reference, and the
  reply is sent to the client tagged by the same reference.} A $\lambda$-calculist will recognize the encoding of
functions as processes of \citet{Milner92}: the server encodes a $\lambda$-abstraction, a
process that receives an argument together with a return channel and sends the result on it,
and \texttt{call} encodes an application, which creates a fresh channel, sends it with the
argument and waits on it. GenServer just adds a state, which the loop passes from a request
to the next. In other terms, Erlang and Elixir programmers use Milner's encoding of the $\lambda$-calculus into the $\pi$-calculus 
daily, packaged in a library.

\paragraph{An example.} Here is the server of Daniel's email, translated into Elixir. It stores
an integer, and its module also contains the public interface (the API) that clients use:
\begin{lstlisting}
defmodule MyServer do
  use GenServer                          # MyServer is a GenServer callback module

  def init(v), do: {:ok, %{current: v}}  # initial state is a map with field current

  def handle_call(:get_int, _from, s), do: {:reply, s.current, s}
  def handle_call({:set_int, v}, _from, s), do: {:reply, :ok, %{s | current: v}}

  # public API
  def start_link(v), do: GenServer.start_link(MyServer, v)
  def get_int(p), do: GenServer.call(p, :get_int)
  def set_int(p, v), do: GenServer.call(p, {:set_int, v})
end
\end{lstlisting}
The state is a map with one field, \texttt{current}; \texttt{s.current} reads the field and
\texttt{\%\{s | current:\ v\}} updates it. The server accepts two kinds of requests, and
\texttt{handle\_call} has one clause for each, selected by pattern matching on its first
argument, the request. The first clause handles the request \texttt{:get\_int}, an atom with
no payload: it replies with the stored integer and leaves the state unchanged. The second
handles \texttt{\{:set\_int, v\}}, a pair formed by the tag \texttt{:set\_int} and a payload
\texttt{v}, the new integer: it replies \texttt{:ok} and stores \texttt{v} in the state. Both
clauses ignore the return address (the underscore in \texttt{\_from} marks an unused
variable), since the library itself sends the reply to it, and both return a triple
\texttt{\{:reply, r, s2\}} with the reply \texttt{r} and the new state \texttt{s2}, as
\texttt{loop} expects. The functions of the API hide the requests: \texttt{get\_int(p)} and
\texttt{set\_int(p, v)} build the request and pass it to \texttt{GenServer.call}, and clients
use the server only through them (\texttt{GenServer.start\_link} and \texttt{GenServer.call} are
the functions \texttt{start\_link} and \texttt{call} above, qualified by the name of their
module). Here is a typical client session, where \texttt{s} is the pid of the server:
\ifarxiv\pagebreak\fi
\begin{lstlisting}
{:ok, s} = MyServer.start_link(42)
MyServer.get_int(s)                      # 42
MyServer.set_int(s, 24)                  # :ok
MyServer.get_int(s) + 1                  # 25
\end{lstlisting}
The callback has the intersection type of the overloaded spec of Section~\ref{sec:elixir}:
\begin{align*}
  \mathtt{handle\_call}:\ &((\at{get\_int},\,\pid(),\,\state)\to\{\at{reply},\Int,\state\})\\
  \land\ &((\{\at{set\_int},\Int\},\,\pid(),\,\state)\to\{\at{reply},\at{ok},\state\})
\end{align*}
where $(t_1,t_2,t_3)\to t$ is the type of the functions of three arguments,
$\{t_1,\ldots,t_n\}$ is the type of the tuples whose components have types
$t_1,\ldots,t_n$, and $\state$ is the type of the state, that is, of the maps whose field
\texttt{current} is an integer (i.e., \texttt{\%\{current: integer()\}}). The second argument, the return address, has type $\pid()$
because no clause uses it. The API is meant to have the types $\mathtt{get\_int}:s\to\Int$
and $\mathtt{set\_int}:(s,\Int)\to\at{ok}$, for some type $s$ of the server.

\paragraph{Typing \texttt{call}.} What is the type $s$ of a server, that is, of the pid of the
process spawned by \texttt{start\_link} (the \texttt{s} of the client session above), which
is the first argument of \texttt{call}? This process runs \texttt{loop}: if it accepts requests
of type $c$ (for \emph{call}, the name that GenServer gives to the requests that expect a
reply) and it replies with values of type $r$ (for \emph{reply}), then it receives pairs formed
by a request and by a return address that accepts $r$. I write $\srv(c,r)$ for the type of its pid:
\[
  \srv(c,r)=\pid(c\times\pid(r))
\]
where $c\times t$ is the type of the pairs formed by a request of type $c$ and a return address of type
$t$. $\srv(c,r)$ is an abbreviation, not a new type constructor: precisely, $\srv(c,r)$ is the type of the
processes to which one can send a request of type $c$ together with an address on which values
of type $r$ can be sent. Since the return address in a request has type $\pid(r)$, the server can send on it only values
of type $r$: the type does not promise that the server replies, but that, if it replies, the
reply has type $r$. By the contravariance law~\eqref{eq:contra}, applied twice, $\srv$ is
contravariant in $c$ and covariant in $r$, as the arrow $c\to r$: for $c'\neq\Tempty$,
$\srv(c,r)\le\srv(c',r')$ if and only if $c'\le c$ and $r\le r'$. A server is a function that
one applies by sending a message.

The function \texttt{call} is then polymorphic:
\[
  \mathtt{call}:\forall c,r.\ (\srv(c,r),\,c)\to r
\]
Given the pid \texttt{p} of a server for requests $c$ and replies $r$, and a request \texttt{c} of type $c$,
\texttt{call(p, c)} returns a value of type $r$. The client owns the fresh address \texttt{w}, so it can
give it the type $\pid(r)$; the pair \texttt{\{c, w\}} then has type $c\times\pid(r)$, which is what
\texttt{p} accepts; and since only values of type $r$ can be sent to \texttt{w}, \texttt{receive\_on(w)}
returns a value of type $r$\techthm{Theorem}{ta-thm:call}. The type of the result of
\texttt{call} is thus fixed by the type of the server: \texttt{call} is exactly as precise as
the type given to \texttt{p}.

\paragraph{The gap.} Today the library types the pid of a server as $\pid()$ and the result of
\texttt{call} as $\Tany$, the type of all values (\texttt{term()} in Erlang and in Elixir). The API of
\texttt{MyServer} is therefore not checked: a spec $\mathtt{get\_int}:(\pid())\to\Int$ is
taken on trust. Daniel's email shows the danger. Suppose that the stored value may now be
\texttt{nil}, so that the first arrow of \texttt{handle\_call} becomes
\[
  (\at{get\_int},\,\pid(),\,\state)\to\{\at{reply},\Int\lor\Nil,\state\}
\]
and that the programmer forgets to update the API. Nothing complains, but a client that
evaluates \texttt{MyServer.get\_int(s) + 1} crashes when the server holds \texttt{nil}. As
Daniel puts it, a sound checker ``MUST catch this'' (the capitals are his). He also added:
``My intuition is that using a union of LHS-types and a union of RHS-types won't be enough
to capture this, as you need to keep the `valid' LHS/RHS pairs somehow in sync''. His
intuition is right: typing the server, before the change, as
$\srv(\at{get\_int}\lor\{\at{set\_int},\Int\},\,\Int\lor\at{ok})$, with the union of the
requests and the union of the replies, is sound, but then \texttt{call} returns
$\Int\lor\at{ok}$ on \texttt{:get\_int}, and the pairs are lost. The type of a server must keep
the correspondence between requests and replies that the intersection type of
\texttt{handle\_call} expresses.

\section{Functions as processes, or how an axiom gets lost}\label{sec:lost}

Since GenServer is Milner's encoding of functions, the natural idea is to type the server by
the translation of types that accompanies this encoding: the Milner--Turner translation
$\MT$ \citep{SW01}. As defined by \citet[\S4.1]{CDV06}, it maps an arrow to the type of the channels that
receive an argument together with a return channel for the result,
\[
  \MT(c\to r)=\pid(c\times\pid(r))=\srv(c,r)
\]
and commutes with the other type constructors, so that the translation of an intersection of
arrows is the intersection of their translations. Leaving aside the state and the return
address, which play no role here\techref{ta-sec:problem}, the type of \texttt{handle\_call} is
an intersection $\bigwedge_{i\in I}(c_i\to r_i)$, one arrow for each kind of request, and
$\MT$ translates it into the intersection of servers $\bigwedge_{i\in I}\srv(c_i,r_i)$.

Let us try it on \texttt{MyServer}, where
\(
 (\at{get\_int}\to\Int)\land(\{\at{set\_int},\Int\}\to\at{ok})
\)  is the type we obtain for \texttt{handle\_call} leaving aside state and return address. Its translation  gives the type:
\begin{multline*}
  \srv(\at{get\_int},\Int)\land\srv(\{\at{set\_int},\Int\},\at{ok})\\
  =\pid((\at{get\_int}\times\pid(\Int))\ \lor\ (\{\at{set\_int},\Int\}\times\pid(\at{ok})))
\end{multline*}
where the equality is given by the law in \eqref{eq:pidand}. To type \texttt{call(p, :get\_int)} we look for an
$r$ such that this type is below $\srv(\at{get\_int},r)=\pid(\at{get\_int}\times\pid(r))$. By
contravariance this holds if and only if
$\at{get\_int}\times\pid(r)\le(\at{get\_int}\times\pid(\Int))\lor(\{\at{set\_int},\Int\}\times\pid(\at{ok}))$,
that is, since the two requests are disjoint, if and only if $\pid(r)\le\pid(\Int)$, that is,
$\Int\le r$. The least solution is $r=\Int$, so \texttt{call} returns an integer and \texttt{get\_int} type-checks. After
Daniel's change the least solution is $r=\Int\lor\Nil$, so \texttt{call} returns a value of type $\Int\lor\Nil$, and the declared result $\Int$ of
\texttt{get\_int} is rejected, as Daniel wants. So far, so good: the translation settles
Daniel's example.

But it does not settle every server, because the requests of different arrows may overlap.
Consider a callback of type
\begin{equation}\label{eq:overlap}
  h:(\at{a}\lor\at{b}\to\Int\lor\Nil)\land(\at{b}\lor\at{c}\to\Int\lor\at{ok})
\end{equation}
Applied to \texttt{:b}, $h$ returns an integer: by contravariance $h$ has type
$\at{b}\to\Int\lor\Nil$ and type $\at{b}\to\Int\lor\at{ok}$, hence, by \eqref{eq:bcd}, type
$\at{b}\to(\Int\lor\Nil)\land(\Int\lor\at{ok})=\at{b}\to\Int$. The translation gives
$\srv(\at{a}\lor\at{b},\Int\lor\Nil)\land\srv(\at{b}\lor\at{c},\Int\lor\at{ok})$, and
\texttt{call} on \texttt{:b} now requires
$\pid(r)\le\pid(\Int\lor\Nil)\lor\pid(\Int\lor\at{ok})$, which by the union law
\eqref{eq:union} holds if and only if $\Int\lor\Nil\le r$ or $\Int\lor\at{ok}\le r$. There are
two incomparable minimal solutions, and neither of them is $\Int$. The point is that the
client sends a single return address, and the union of the two kinds of pairs offers either
an address for $\Int\lor\Nil$ or an address for $\Int\lor\at{ok}$, never one address for their
intersection.

The reason is general: $\MT$ respects \eqref{eq:bddl} but not \eqref{eq:bcd}. By
\eqref{eq:pidand} and the distributivity of products over unions,
\begin{align*}
  \MT((\sigma\to\rho)\land(\tau\to\rho))
    &=\pid((\sigma\times\pid(\rho))\lor(\tau\times\pid(\rho)))\\
    &=\pid((\sigma\lor\tau)\times\pid(\rho))=\MT((\sigma\lor\tau)\to\rho)\\
  \MT((\sigma\to\tau)\land(\sigma\to\rho))&=\pid(\sigma\times(\pid(\tau)\lor\pid(\rho)))\\
  \MT(\sigma\to\tau\land\rho)&=\pid(\sigma\times\pid(\tau\land\rho))
\end{align*}
For $\sigma\neq\Tempty$, by contravariance
$\MT((\sigma\to\tau)\land(\sigma\to\rho))\le\MT(\sigma\to\tau\land\rho)$ holds if and only if
$\pid(\tau\land\rho)\le\pid(\tau)\lor\pid(\rho)$, which by the union law holds if and only if
$\tau\le\rho$ or $\rho\le\tau$. In words, the translation maps the two sides of \eqref{eq:bcd},
which are equivalent types, to different types whenever $\tau$ and $\rho$ are incomparable.

Twenty years ago, Mariangiola Dezani-Ciancaglini, Daniele Varacca and I met this very problem
when we were trying to encode the $\lambda$-calculus with semantic subtyping (i.e., CDuce) into the $\pi$-calculus with semantic subtyping, called C$\pi$ \citep{CDV06} (at the
time, I confess, Erlang servers were not on our minds, and WhatsApp was still three years away). We singled out the equation
$(\sigma\to\tau)\land(\sigma\to\tau')=\sigma\to\tau\land\tau'$, the form that \eqref{eq:bcd}
takes in CDuce, as ``for the goals of this work an utmostly interesting equation'', and
observed that ``the Milner-Turner translation of arrow types respects the subtyping relation
in the context of the simply typed $\lambda$-calculus, but it breaks down in the presence of
intersection types''. We also gave the operational reason, which we called
\emph{output-driven overloading}. A process that receives a request together with a return
channel can choose its reply not only according to the type of the request, but also according to the type of the return channel it received, which the $\pi$-calculus
of \citet{CDV08} can test: if it replies with a $\tau$ on a channel of type $\pid(\tau)$ and
with a $\tau'$ on a channel of type $\pid(\tau')$, then it has type
$\MT((\sigma\to\tau)\land(\sigma\to\tau'))$ but not type $\MT(\sigma\to\tau\land\tau')$, unless $\tau\le\tau'$ or $\tau'\le\tau$
\citep[\S4.2]{CDV06}.

Our solution was to scramble the return channel. The return channel of the encoding has the
type of a \emph{$\lambda$-channel}
\[
  \chl(t)=(\neg\pid(\neg t)\lor\pid(\Tany))\land\pid(\Tempty)
\]
which contains $\pid(t)$ together with ``noise'', the channels that accept neither all of $t$
nor all of $\neg t$. Thanks to the noise, $\lambda$-channels satisfy
$\chl(t_1\land t_2)=\chl(t_1)\lor\chl(t_2)$: as we
wrote, the return channel ``must be scrambled by introducing some noise at the type level so
that the receiver cannot gain information by testing the type of the return channel''. The
resulting encoding of CDuce into the $\pi$-calculus preserves typing, subtyping, and reduction
\citep{CDV06}.

Unfortunately, this solution is not available on the BEAM. To reply on a $\lambda$-channel, a
process must first check that the channel is a genuine $\pid(r)$, and the encoding contains a
branch, the ``functional garbage'', that receives the noise and does nothing
\citep[\S5.3]{CDV06}. A BEAM pid carries no type at run time, so this check cannot be
programmed, and a server typed with $\lambda$-channels could reply only where $\lambda$-channels add no noise, that is, where they coincide with pid types\techref{ta-sec:cdv-details}. For the same reason,
output-driven overloading cannot be programmed either, but the types cannot exclude it: the
law $\pid(t_1\land t_2)\le\pid(t_1)\lor\pid(t_2)$ that would restore \eqref{eq:bcd} is false,
since a pid that accepts only the atom \texttt{:foo} is in $\pid(\Int\land\Bool)=\pid(\Tempty)$ but
neither in $\pid(\Int)$ nor in $\pid(\Bool)$.

However, there is one more idea in \citet{CDV06}, and it is the one we need. The rule that
types functions requires the arrows of their interfaces to have disjoint domains, a
restriction that ``is necessary (but not sufficient) in order to avoid the problem of
output-driven overloading''. It causes no loss of generality, since every interface can be put
in this form by iterating the rewriting
\[
  (\sigma_h\to\tau_h)\land(\sigma_k\to\tau_k)\ \leadsto\
  (\sigma_h\land\sigma_k\to\tau_h\land\tau_k)\land(\sigma_k\setminus\sigma_h\to\tau_k)\land(\sigma_h\setminus\sigma_k\to\tau_h)
\]
for every pair of arrows whose domains intersect. The rewriting applies \eqref{eq:bcd} to the
common part of the two domains, and the next section applies it to the type of
\texttt{handle\_call}.

\section{Overloaded servers, or how the axiom is regained}\label{sec:server-of}

So let us apply the law before the encoding. Let the type of the \texttt{handle\_call} of a
module $m$ be $\bigwedge_{i\in I}(c_i\to r_i)$. For every nonempty $P\subseteq I$, let
\[
  \textstyle C_P=\bigwedge_{i\in P}c_i\setminus\bigvee_{j\in I\setminus P}c_j
  \qquad\qquad
  R_P=\bigwedge_{i\in P}r_i
\]
The \emph{region} $C_P$ contains the requests that are in the domains of exactly the arrows in
$P$, and $R_P$ is the reply that these arrows guarantee together. The regions are pairwise
disjoint, and $c_i=\bigvee_{P\ni i}C_P$ for every $i$. Hence, by \eqref{eq:bddl} and then by
\eqref{eq:bcd},
\[
  \textstyle
  \bigwedge_{i\in I}(c_i\to r_i)
  =\bigwedge_{i\in I}\bigwedge_{P\ni i}(C_P\to r_i)
  =\bigwedge_{P}\bigwedge_{i\in P}(C_P\to r_i)
  =\bigwedge_{P}(C_P\to R_P)
\]
This is the normal form that the rewriting of \citet{CDV06} computes: its domains, the
regions, are pairwise disjoint, and the last equality above, which uses \eqref{eq:bcd},
gathers in $R_P$ the replies that all the arrows of $P$ guarantee on $C_P$.

Consider now a server started from $m$, that is, the process spawned by
\texttt{start\_link($m$, $v$)}, whose pid \texttt{start\_link} returns. I write $\so(m)$ for the
type of this pid, and I define it as the translation of the normal form
$\bigwedge_P(C_P\to R_P)$ of the type of \texttt{handle\_call}, where the empty regions can be
dropped:\footnote{~If $C_P=\Tempty$ then
  $\srv(C_P,R_P)=\pid(\Tempty\times\pid(R_P))=\pid(\Tempty)$, the largest pid type, which
  leaves the intersection unchanged as long as some region is nonempty, that is, $\bigvee_ic_i\neq\Tempty$.}
\[
  \textstyle\so(m)=\MT\big(\bigwedge_P(C_P\to R_P)\big)=\bigwedge_{P\,:\,C_P\neq\Tempty}\srv(C_P,R_P)
\]
The order of the two steps is the point. The translation cannot apply
\eqref{eq:bcd} afterwards: as we saw in Section~\ref{sec:lost}, for incomparable $\tau$ and
$\rho$ it maps $(\sigma\to\tau)\land(\sigma\to\rho)$ to the type of a server that accepts a
return address for $\tau$ or one for $\rho$, but not one for $\tau\land\rho$. Applying
\eqref{eq:bcd} first computes the intersections $R_P$ while they are still intersections of
replies, and leaves to the translation arrows with disjoint domains, on which, as we shall
see, it loses nothing. Thus \texttt{start\_link}, applied to $m$, returns a value of type
$\{\at{ok},\so(m)\}$, a type that depends on $m$ and that the module types of
\citet{Bretagne26} and \citet{BC26} can express\techref{ta-sec:op}.

Take a module $m$ whose \texttt{handle\_call} has the type of $h$ in equation
\eqref{eq:overlap} of Section~\ref{sec:lost}. The regions are \texttt{:a} (in the first
domain only), \texttt{:b} (in both) and \texttt{:c} (in the second only), and
\[
  \so(m) = \srv(\at{a},\Int\lor\Nil)\land\srv(\at{b},\Int)\land\srv(\at{c},\Int\lor\at{ok})
\]
and \texttt{call} on \texttt{:b} returns an integer, as $h$ applied to \texttt{:b} does. For
\texttt{MyServer}, and in general when the domains are pairwise disjoint (one arrow per clause
with pairwise disjoint request patterns, which is the common case), the regions are the clauses and
$\so(m)$ is the translation $\bigwedge_i\srv(c_i,r_i)$ of Section~\ref{sec:lost}.

Two claims justify the definition. First, $\so$ is \emph{exact}: if the $C_j$ ($j\in J$) are
pairwise disjoint, then
\[
  \textstyle\bigwedge_{j\in J}(C_j\to R_j)\le c\to r
  \quad\iff\quad
  \bigwedge_{j\in J}\srv(C_j,R_j)\le\srv(c,r)
\]
because both sides hold if and only if $c\le\bigvee_{j\in J}C_j$ and $R_j\le r$ for every $j$ such
that $c\land C_j\neq\Tempty$: on the left by the decomposition of arrows of \citet{FCB08}, on the
right by the decomposition of products and the union law. The regions are pairwise disjoint, so
the equivalence applies to them, and its right-hand side is then $\so(m)\le\srv(c,r)$.
Consequently \texttt{call}, on a request of type $c\le\bigvee_ic_i$, returns exactly $\bigvee_{P\,:\,c\land C_P\neq\Tempty}R_P$, the
type of the application of \texttt{handle\_call} to $c$\techthm{Claim}{ta-claim:exact}; and on a
request outside the domain of \texttt{handle\_call}, \texttt{call} is rejected, as it
should be. Second, $\so$ is \emph{optimal}: $\so(m)$ is below the translation of every
intersection of arrows equivalent to the type of \texttt{handle\_call}, so it is the most
precise type that the translation of such an intersection can give to the
server\techthm{Claim}{ta-claim:least}.\footnote{~\label{fn:optimality}The restriction to intersections of arrows
  matters, since $\MT$ also commutes with negation. For instance, if
  $\mathtt{handle\_call}:\at{a}\to\Int$, then the type
  $(\at{a}\to\Int)\land\neg((\at{a}\to\at{x})\land(\at{a}\to\at{y})\land\neg(\at{a}\to\Tempty))$ is
  equivalent to $\at{a}\to\Int$ (the negated part is empty), but its translation is strictly
  smaller than $\so(m)=\srv(\at{a},\Int)$, as the union law shows.}

Exactness is what checks the API against the implementation. The API declares the type of
the server, either as $\so(\mathtt{MyServer})$ or by hand, and \texttt{start\_link} checks that
$\so(m)$ is below it. A type written by hand with the old reply,
$\srv(\at{get\_int},\Int)\land\cdots$, is rejected after Daniel's change: since the two
requests are disjoint, exactness reduces
$\so(\mathtt{MyServer})\le\srv(\at{get\_int},\Int)$ to $\Int\lor\Nil\le\Int$, which is false.

What is, then, an overloaded process? Appendix~D of the extended version of \citet{CDV06}
translates channel types back into function types, reading $\pid(t)$ as $t\to\unit$, the type
of a continuation, and \citet{YHM26} build pid types on this reading. Then $\srv(c,r)$ reads
$(c\times(r\to\unit))\to\unit$, the type of $c\to r$ in continuation-passing style. Law
\eqref{eq:pidand}, read on continuations, becomes
$\bigwedge_i(t_i\to\unit)=(\bigvee_it_i)\to\unit$; applied with $t_i=c_i\times(r_i\to\unit)$,
it turns the translation of an intersection of arrows into a single arrow:
\[
  \textstyle
  \bigwedge_{i\in I}\srv(c_i,r_i)
  =\bigwedge_{i\in I}\big((c_i\times(r_i\to\unit))\to\unit\big)
  =\big(\bigvee_{i\in I}(c_i\times(r_i\to\unit))\big)\to\unit
\]
A process is therefore never overloaded on its result: its overloading lies in its domain
$\bigvee_{i\in I}(c_i\times(r_i\to\unit))$, a union of products in which the type of the request
determines the type of the continuation. Applied to a request in $c_1\land c_2$, a function
returns a value of type $r_1\land r_2$; a process receives the request with a single
continuation, which this union requires to accept all of $r_1$ or all of $r_2$. The regions
add the summand $C_{\{1,2\}}\times((r_1\land r_2)\to\unit)$, which restores the
intersection.

A real \texttt{handle\_call} may also reply later. In a second email (received when most of this essay was already written), Daniel pointed out that
this is common: the server may not have the answer yet, because it waits for another message,
or the answer may be expensive to compute, and the server hands the request to a worker
process. The clause then returns \texttt{\{:noreply, s\}} and keeps the return address
\texttt{from}, in its state or in the message to the worker, and eventually some process calls
\texttt{GenServer.reply(from, v)}, which sends the reply \texttt{v} to \texttt{from} and makes it
the result of the client's \texttt{GenServer.call}. The type of
\texttt{from} must then carry the type of the reply, ``or else we can get unsoundness
again'', as Daniel wrote: today \texttt{GenServer.reply} accepts any reply. In our setting
the type of \texttt{from} does carry it: the arrow of such a clause types \texttt{from} as
$\pid(r_i)$, so that \texttt{GenServer.reply(from, v)} type-checks only if \texttt{v} has type
$r_i$, wherever \texttt{from} has been stored or sent\techref{ta-sec:deferred}. Daniel's email speaks of
a phantom type, and $\pid(r_i)$ is one.\footnote{~A \emph{phantom type} is a parameterized type whose parameter
  has no counterpart at run time: the type checker tracks it, but no run-time test can
  recover it \citep{LM99}. Such is the parameter $t$ of $\pid(t)$: at run time we can test
  whether a value is a pid, but not which messages the process it denotes accepts. The laws of
  Section~\ref{sec:servers} add contravariance on top of it.}

What matters about this phantom type is its variance. A
clause of \texttt{handle\_call} that does not use \texttt{from} is typed by an arrow that
types \texttt{from} as $\pid()$, as in Section~\ref{sec:servers}, so
the address of a client that waits for an integer, of type $\pid(\Int)$, must also have type
$\pid()$. Contravariance gives exactly this: $\pid(\Int)\le\pid()$ since $\Tempty\le\Int$, but
$\pid(\Int)\not\le\pid(\Tany)$. If $\pid()$ were covariant this would be unsound: $\pid(\Int)$ would be
a subtype of $\pid(\Tany)$, an address for integers would be an address for any value, and
\texttt{GenServer.reply(from, :foo)} would type-check even if the client waits for an integer. An invariant parameter would be sound
but too rigid: $\pid(\Int)$ would not be a subtype of $\pid()$, and such a clause could not
accept the address. With disjoint domains the $\pid(r_i)$
form gives $\so(m)$ again; with overlapping domains it gives only
$\bigwedge_i\srv(c_i,r_i)$, unless the programmer also declares the arrows of the regions
that are lost\techref{ta-sec:deferred}.

\section{\texorpdfstring{$\so$}{server\_of} for free: expansions and intersections \`a la
  Church}\label{sec:poly}

At this point a skeptical reader may object that $\so$ is an ad hoc operator, tailored for
GenServer: the checker must know it, read the type of \texttt{handle\_call}, compute the
regions and build a server type, and another behaviour would need another operator. It is
not: $\so$ need not be defined, since it is what polymorphism gives to the function that
starts a server, provided that instantiation may produce intersections. The library only has
to declare how the type of a callback determines the type of a server, and an ordinary
polymorphic type suffices for this. Take the function \texttt{serve} that starts a server
from the callback itself and from the initial state, that is, \texttt{start\_link} with the
callback in place of the module and without \texttt{init}:\footnote{~The library does not
  need to offer \texttt{serve}, which is used here only to compute a type:
  \texttt{start\_link(m, v)} returns \texttt{\{:ok, serve(\&m.handle\_call/3, s)\}}, where
  \texttt{s} is the state computed by \texttt{m.init(v)}.}
\begin{lstlisting}
def serve(h, s), do: spawn(fn -> loop(h, s) end)
\end{lstlisting}
where \texttt{loop} is the loop of Section~\ref{sec:servers}. Its type is
\[
  \mathtt{serve}:\forall c,r,s.\ \big(\,((c,\pid(),s)\to\{\at{reply},r,s\})\,,\ s\big)\to\srv(c,r)
\]
and it says nothing about regions: a callback for requests $c$ and replies $r$ gives a server
for requests $c$ and replies $r$. The regions come from the instantiation. In the
polymorphic set-theoretic types of \citet{CNXL14}, an application may instantiate a
polymorphic function with a \emph{set} of type substitutions, and the function then has the
intersection of the corresponding instances. This is how one deduces, for instance, that the
polymorphic identity function \texttt{id}, of type $\forall\alpha.\ \alpha\to\alpha$, also has
type $(\Int\to\Int)\land(\Bool\to\Bool)$: instantiating $\alpha$ with $\Int$ gives
$\Int\to\Int$, instantiating $\alpha$ with $\Bool$ gives $\Bool\to\Bool$, and since \texttt{id}
has all the types of its instances, it has their intersection. The same type is obtained by \emph{expanding} the
polymorphic type, in the sense of Section~\ref{sec:intersection}, into
$\forall\alpha_1,\alpha_2.\ (\alpha_1\to\alpha_1)\land(\alpha_2\to\alpha_2)$, and then
instantiating the two fresh variables with $\Int$ and $\Bool$. Now let us apply this to \texttt{serve}:
take one substitution for each region,
$[c:=C_P,\ r:=R_P,\ s:=\state]$, and give \texttt{serve} the intersection of the instances that these substitutions produce:
\[
  \textstyle\mathtt{serve}:\bigwedge_P\big(((C_P,\pid(),\state)\to\{\at{reply},R_P,\state\},\ \state)\to\srv(C_P,R_P)\big)
\]
The type of \texttt{handle\_call} is equivalent to
$\bigwedge_P((C_P,\pid(),\state)\to\{\at{reply},R_P,\state\})$, so it is below the first
component of every domain of this intersection, and, for an initial state \texttt{s} of type
$\state$, $\mathtt{serve}(\mathtt{handle\_call},\mathtt{s})$ has type
$\bigwedge_P\srv(C_P,R_P)$, that is, $\so(m)$.\footnote{~Instantiating \texttt{call} in the
  same way would bring back the axiom~\eqref{eq:bcd}, but with processes this is not sound in
  general: all the instances share the return address that \texttt{call}
  creates\techparref{ta-par:subst}. For \texttt{serve} this is not a
  problem\techthm{Theorem}{ta-thm:start}.} No other set of substitutions does better:
every valid set gives a supertype of $\so(m)$\techref{ta-sec:poly}.

Therefore $\so(m)$ need not be defined: it is the type that the rule for polymorphic
application gives to $\mathtt{serve}(\mathtt{handle\_call},\mathtt{s})$ with the best set of
substitutions. What is specific to GenServer is only the type of \texttt{serve}, that is, the
protocol, which the library declares once and the checker trusts, since the library is
untyped. The regions are not specific to GenServer: they are the normalization of an
overloaded arrow into arrows with disjoint domains, a property of set-theoretic types, and
they apply in the same way to any other signature that derives the type of a server from the
type of a callback from requests to replies. The checker must still
find the substitutions, but this is a generic problem, to which I come back below. A new
behaviour then needs a new signature, not a new operator in the checker.

Sets of substitutions have a history, and this is where Simona's work comes into play.
\citet{CNXL14} describe the problem of choosing them as an instance of ``the problem of
determining expansions for intersection type systems'', with a reference to \citet{CDV80}:
the instantiation of \texttt{serve} at the regions is an expansion of its type, which
duplicates the type into one copy for each region before substituting in each copy. The
calculus of \citet{CNXL14} is explicitly typed, and its related work compares it with the
explicitly typed calculi for intersection types of Simona Ronchi Della Rocca. In
\citet{RDR02}, typing derivations are written as terms, and the typed versions of the same
untyped term are joined by an intersection; \citet{LRDR07} separate the computational content,
an untyped term with marks, from the logical content, a proof term. In the style of
\citet{RDR02}, $\mathtt{serve}(\mathtt{handle\_call},\mathtt{s})$ would be typed as the
intersection of one typed copy of \texttt{serve} for each region; \citet{CNXL14} decorate a single copy with
the set of substitutions $[c:=C_P,\ r:=R_P,\ s:=\state]_{P}$. The two lines that
\citet{CNXL14} join, explicitly typed intersections and semantic subtyping, had already met
once: Simona's paper and the paper on the relevance of semantic subtyping of
Section~\ref{sec:union} appeared in the same volume, the proceedings of ITRS 2002. Finally,
to infer such
substitutions the type reconstruction of \citet{CNXA15} solves \emph{tallying} problems, which
play for set-theoretic types the role that unification \citep{RDR88} plays for intersection
types.

One obstacle remains: unlike \texttt{serve}, \texttt{start\_link} does not receive
\texttt{handle\_call} but a module. With the module types of \citet{Bretagne26} and
\citet{BC26}, the behaviour GenServer is a module type with parameters, among which the type
of the requests and the type of the replies, and a module has this type only at its declared
parameters, here $\bigvee_ic_i$ and $\bigvee_ir_i$, since parameters are invariant\iftechapp\ (see the types for \N{\tt reply:} and \N{\tt call\_msg:} in the code of Appendix~\ref{ta-sec:problem})\fi. The only
valid instantiation then gives $\srv(\bigvee_ic_i,\bigvee_ir_i)$, the union that Daniel's
intuition rejected. Typing \texttt{start\_link} by quantifying over the types of the
callbacks of the module, rather than over the parameters of the behaviour, or changing the
typing of applications, so that a module is accepted at every instance of the behaviour that
its declarations satisfy, restores the
instantiation at the regions, and then \texttt{start\_link} does not need $\so$ either; but
these are sketches that I have not yet checked against the module calculus\techref{ta-sec:open}.
Until they are checked, \texttt{start\_link} keeps the type of Section~\ref{sec:server-of}, with
$\so$, which polymorphism explains but does not yet replace.

\section{Back to WhatsApp}\label{sec:whatsapp}

So, can an Erlang checker close the gap? Nothing in the solution is specific to Elixir, and a
checker such as eqWAlizer needs:
\begin{itemize}
\item the intersection type of \texttt{handle\_call}, which programmers already write as
  overloaded specs;
\item a type $\srv(c,r)$ for the pids of servers, and the polymorphic type of
  \texttt{gen\_server:call};
\item a type for the pid returned by \texttt{start\_link}: $\so$ of the callback module, or a
  server type declared by the API and checked against it;
\item the laws of pid types of Section~\ref{sec:servers}, union law included.
\end{itemize}
eqWAlizer is not built on the set-theoretic types of Section~\ref{sec:union}: it uses part of
the theory of \citet{Lanvin21}, but only for its gradual typing. These ingredients, therefore,
do not all belong to the theory that eqWAlizer currently implements, but they are not too far from it,
and none of them requires a change in
the code of \texttt{gen\_server}: besides the specs that programmers write, they concern the
signatures of the library, that is, those of \texttt{start\_link},
\texttt{call} and \texttt{reply}, which the checker must trust since the library is untyped,
and the checker itself\techref{ta-sec:otp}. In the common case in which the request types of
the arrows are pairwise disjoint (one arrow per clause, with disjoint patterns), moreover, no
region has to be computed: $\so(m)$ is then the plain intersection
$\bigwedge_i\srv(c_i,r_i)$, which can be read directly from the overloaded spec.

Of course, some gaps remain\techref{ta-sec:conditions}. The main one in practice is that
servers are usually registered under a name (an atom, often the name of the module, that the
runtime associates with the pid), and the API calls \texttt{GenServer.call} with the name
instead of the pid, so that the pid of type $\so(m)$ returned by \texttt{start\_link} never
reaches the API; giving types to names would close this gap, for instance through a typed
\texttt{name:} option. Calls on a pid obtained from \texttt{whereis}, which returns the pid
registered under a name, are untyped, and a hot code upgrade, which replaces the code of a
running server, must preserve the type of the server: the server type of the new code must be
below that of the old one, and \texttt{code\_change}, the callback that converts the state, must
map the old state into the new state type. The other callbacks (\texttt{handle\_cast}, for requests without reply, and
\texttt{handle\_info}, for the other messages) raise no new question. The generic state machine \texttt{gen\_statem}
\citep{GenStatemErl} does: besides its data, which plays the role of the state of a GenServer,
it keeps a control state that determines how each event is handled, so its replies depend on
this state: its servers would need types indexed by states, closer to the session types
that Mario Coppo and Mariangiola Dezani-Ciancaglini have studied in the last twenty years
\citep{CDY07,BCDDDY08} than to intersections. But this is another story, maybe for the next
birthday.

\section{Conclusion}\label{sec:conclusion}

All these pages tell a very simple story. \texttt{GenServer} is a library to create servers that implement functions. For that it provides two functions: \textsf{serve($f$)}, which creates a server that implements the function $f$ (it is the function \texttt{serve} of Section~\ref{sec:poly}, stripped of the specificities of \texttt{GenServer}), and \textsf{call($p$, $e$)}, which sends the argument $e$ to the server $p$ and returns the server's result (it is the \textsf{apply} function for servers rather than for functions). \texttt{GenServer} implements the ``functions as processes'' transformation of Robin~\citet{Milner92}: the server \textsf{serve($f$)} is an address on which you deposit a request together with an address on which the server puts its answer; \textsf{call($p$, $e$)} handles the protocol for you, as it sends to the server $p$ the request $e$ with a fresh return address, and returns the result it reads on the latter. In an untyped setting this is all you need.

In a typed world, your $f$'s have a type and your servers have typed mailboxes: $\pid(t)$ is the type of the addresses that accept the messages of type $t$. If $f$ is of type $c\to r$, from calls to replies, then \textsf{serve($f$)} is an address of type $\pid(c\times\pid(r))$: you deposit in it a request of type $c$ and an address that accepts $r$; and since \textsf{serve} works not just with one function but with any function, and \textsf{call} not just with one server but with any server, they have polymorphic types: $\mathsf{serve}:\forall c,r.\ (c\to r)\to\pid(c\times\pid(r))$ and $\mathsf{call}:\forall c,r.\ (\pid(c\times\pid(r)),\,c)\to r$. If you are in a simply typed world, this is all you need: your \textsf{call}s to a server are typed as the applications of the function that the server implements.

But in the world of \texttt{GenServer}, your $f$'s are overloaded: the type of the result depends on the type of the request. Transposed into the world of Mario, Mariangiola and Simona, your $f$ has an intersection type as in \citet{CD78,CD80}, and it is defined by multiple clauses, one for each arrow, as in the multiple typed copies that \citet{RDR02} joins in an intersection. If $f$ is of type $\bigwedge_i(c_i\to r_i)$, then, by the magic of expansion \citep{CDV80}, which turns the type of \textsf{serve} into the intersection of its instances $(c_i\to r_i)\to\pid(c_i\times\pid(r_i))$, \textsf{serve($f$)} has type $\bigwedge_i\pid(c_i\times\pid(r_i))$, that is, by \eqref{eq:pidand}, $\pid(\bigvee_i(c_i\times\pid(r_i)))$. If all $c_i$ are pairwise disjoint, then \textsf{call($p$, $e$)}, with $p$ of this type and $e$ of type $c_k$, has type $r_k$, as the application of $f$ to $e$ does. If the $c_i$ are not pairwise disjoint, this is no longer true: with a request in both $c_h$ and $c_k$, the server accepts a return address for $r_h$ or one for $r_k$, never one for $r_h\land r_k$, which is the type of the reply of $f$. The encoding has lost the axiom \eqref{eq:bcd}, as Mariangiola, Daniele and I found twenty years ago.

But the heroes of this story come once more to the rescue: we can split each arrow of $\bigwedge_i(c_i\to r_i)$ by the law \eqref{eq:bddl}, so that any two arrows have either disjoint domains or the same domain, and then use \eqref{eq:bcd} to merge the arrows with the same domain, so that all the domains are pairwise disjoint. So all you have to do is to expand the type of \textsf{serve} into one copy for each of these new arrows, rather than for each arrow of $f$ (for \texttt{start\_link}, which receives a module, the operator $\so$ applies the two laws directly), and let union, intersection and expansion do the rest: the server gets the type that makes every \textsf{call} as precise as the corresponding application of $f$. Both born in the 1980s, Erlang and the axiom \eqref{eq:bcd} finally meet in WhatsApp.

\section*{Acknowledgments}

My first thanks go to Mario, Mariangiola and Simona, whose work this essay has followed from
Turin to WhatsApp. For their sixtieth birthdays, Rocco De Nicola, Daniele Varacca and I
contributed the semantic subtyping of the $\pi$-calculus \citep{CDV08}, whose channel types
are now the types of pids in the type system that we are designing for Elixir; this essay is
its sequel. With Mariangiola I have written several
papers, and she has always inspired me. Together we encoded the $\lambda$-calculus into the $\pi$-calculus, both with semantic subtyping
\citep{CDV06}, the paper on which a large part of this one rests. We jointly supervised Elena
Giachino, and with her and Luca Padovani we wrote \emph{Foundations of session types}
\citep{CDGP09}. With Luca Padovani, Mariangiola and I studied global types and multiparty sessions
\citep{CDP12}. Earlier, as an editor of Information and Computation, she had handled the
paper by Gang Chen and me on dependent types with subtyping. It was the first time that all
the reports on one of my papers were of such quality, and we dared to thank her in the paper
``for such a wise choice'' of reviewers \citep{CC01} (I was young). At 80, Mariangiola is still an
indefatigable young researcher.

I also thank Daniel Gorín for the problem and for the permission to quote his emails; Luca Padovani and Daniele Varacca, with whom the typing of processes in
Elixir is being developed; Aghilas Boussaa, Antonin Bretagne, and José
Valim, for the module types;
Guillaume Duboc and again José Valim for the type system of Elixir. Finally, I
thank Victor Lanvin, who put me in touch with Roberto Aloi, and Roberto, who invited me to the
Stockholm meetup that he was organizing: this essay was born from the discussions at that
meetup. Roberto Aloi, Mickaël Laurent and José Valim gave me useful suggestions on a draft version of this essay.

Happy eightieth birthday, Mario, Mariangiola and Simona.

\bigskip
\hfill \emph{Paris, le 3 novembre 2026}

\bibliographystyle{plainnat}
\bibliography{cdr80}

@Article{CD78,
  author = 	 {Mario Coppo and Mariangiola Dezani-Ciancaglini},
  title = 	 {A new type assignment for $\lambda$-terms},
  journal = 	 {Archiv f{\"u}r mathematische Logik und Grundlagenforschung},
  year = 	 {1978},
  volume = 	 {19},
  OPTnumber = 	 {19},
  pages = 	 {139--156},
}

@article{CD80,
  author    = {Mario Coppo and
               Mariangiola Dezani-Ciancaglini},
  title     = {An extension of the basic functionality theory for the {\(\lambda\)}-calculus},
  journal   = {Notre Dame Journal of Formal Logic},
  volume    = {21},
  number    = {4},
  pages     = {685--693},
  year      = {1980},
  }

@InCollection{CDV80,
  author = 	 {Mario Coppo and Mariangiola Dezani-Ciancaglini and Betti Venneri},
  title = 	 {Principal Type Schemes and Lambda-calculus Semantics},
  editor = 	 {J. R. Hindley and J. P. Seldin},
  booktitle = 	 {To {H. B.} {C}urry: Essays on Combinatory Logic, Lambda Calculus and Formalism},
  publisher = 	 {Academic Press},
  year = 	 1980,
}

@article{CDV81,
author = {Coppo, Mario and Dezani-Ciancaglini, Mariangiola and Venneri, Betti},
title = {Functional Characters of Solvable Terms},
journal = {Mathematical Logic Quarterly},
volume = {27},
number = {2-6},
pages = {45-58},
doi = {10.1002/malq.19810270205},
year = {1981}
}

@article{BCD83,
author       =   "Henk Barendregt and Mario Coppo and Mariangiola Dezani-Ciancaglini",
title        =   "A filter lambda model and the completeness of type assignment",
journal      =   "Journal of Symbolic Logic", volume=48, number=4, year=1983,
pages        =   "931-940",
checked      =   "Not checked",
source       =   "Cribbed from rwhbib (corroborated by jcmbib)"}

@InCollection{Pottinger80,
  title={A type assignment for the strongly normalizable $\lambda$-terms},
  author={Garrel Pottinger},
  editor={J. R. Hindley and J. P. Seldin},
  year=1980,
  booktitle = 	 {To {H. B.} {C}urry: Essays on Combinatory Logic, Lambda Calculus and Formalism},
  publisher = {Academic Press},
  pages = 	 {561--577}}

@article{RDR88,
	Author = {{Ronchi Della Rocca}, Simona},
	Journal = {Theoretical Computer Science},
	Number = {1-2},
	Pages = {181--209},
	Title = {Principal type scheme and unification for intersection type discipline},
	Volume = {59},
	Year = {1988}}

@article{RDR02,
  author    = {Simona {Ronchi Della Rocca}},
  title     = {Intersection typed $\lambda$-calculus},
  journal   = {Electronic Notes in Theoretical Computer Science},
  volume    = {70},
  number    = {1},
  year      = {2002},
  pages     = {163-181},
  note      = {ITRS\,'02, Intersection Types and Related Systems},
}

@article{LRDR07,
  author    = {Luigi Liquori and Simona {Ronchi Della Rocca}},
  title     = {Intersection-types {\`a} la {Church}},
  journal   = {Inf. Comput.},
  volume    = {205},
  number    = {9},
  year      = {2007},
  pages     = {1371-1386},
}

@article{BDdL95,
author = {Franco Barbanera and Mariangiola Dezani-Ciancaglini and Ugo de'Liguoro},
title = {Intersection and union types: syntax and semantics},
year = {1995},
issue_date = {June 1995},
publisher = {Academic Press, Inc.},
address = {USA},
volume = {119},
number = {2},
issn = {0890-5401},
journal = {Inf. Comput.},
month = jun,
pages = {202--230},
doi       = {10.1006/inco.1995.1086},
}

@article{MPS86,
title = {An ideal model for recursive polymorphic types},
journal = {Information and Control},
volume = {71},
number = {1},
pages = {95-130},
year = {1986},
issn = {0019-9958},
doi = {10.1016/S0019-9958(86)80019-5},
author = {David MacQueen and Gordon Plotkin and Ravi Sethi}
}

@INPROCEEDINGS{BCF03,
	AUTHOR = {Véronique Benzaken and Giuseppe Castagna and Alain Frisch},
	TITLE = {{CD}uce: an {XML}-Centric General-Purpose Language},
        BOOKTITLE = {ICFP~'03, 8th ACM International Conference on Functional Programming},
        PAGES ={51--63},
        PUBLISHER = {ACM Press},
	YEAR = {2003},
        doi = {10.1145/944705.944711},
}

@INPROCEEDINGS{FCB02,
   author = "Alain Frisch and Giuseppe Castagna and V\'eronique Benzaken",
   title = "Semantic {S}ubtyping",
   booktitle={LICS '02, 17th Annual IEEE Symposium on Logic in
     Computer Science},
   doi = {10.1109/LICS.2002.1029823},
   DMI-CATEGORY = {intc},
   PUBLISHER = {IEEE Computer Society Press},
   pages = "137--146",
   year = "2002",
}

@article{FCB08,
 author = {Frisch, Alain and Castagna, Giuseppe and Benzaken, V{\'e}ronique},
 title = {Semantic subtyping: dealing set-theoretically with function, union, intersection, and negation types},
 journal = {Journal of the \acro{ACM}},
 issue_date = {September 2008},
 volume = {55},
 number = {4},
 month = sep,
 year = {2008},
 issn = {0004-5411},
 pages = {19:1--19:64},
 articleno = {19},
 numpages = {64},
 doi = {10.1145/1391289.1391293},
 acmid = {1391293},
 publisher = {\acro{ACM}},
 address = {New York, NY, \acro{USA}},
}

@article{DFGM03,
title = "The Relevance of Semantic Subtyping",
journal = "Electronic Notes in Theoretical Computer Science",
volume = "70",
number = "1",
pages = "88 - 105",
year = "2002",
note = "ITRS '02, Intersection Types and Related Systems",
issn = "1571-0661",
doi = "10.1016/S1571-0661(04)80492-4",
author = "Mariangiola Dezani-Ciancaglini and Alain Frisch and Elio Giovannetti and Yoko Motohama",
}

@TECHREPORT{Forsythe,
    author = {John C. Reynolds},
    title = {Design of the Programming Language {F}orsythe},
    institution = {Carnegie Mellon University},
    number = {CMU-CS-96-146},
    year = {1996}
}

@phdthesis{Pierce92,
author = {Pierce, Benjamin C.},
title = {Programming with Intersection Types and Bounded Polymorphism},
year = {1992},
school = {Carnegie Mellon University},
url = {https://www.cis.upenn.edu/~bcpierce/papers/thesis.pdf},
address = {USA},
}

@article{HP03,
  author    = {Haruo Hosoya and
               Benjamin C. Pierce},
  title     = {{XD}uce: A statically typed {XML} processing language},
  journal   = {ACM Trans. Internet Techn.},
  volume    = {3},
  number    = {2},
  year      = {2003},
  pages     = {117-148},
  doi = {10.1145/767193.767195},
}

@article{CLPS19,
	AUTHOR = {Giuseppe Castagna and Victor Lanvin and Tommaso Petrucciani and Jeremy G.~Siek},
	TITLE = {Gradual Typing: a New Perspective},
        JOURNAL = {Proc. ACM Program. Lang.},
        doi = {10.1145/3290329},
        VOLUME = {3},
	ARTICLENO = {16},
        NUMBER = {{POPL}},
	MONTH = jan,
  	YEAR = {2019},
	NUMPAGES = {32},
	}

@PhdThesis{Lanvin21,
  author = 	 {Victor Lanvin},
  title = 	 {A Semantic Foundation for Gradual Set-Theoretic Types},
  school = 	 {Université de Paris},
  month = nov,
  url =  {https://theses.hal.science/tel-03853222},
  year = 	 2021,
  }

@article{Design24,
  title   = {The {D}esign {P}rinciples of the {E}lixir {T}ype {S}ystem},
  author  = {Giuseppe Castagna and Guillaume Duboc and José Valim},
  journal = {The Art, Science, and Engineering of Programming},
  year    = {2024},
  volume  = {8},
  doi     = {10.22152/programming-journal.org/2024/8/4},
  number  = {2}
}

@article{CLN24,
author = {Giuseppe Castagna and Mickaël Laurent and Kim Nguyen},
title = {Polymorphic Type Inference for Dynamic Languages},
year = {2024},
issue_date = {January 2024},
publisher = {Association for Computing Machinery},
volume = {8},
number = {POPL},
doi = {10.1145/3632882},
journal = {Proc. ACM Program. Lang.},
month = jan,
articleno = {40},
numpages = {},
        SLIDES = {},
	DMI-CATEGORY = {intc}
}

@article{CLNL22,
author = {Giuseppe Castagna and Mickaël Laurent and Kim Nguyen and Matthew Lutze},
title = {On Type-Cases, Union Elimination, and Occurrence Typing},
year = {2022},
issue_date = {January 2022},
publisher = {Association for Computing Machinery},
volume = {6},
number = {POPL},
doi = {10.1145/3498674},
journal = {Proc. ACM Program. Lang.},
month = jan,
articleno = {13},
numpages = {31},
        SLIDES = {},
	DMI-CATEGORY = {intc}
}

@inproceedings{THF08,
 author = {Tobin-Hochstadt, Sam and Felleisen, Matthias},
 title = {The Design and Implementation of {T}yped {S}cheme},
 booktitle = {Proc.\  of the 35th Annual ACM SIGPLAN-SIGACT Symposium on Principles of Programming Languages},
 series = {POPL '08},
 year = {2008},
 doi = {10.1145/1328438.1328486},
 acmid = {1328486},
 publisher = {ACM},
 
}

@phdthesis{Duboc26,
  author = {Guillaume Duboc},
  school = {Université Paris Cité},
  title = {Typing Dynamic Languages with Set-Theoretic Types: The Case of {Elixir}},
  year = {2026},
  url = {https://gldubc.github.io/assets/duboc-phd-thesis-typing-elixir.pdf},
  month = jan,
}

@inproceedings{LS06,
  title     = {Practical type inference based on success typings},
  author    = {Tobias Lindahl and Konstantinos Sagonas},
  booktitle = {ACM-SIGPLAN International Conference on Principles and Practice of Declarative Programming},
  publisher = {Association for Computing Machinery},
  address   = {New York, NY, USA},
  year      = {2006},
  pages     = {167--178}
}

@Article{Milner92,
  author =   {Milner, Robin},
  title =    {Functions as processes},
  journal =      {Mathematical Structures in Computer Science},
  year =     1992,
  volume =   2,
  number =   2,
  pages =    {119--141}
}

@inproceedings{CDV06,
	author = {Giuseppe Castagna and Mariangiola Dezani-Ciancaglini and Daniele Varacca},
	title = {Encoding {CDuce} in the {C}$\pi$-calculus},
	booktitle = {{CONCUR} 2006, 17th.\ International 
                     Conference on Concurrency Theory},
	year = {2006},
	number = {4137},
	series = {LNCS},
	pages = {310-326},
	publisher = {Springer},
  note = {Extended version at \url{http://www.di.unito.it/~dezani/papers/cdv.pdf}}
}

@article{CDV08,
  author  = {Giuseppe Castagna and Rocco {De Nicola} and Daniele Varacca},
  title   = {Semantic subtyping for the $\pi$-calculus},
  journal = {Theoretical Computer Science},
  volume  = {398},
  number  = {1--3},
  pages   = {217--242},
  year    = {2008},
  doi     = {10.1016/j.tcs.2008.01.049},
  note    = {Essays in honour of Mario Coppo, Mariangiola Dezani-Ciancaglini and Simona Ronchi della Rocca. Extended abstract in {LICS}\,'05}
}

@inproceedings{CNXL14,
	AUTHOR = {Giuseppe Castagna and Kim Nguyen and Zhiwu Xu and Hyeonseung Im and Sergueï Lenglet and Luca Padovani},
	TITLE = {Polymorphic Functions with Set-Theoretic Types. {Part 1}: Syntax, Semantics, and Evaluation},
	BOOKTITLE = {Proceedings of the 41st Annual {ACM} {SIGPLAN}-{SIGACT} Symposium on Principles of Programming Languages},
	series = {{POPL}\,'14},
	YEAR = {2014},
        PAGES = {5--17},
	MONTH = jan,
}

@inproceedings{CNXA15,
 author = {Castagna, Giuseppe and Nguyen, Kim and Xu, Zhiwu and Abate, Pietro},
 title = {Polymorphic functions with set-theoretic types. {Part 2}: local type inference and type reconstruction},
 booktitle = {Proceedings of the 42nd Annual \acro{ACM} \acro{SIGPLAN}-\acro{SIGACT} Symposium on Principles of Programming Languages},
 series = {\acro{POPL}\,'15},
 year = {2015},
 pages = {289--302},
 doi = {10.1145/2676726.2676991},
 month = jan,
}

@inproceedings{CDGP09,
  booktitle = {{PPDP'09}},
  OPurl = {http://www.di.unito.it/~dezani/papers/cdgp.pdf},
  title = {Foundations of Session Types},
  author = {Giuseppe Castagna and Dezani-Ciancaglini, Mariangiola and Giachino,
            Elena and Padovani, Luca},
  publisher = {{ACM Press}},
  year = {2009},
  pages={219--230},
  OPnote = {Full version:http://www.di.unito.it/~dezani/papers/cdgpFull.pdf},
}

@inproceedings{CDY07,
  author = {Mario Coppo and Mariangiola Dezani-Ciancaglini and Nobuko Yoshida},
  series = {LNCS 4468},
  booktitle = {FMOODS\,'07},
  title = {{Asynchronous Session Types and Progress for Object-Oriented
           Languages}},
  publisher = {Springer},
  year = {2007},
 }

@inproceedings{BCDDDY08,
  booktitle = {CONCUR\,'08},
  title = {{Global Progress in Dynamically Interleaved Multiparty Sessions}},
  author = {Lorenzo Bettini and Mario Coppo and Loris D'Antoni and Marco {De Luca} and
            Mariangiola Dezani-Ciancaglini and Nobuko Yoshida},
  year = {2008},
  series = {LNCS 5201},
}

@inproceedings{CGL92,
  author    = {Giuseppe Castagna and Giorgio Ghelli and Giuseppe Longo},
  title     = {A calculus for overloaded functions with subtyping},
  booktitle = {Proceedings of the 1992 {ACM} Conference on {LISP} and Functional Programming ({LFP} '92)},
  pages     = {182--192},
  publisher = {ACM},
  address   = {San Francisco},
  year      = {1992},
  doi       = {10.1145/141471.141537}
}

@article{CGL95,
  author  = {Giuseppe Castagna and Giorgio Ghelli and Giuseppe Longo},
  title   = {A calculus for overloaded functions with subtyping},
  journal = {Information and Computation},
  volume  = {117},
  number  = {1},
  pages   = {115--135},
  year    = {1995},
  doi     = {10.1006/inco.1995.1033}
}

@incollection{FSOP,
  author    = {Giuseppe Castagna},
  title     = {Programming with union, intersection, and negation types},
  note      = {Preprint at  \href{https://arxiv.org/abs/2111.03354}{arXiv:2111.03354}},
  month = mar,
  year = 	 2024,
  isbn = {978-3-031-34517-3},
  doi=           {10.1007/978-3-031-34518-0_12},
  publisher = 	 {Springer},
  editor = {Bertrand Meyer},
  pages=         {309--378},
  booktitle = {The French School of Programming}
}

@misc{eqWAlizer,
  author       = {{WhatsApp}},
  title        = {{eqWAlizer}: a type-checker for {Erlang}},
  howpublished = {\url{https://github.com/WhatsApp/eqwalizer}},
  year         = {2022},
  note         = {Since 2026 maintained within the Erlang Language Platform}
}

@misc{Elixir117,
  author       = {{Elixir}},
  title        = {Elixir v1.17 released: set-theoretic types in patterns, calendar durations, and Erlang/OTP 27 support},
  key          = {Elixir documentation},
  howpublished = {\url{https://elixir-lang.org/blog/2024/06/12/elixir-v1-17-0-released}},
  note         = {Accessed June 12, 2024},
  year         = {2024}
}

@misc{Elixir120,
  author       = {{Elixir}},
  title        = {Elixir v1.20 released: now a gradually typed language},
  key          = {Elixir documentation},
  howpublished = {\url{https://elixir-lang.org/blog/2026/06/03/elixir-v1-20-0-released/}},
  note         = {Accessed October 19, 2026},
  year         = {2026}
}

@article{RDRV84,
  author  = {Simona {Ronchi Della Rocca} and Betti Venneri},
  title   = {Principal type schemes for an extended type theory},
  journal = {Theoretical Computer Science},
  volume  = {28},
  number  = {1--2},
  pages   = {151--169},
  year    = {1984}
}

@article{Cas15,
	author = {Giuseppe Castagna},
	title = {Covariance and Contravariance: a fresh look at an
                  old issue (a primer in advanced type systems for
                  learning functional programmers)},
	journal = {Logical Methods in Computer Science},
	year = {2020},
	volume = {16},
	number = {1},
	pages = {15:1--15:58},
}

@article{CC01,
  author  = {Giuseppe Castagna and Gang Chen},
  title   = {Dependent types with subtyping and late-bound overloading},
  journal = {Information and Computation},
  volume  = {168},
  number  = {1},
  pages   = {1--67},
  year    = {2001}
}

@article{CDP12,
  author  = {Giuseppe Castagna and Mariangiola Dezani-Ciancaglini and Luca Padovani},
  title   = {On global types and multi-party sessions},
  journal = {Logical Methods in Computer Science},
  volume  = {8},
  number  = {1:24},
  pages   = {1--45},
  year    = {2012}
}

@unpublished{YHM26,
  author = {Giuseppe Castagna and Luca Padovani and Daniele Varacca and Amine Nasri},
  title  = {You Have Mail: Set-Theoretic Types for Actors},
  note   = {Unpublished preprint},
  url    = {https://www.irif.fr/~gc/papers/elixir-type-processes.pdf},
  year   = {2026}
}

@mastersthesis{Bretagne26,
  author = {Antonin Bretagne},
  title  = {Semantic typing for {Elixir}'s module type system},
  school = {Master Parisien de Recherche en Informatique (MPRI), ENS Paris-Saclay},
  note   = {Report available at \href{https://codeberg.org/EMan/elixir-module-system/raw/commit/ecb07acd804d0cadf49a6dc4e5c7dd5aaa1b92d7/report.pdf}{this link}},
  year   = {2026},
  month  = sep
}

@unpublished{BC26,
  author = {Aghilas Y. Boussaa and Antonin Bretagne and Giuseppe Castagna},
  title  = {Typing modules in {Elixir}},
  note   = {In preparation},
  year   = {2026}
}

@article{CD26,
  author = {Giuseppe Castagna and Guillaume Duboc},
  title  = {Guard analysis and safe erasure gradual typing: a type system for {Elixir}},
  journal = {Logical Methods in Computer Science},
  note   = {To appear. Preprint at \href{https://arxiv.org/abs/2408.14345}{arXiv:2408.14345}},
  year   = {2026}
}

@misc{GenServerErl,
  author       = {{Ericsson}},
  title        = {{gen\_server}: generic server behaviour, {Erlang/OTP} {STDLIB} reference manual},
  howpublished = {\url{https://www.erlang.org/doc/apps/stdlib/gen_server.html}},
  year         = {2026}
}

@misc{GenStatemErl,
  author       = {{Ericsson}},
  title        = {{gen\_statem}: generic state machine behaviour, {Erlang/OTP} {STDLIB} reference manual},
  howpublished = {\url{https://www.erlang.org/doc/apps/stdlib/gen_statem.html}},
  year         = {2026}
}

@misc{GenServerDoc,
  author       = {{Elixir}},
  title        = {{GenServer}: a behaviour module for implementing the server of a client-server relation},
  howpublished = {\url{https://hexdocs.pm/elixir/GenServer.html}},
  year         = {2026}
}

@techreport{Pierce91,
  author      = {Benjamin C. Pierce},
  title       = {Programming with Intersection Types, Union Types, and Polymorphism},
  institution = {Carnegie Mellon University},
  number      = {CMU-CS-91-106},
  year        = {1991}
}

@article{CL17,
  author  = {Giuseppe Castagna and Victor Lanvin},
  title   = {Gradual Typing with Union and Intersection Types},
  journal = {Proc. ACM Program. Lang.},
  volume  = {1, Article 41},
  number  = {ICFP\,'17},
  doi     = {10.1145/3110285},
  month   = sep,
  year    = {2017}
}

@article{Scott76,
  author    = {Dana S. Scott},
  title     = {Data Types as Lattices},
  journal   = {SIAM Journal on Computing},
  volume    = {5},
  number    = {3},
  pages     = {522--587},
  year      = {1976},
  doi       = {10.1137/0205037}
}

@incollection{Scott75,
  author    = {Dana S. Scott},
  title     = {Open problem},
  editor    = {C. B{\"o}hm},
  booktitle = {$\lambda$-Calculus and Computer Science Theory (Proc. Sympos., Rome, 1975)},
  series    = {LNCS},
  volume    = {37},
  publisher = {Springer},
  pages     = {368},
  year      = {1975}
}

@book{SW01,
  author    = {Davide Sangiorgi and David Walker},
  title     = {The $\pi$-calculus: A Theory of Mobile Processes},
  publisher = {Cambridge University Press},
  year      = {2001}
}

@article{MS04,
  author    = {Massimo Merro and Davide Sangiorgi},
  title     = {On Asynchrony in Name-Passing Calculi},
  journal   = {Mathematical Structures in Computer Science},
  volume    = {14},
  number    = {5},
  pages     = {715--767},
  year      = {2004}
}

@inproceedings{LM99,
  author    = {Leijen, Daan and Meijer, Erik},
  title     = {Domain Specific Embedded Compilers},
  booktitle = {Proceedings of the 2nd Conference on Domain-Specific Languages (DSL '99)},
  publisher = {USENIX},
  pages     = {109--122},
  year      = {1999}
}

@misc{ElixirRoadmap,
  author       = {{Elixir}},
  title        = {Gradual set-theoretic types},
  key          = {Elixir documentation},
  howpublished = {\url{https://hexdocs.pm/elixir/gradual-set-theoretic-types.html}},
  note         = {Section ``Roadmap''. Accessed October 29, 2026},
  year         = {2026}
}

\iftechapp
\pagebreak
\appendix
\section{Technical development}\label{app:tech}
\setlength{\emergencystretch}{3em}

This appendix is a technical companion to Sections~\ref{sec:servers}--\ref{sec:poly}: it
gives the definitions, the claims and their proofs that the main text only
states, and a minimal model of GenServer in which the claims become theorems. I assume that
the reader has read the main text.

The running example is the server \texttt{MyServer} of Section~\ref{sec:servers}, decorated
with type annotations. Appendix~\ref{ta-sec:problem} presents it, together with the problem;
Appendix~\ref{ta-sec:op} defines $\so$ and states its properties, and Appendix~\ref{ta-sec:poly}
relates it to polymorphism; Appendix~\ref{ta-sec:deferred} deals with the clauses that reply
later, and Appendix~\ref{ta-sec:conditions} lists the conditions under which all this works.
Appendix~\ref{ta-app:formal} fixes a minimal model of the gen\_server loop on which the claims
become theorems, Appendix~\ref{ta-sec:otp} compares this model with OTP, and
Appendix~\ref{ta-sec:open} lists the questions that I leave open.
Appendix~\ref{ta-sec:proofs} gives the proofs and the technical details.

\subsection{The running example and the problem}\label{ta-sec:problem}

Today the server \texttt{MyServer} of Section~\ref{sec:servers} is ``typed'' with one spec per
clause of \texttt{handle\_call}, that is, with an intersection type, and with \texttt{pid()}
for the server, so that nothing relates its API to its implementation. Here is the same
server with the types that a programmer should write:
\begin{lstlisting}
defmodule MyServer do
  (*@\N{\$behaviour}@*) GenServer(*@\N{[}@*)(*@\N{init\_arg:}@*) integer(), 
                       (*@\N{reply:}@*) integer() or :ok,
                       (*@\N{call\_msg:}@*) :get_int or {:set_int, integer()}(*@\N{]}@*)
  (*@\N{\$opaque}@*) state = %{current: integer()}

  def init(v), do: {:ok, %{current: v}}

  (*@\N{\$}@*) ((:get_int, pid(), state) -> {:reply, integer(), state}) and
    (({:set_int, integer()}, pid(), state) -> {:reply, :ok, state})
  def handle_call(:get_int, _from, s), do: {:reply, s.current, s}
  def handle_call({:set_int, v}, _from, s), do: {:reply, :ok, %{s | current: v}}

  # public API
  (*@\N{\$type}@*) server = GenServer.(*@\N{server\_of}@*)(MyServer)
  # = GenServer.server(:get_int, integer())
  #   and GenServer.server({:set_int, integer()}, :ok)

  (*@\N{\$}@*) (integer()) -> {:ok, server}
  def start_link(v), do: GenServer.start_link(MyServer, v)

  (*@\N{\$}@*) (server) -> integer()
  def get_int(pid), do: GenServer.call(pid, :get_int)            # r = integer()

  (*@\N{\$}@*) (server, integer()) -> :ok
  def set_int(pid, v), do: GenServer.call(pid, {:set_int, v})    # r = :ok
end

{:ok, s} = MyServer.start_link(42)
MyServer.get_int(s)              # 42
:ok = MyServer.set_int(s, 24)
MyServer.get_int(s) + 1          # 25
\end{lstlisting}
In the code, purple marks the syntax that is not in Elixir today, which comes from two
proposals. A line \texttt{\$ $t$} declares the type $t$ of
the function definition that follows: it is the syntax of the type signatures of \citet{Design24},
which the roadmap of Elixir plans to introduce \citep{ElixirRoadmap}. The type of
\texttt{handle\_call} is an intersection, written \texttt{and}, of two arrows, one for each
kind of request, whose domains list the types of the three arguments (\texttt{or}, in the
other lines, is union). The other constructs come from the module types of
\citet{Bretagne26} and \citet{BC26}. In them a behaviour is a module type with type
parameters, and \texttt{\$behaviour} \inst{GenServer}{\prm{$p$}~$t$, \ldots} declares that
the module implements the behaviour GenServer, with its parameters $p$ instantiated at the
types $t$: the type \texttt{init\_arg} of the argument of \texttt{init}, the type
\texttt{reply} of the replies, and the type \texttt{call\_msg} of the requests.
\texttt{\$opaque state = \ldots} defines \texttt{state}, the type of the internal state of
the server, as an abstract type: inside the module it denotes the maps whose field
\texttt{current} is an integer; outside, it is opaque, so that clients cannot inspect the
state. \texttt{\$type} defines a type: here \texttt{server}, the type of the pids of the
servers started from \texttt{MyServer}, which the operator \texttt{GenServer.server\_of}
computes from the type of \texttt{handle\_call} (Appendix~\ref{ta-sec:op}). The comment
under it gives the result, written with the type \texttt{GenServer.server(c, r)}, a parameterized type that the
library declares, written $\srv(c,r)$ below, and the comments of the API give the type $r$
at which \texttt{call} is instantiated.

In general, the reply of a server depends on the request, and \texttt{handle\_call} has an
intersection type
\begin{equation}\label{ta-eq:handle}
  \textstyle\bigwedge_{i\in I}\,((c_i,\ \pid(),\ \state) \to \{\at{reply},\ r_i,\ \state\})
\end{equation}
where $c_i$ is a call's request type and $r_i$ the corresponding reply type: for
\texttt{MyServer}, $c_1=\at{get\_int}$ and $r_1=\Int$, $c_2=\{\at{set\_int},\Int\}$ and
$r_2=\at{ok}$. We want \texttt{call} to return $r_i$ on a request of type $c_i$, so that the
API is checked against the implementation: if the reply of a request changes and the API
does not, the checker must reject the API. This excludes the simplest type for the server,
the one with the union of the requests and the union of the replies: it is sound, but with
it \texttt{call} on \texttt{:get\_int} returns $\Int\lor\at{ok}$ (Section~\ref{sec:servers}).
This was also pointed out in Daniel's message: ``using a union of LHS-types and a union of RHS-types won't be enough
to capture this, as you need to keep the `valid' LHS/RHS pairs somehow in sync''. The second
argument of \texttt{handle\_call} is the return address \texttt{from}. Until
Appendix~\ref{ta-sec:deferred} every arrow types it as $\pid()$, that is, $\pid(\Tempty)$, the
largest pid type, which suits the clauses that \emph{do not} use \texttt{from}, such as those of
\texttt{MyServer}, whose replies the library sends (no message can be sent to a pid of this type). Then the return address and the state
play no role in the type of the server: what matters are the request types $c_i$ and the
reply types $r_i$. Appendix~\ref{ta-sec:deferred} deals with the clauses that \emph{do} use
\texttt{from}.

Let me recall that $\pid(t)$ is the type of the pids that accept the messages of type $t$, and that it satisfies the laws \eqref{eq:contra} (contravariance), \eqref{eq:pidand} (interface views), and \eqref{eq:union} (the union law), stated in Section~\ref{sec:servers}.
The pid of a server that accepts requests of type $c$ and replies of type $r$,
and the function \texttt{call} of the library, have the types
\[
  \srv(c,r) = \pid(c\times\pid(r))
  \qquad\qquad
  \mathtt{call} : \forall c,r.\ (\srv(c,r),\ c)\to r
\]
In OTP the request also carries the atom \texttt{:"\$gen\_call"}, the return address a fresh
reference, and \texttt{start\_link} may also return \texttt{:ignore} or an error; as in
Section~\ref{sec:servers}, I omit them until Appendix~\ref{ta-sec:otp}, which restores them.

\paragraph{The requirement.} Daniel's email continues with a requirement. The stored value
may now be \texttt{nil} (Erlang's \texttt{undefined}); the implementation is updated, but the
API is not. Then a sound checker ``MUST catch this'' (the capitals are his), since a client
that evaluates \texttt{MyServer.get\_int(s) + 1} crashes when the server holds \texttt{nil}.
Here is the second version of \texttt{MyServer}, where both \texttt{init} and
\texttt{set\_int} accept \texttt{nil} (the clauses do not change, just the types):
\begin{lstlisting}
  (*@\N{\$behaviour}@*) GenServer(*@\N{[}@*)(*@\N{init\_arg:}@*) integer() or nil,
                       (*@\N{reply:}@*) integer() or nil or :ok,
                       (*@\N{call\_msg:}@*) :get_int or {:set_int, integer() or nil}(*@\N{]}@*)
  (*@\N{\$opaque}@*) state = %{current: integer() or nil}
  (*@\N{\$}@*) ((:get_int, pid(), state) -> {:reply, integer() or nil, state}) and
    (({:set_int, integer() or nil}, pid(), state) -> {:reply, :ok, state})
  ...
  (*@\N{\$}@*) (server) -> integer()                                       # not updated
  def get_int(pid), do: GenServer.call(pid, :get_int)           # rejected
\end{lstlisting}
The two requests are still disjoint, so there is one region per clause
(Section~\ref{sec:server-of}):
\[
  \so(\mathtt{MyServer})=\srv(\at{get\_int},\Int\lor\Nil)\land\srv(\{\at{set\_int},\Int\lor\Nil\},\at{ok})
\]
Therefore \texttt{call} on \texttt{:get\_int} returns $\Int\lor\Nil$, which is not a subtype
of the declared result \texttt{integer()} of \texttt{get\_int}: this function is rejected, and
the expression \texttt{MyServer.get\_int(s) + 1} cannot crash on
\texttt{nil}. The function \texttt{set\_int}, instead, is still accepted: its requests
\texttt{\{:set\_int, v\}}, with \texttt{v} an integer, are in the region of the second
arrow, whose reply is \texttt{:ok}.

There are two ways to fix this second version. The first one updates the API, declaring for \texttt{get\_int} the type
\texttt{\$ (server) -> integer() or nil} : then \texttt{get\_int} is
accepted, and the error moves to the client, where \texttt{MyServer.get\_int(s) + 1} is
rejected, since \texttt{+} does not accept \texttt{nil}. The second one keeps the API and
changes the server, so that \texttt{get\_int} waits while the stored value is \texttt{nil}:
this is the third version of the example, that we discuss in Appendix~\ref{ta-sec:deferred}.

\subsection{The operator \texorpdfstring{$\so$}{server\_of}}\label{ta-sec:op}

Forget the return address \texttt{\_from} and the state \texttt{s}: through \texttt{handle\_call}, a server started
from a module $m$ implements a function of type $\bigwedge_{i\in I}(c_i\to r_i)$ from requests
to replies, and we want \texttt{call} to return what an application of this function
returns. Section~\ref{sec:lost} shows that the Milner--Turner translation
$\bigwedge_{i\in I}\srv(c_i,r_i)$ may fail to do it when the $c_i$ overlap (see also the
paragraph ``\nameref{ta-par:overlap}'' below), and Section~\ref{sec:server-of} that
applying~\eqref{eq:bcd} before the translation repairs it: we split the requests into
regions, map each region to the intersection of the replies of the arrows whose domains
contain it, and only then translate.

For $P\subseteq I$ nonempty, let $C_P$ be the requests in the domains of exactly the arrows
in $P$, and $R_P$ the reply they guarantee:
\[
  \textstyle C_P = \bigwedge_{i\in P}c_i \setminus \bigvee_{j\in I\setminus P}c_j
  \qquad
  R_P = \bigwedge_{i\in P}r_i
  \qquad
  \so(m) = \bigwedge_{P\,:\,C_P\neq\Tempty}\srv(C_P,R_P)
\]
where $m$ is the module whose \texttt{handle\_call} has the type \eqref{ta-eq:handle} of
Appendix~\ref{ta-sec:problem}.\footnote{I assume, here and in the rest of the appendix, that
$\bigvee_{i\in I}c_i\neq\Tempty$: otherwise \texttt{handle\_call} accepts no request, every
region is empty, and $\so(m)$ is the empty intersection $\Tany$, which is not even a pid type.}
The regions $C_P$ are pairwise disjoint and form a partition of the request domain
$\bigvee_{i\in I}c_i$ of \texttt{handle\_call}: every request in this domain is in exactly one
region, and a server of type $\so(m)$ replies to a request in a region $C_P$ with a value of
type $R_P$, exactly as an application of \texttt{$m$.handle\_call} would do.
Claim~\ref{ta-claim:least} below shows that $\so(m)$ is
the \emph{most precise} type that the Milner--Turner translation can give to such a server,
when it translates an intersection of arrows equivalent to the type of
\texttt{handle\_call} 
(the restriction to intersections of arrows matters: see Footnote~\ref{fn:optimality}).

Then \texttt{start\_link} returns $\so(m)$:
\[
  \mathtt{start\_link} : \forall a.\ (m \named \inst{GenServer}{\prm{init\_arg}~$a$},\ a)\to\{\at{ok},\ \so(m)\}
\]
This signature uses the module types of Appendix~\ref{ta-sec:problem}:
\inst{GenServer}{\prm{init\_arg}~$a$} is the type of the modules that implement GenServer
with the parameter \texttt{init\_arg} instantiated at $a$, whatever the other two. A behaviour
is a type only when all its parameters are instantiated: a parameter that a signature does not
mention stands for a type variable, quantified by the signature and occurring nowhere else. The binder
$m\named T$ says that the first argument has type $T$ and gives it the name $m$, which the
result type uses: $\so(m)$ is computed from the type of the \texttt{handle\_call} of $m$,
which differs from module to module, so \texttt{start\_link} has a dependent type, the type of
a function whose result type depends on the module it receives (it depends only on the type
of this module, which is known statically). The dependency cannot be avoided by writing
the result type with the variables of the parameters \texttt{call\_msg} and \texttt{reply}: since
parameters are invariant, this gives only $\srv(\bigvee_ic_i,\bigvee_ir_i)$, the type with the union of
the requests and the union of the replies of Appendix~\ref{ta-sec:problem}
(see Appendix~\ref{ta-sec:poly}).

The first property of $\so$ is that the translation loses nothing on arrows with pairwise
disjoint domains.

\begin{restatable}{claim}{claimexact}\label{ta-claim:exact}
Let $J\neq\emptyset$, and let the $C_j$ ($j\in J$) be pairwise disjoint. Then the following
conditions are equivalent:
\begin{enumerate}
\item[(i)] $\bigwedge_{j\in J}(C_j\to R_j)\le c\to r$;
\item[(ii)] $\bigwedge_{j\in J}\srv(C_j,R_j)\le\srv(c,r)$;
\item[(iii)] $c\le\bigvee_{j\in J}C_j$, and $R_j\le r$ for every $j\in J$ such that
  $c\land C_j\neq\Tempty$.
\end{enumerate}
\end{restatable}
In words, the subtyping inequalities (i) and (ii) both say what (iii) states: the $C_j$ cover
$c$, and every $C_j$ that $c$ meets guarantees a reply in $r$.

Let us see the claim at work on the second version of \texttt{MyServer}
of Appendix~\ref{ta-sec:problem}, that is the one with the type of \texttt{handle\_call} updated with the \texttt{nil} reply. Suppose that the API had written the type \texttt{server} by
hand, with the old reply that does not include \texttt{nil}:
\[
  \srv(\at{get\_int},\Int)\land\srv(\{\at{set\_int},\Int\},\at{ok})
\]
Then the error moves to \texttt{start\_link}, which checks that $\so(\texttt{MyServer})$ is
below this type, hence below $\srv(\at{get\_int},\Int)$. The regions of the second version are
$C_1=\at{get\_int}$ and $C_2=\{\at{set\_int},\Int\lor\Nil\}$, with replies $R_1=\Int\lor\Nil$
and $R_2=\at{ok}$; the request \texttt{:get\_int} meets $C_1$ only, so the claim reduces
$\so(\mathtt{MyServer})\le\srv(\at{get\_int},\Int)$ to $\Int\lor\Nil\le\Int$, which is false.
The claim is needed here: the contravariance of the return address alone,
$\srv(\at{get\_int},\Int\lor\Nil)\not\le\srv(\at{get\_int},\Int)$, would not suffice to reject
the type written by hand, since an intersection may be below a type even if none of its
components is.

To use the claim on the type of \texttt{handle\_call}, we need it in region form. As
Section~\ref{sec:server-of} shows, the function from requests to replies is equivalent to the
intersection of the arrows of its regions, whose domains are pairwise disjoint:
\begin{equation}\label{ta-eq:regions}
  \textstyle\bigwedge_{i\in I}(c_i\to r_i)\ =\ \bigwedge_{P\,:\,C_P\neq\Tempty}(C_P\to R_P)
\end{equation}
The same holds for the type of \texttt{handle\_call}, whose domains are triples with the same
second and third components:
\begin{multline}\label{ta-eq:regions-h}
  \textstyle\bigwedge_{i\in I}((c_i,\pid(),\state)\to\{\at{reply},r_i,\state\})\\
  \textstyle=\ \bigwedge_{P\,:\,C_P\neq\Tempty}((C_P,\pid(),\state)\to\{\at{reply},R_P,\state\})
\end{multline}
where the domains $(C_P,\pid(),\state)$ are pairwise disjoint.

We can now compute what \texttt{call} returns. Let $\mathtt{p}$ be a server of type $\so(m)$
and $\mathtt{c}$ a request of type $c$. The application $\mathtt{call}(\mathtt{p},\mathtt{c})$
can be given the type $r$ if and only if $\so(m)\le\srv(c,r)$, that is, by
Claim~\ref{ta-claim:exact} applied to the regions, if and only if $c\le\bigvee_ic_i$ and
$R_P\le r$ for every $P$ such that $c\land C_P\neq\Tempty$. Therefore, on a request of type
$c\le\bigvee_ic_i$, \texttt{call} returns the least such $r$,
\[
  \textstyle\bigvee_{P\,:\,c\land C_P\neq\Tempty}R_P
\]
and on a request outside the domain of \texttt{handle\_call} it is rejected. This is exactly
the result of applying the function $\bigwedge_P(C_P\to R_P)$ to $c$: when the domains are
pairwise disjoint, the type of an application is the union of the codomains of the arrows
whose domain meets the argument (a special case of the type of applications in semantic subtyping: cf.\ \cite[\S4.4.3]{Cas15}).

The second property of $\so$ is optimality.

\begin{restatable}{claim}{claimleast}\label{ta-claim:least}
Let $m$ be a module whose \texttt{handle\_call} has the type~\eqref{ta-eq:handle}, so that its
function from requests to replies has type $\bigwedge_{i\in I}(c_i\to r_i)$.
If $\bigwedge_{j\in J}(d_j\to s_j)$ is equivalent to $\bigwedge_{i\in I}(c_i\to r_i)$, then
$\so(m)\le\MT\big(\bigwedge_{j\in J}(d_j\to s_j)\big)$.
\end{restatable}

In other terms, Claim~\ref{ta-claim:least} says that $\so(m)$ is the best type that $\MT$ can
give to the server from an intersection of arrows equivalent to the type of
\texttt{handle\_call}, and Claim~\ref{ta-claim:exact} says that on this type $\MT$ is faithful
with respect to server types, which are all that \texttt{call} and \texttt{start\_link}
compare.

\paragraph{Overlapping domains.}\label{ta-par:overlap} Without the regions, exactness fails.
Take the callback $h$ of~\eqref{eq:overlap} (Section~\ref{sec:lost}), of type
$(\at{a}\lor\at{b}\to\Int\lor\Nil)\land(\at{b}\lor\at{c}\to\Int\lor\at{ok})$: applied to
\texttt{:b} it returns an integer, by~\eqref{eq:bcd}. With the plain translation
$\srv(\at{a}\lor\at{b},\Int\lor\Nil)\land\srv(\at{b}\lor\at{c},\Int\lor\at{ok})$, by the union
law \texttt{call} on \texttt{:b} has two incomparable minimal solutions for $r$,
$\Int\lor\Nil$ and $\Int\lor\at{ok}$, while $\so(m)$ contains the region
$\srv(\at{b},\Int)$, and \texttt{call} returns $\Int$ (Section~\ref{sec:server-of}). The loss
is one of precision only, and it occurs only on such regions.

\begin{restatable}{claim}{claimplain}\label{ta-claim:plain}
Let $\Tm=\bigwedge_{i\in I}\srv(c_i,r_i)$. Then $\so(m)\le\Tm$; for every nonempty region
$C_P$, $\Tm\le\srv(C_P,R_P)$ if and only if $r_i=R_P$ for some $i\in P$; hence $\so(m)$ and
$\Tm$ are equivalent if and only if every nonempty region has such an $i$.
\end{restatable}
Typing the server with the plain translation is thus sound, and so is every solution for $r$
that it gives to \texttt{call}. Since $R_P\le r_i$ for every $i\in P$, it is coarser than
$\so(m)$ exactly on the nonempty regions $C_P$ such that $R_P<r_i$ for every $i\in P$, where the
arrows of $P$ guarantee together strictly more than each of them alone, such as the region
\texttt{:b} of $h$. Overlapping domains are thus not enough for the two types to differ: for
$(\at{a}\lor\at{b}\to\Int)\land(\at{b}\lor\at{c}\to\Int\lor\Nil)$ the region \texttt{:b} has
$R_P=\Int=r_1$, and the two types coincide. With pairwise disjoint domains every nonempty region is some
$C_{\{i\}}=c_i$, with $R_{\{i\}}=r_i$, and the two types coincide.

\subsection{\texorpdfstring{$\so$}{server\_of} from polymorphism, and why \texttt{start\_link} still needs it}\label{ta-sec:poly}

\paragraph{Making $\so$ less ad hoc.} As defined in Appendix~\ref{ta-sec:op}, $\so$ is a
primitive operator that reads the type of \texttt{handle\_call}: it occurs in the type
\texttt{server} of the API (Appendix~\ref{ta-sec:problem}) and in the signature of
\texttt{start\_link}. So it seems tied to GenServer: another behaviour (e.g.,
\texttt{gen\_statem}, a user-defined server) would need a new operator, hence a change of the
checker. The aim of this section is to eliminate this operator, in two steps. First, I consider
the function \texttt{serve} of Section~\ref{sec:poly}, which starts a server from the callback
\texttt{handle\_call} instead of a module: it needs no operator, since its polymorphic type,
instantiated at the regions, gives the type $\so(m)$. Then I discuss how to do the same for
\texttt{start\_link}, whose problem is that it receives a module instead of the callback, and
that the behaviour, as the type of this module, exposes only its parameters, which are
invariant. For the moment this second step fails, so \texttt{start\_link} keeps $\so$, and the
remedies that I propose are only sketches.

\paragraph{A function that takes the callback.} Section~\ref{sec:poly} obtains $\so(m)$ by
polymorphism, from the function that starts a server from the callback itself and from the
initial state:
\begin{lstlisting}
def serve(h, s), do: spawn(fn -> loop(h, s) end)
\end{lstlisting}
where \texttt{loop} is the loop of Section~\ref{sec:servers}. Let me recall the argument. When
\texttt{from} is typed $\pid()$,
\[
  \mathtt{serve} : \forall c,r,s.\ \big((c,\pid(),s)\to\{\at{reply},r,s\},\ s\big)\to\srv(c,r)
\]
and the application rule of \citet{CNXL14} can instantiate it with the \emph{set} of
substitutions $[c:=C_P,\ r:=R_P,\ s:=\state]$, one for each nonempty region, which gives
\texttt{serve} the intersection of the instances:
\[
  \textstyle
  \mathtt{serve} : \bigwedge_P\big(((C_P,\pid(),\state)\to\{\at{reply},R_P,\state\},\ \state)\to\srv(C_P,R_P)\big)
\]
By~\eqref{ta-eq:regions-h}, the type of \texttt{handle\_call} is below the first component of
the domain of every instance, so, for an initial state \texttt{s} of type $\state$,
$\mathtt{serve}(\mathtt{handle\_call},\mathtt{s})$ has type $\bigwedge_P\srv(C_P,R_P)=\so(m)$.
This computes the type, but does not prove it sound: with processes, instantiating a function
with a set of substitutions is unsound in general, since the instances describe different
processes while the application spawns only one, whose mailbox receives the messages of all of
them.\footnote{It is the same phenomenon as
generalizing the type of \texttt{ref []} in ML, where all the instances share one cell. See \parref{ta-par:spawn} for a counterexample.} For \texttt{serve} the type is sound, as Theorem~\ref{ta-thm:start}
shows by typing the loop directly. Moreover, no other set of substitutions does better: every
set whose instances all accept \texttt{handle\_call} and the initial state gives a supertype of
$\so(m)$ (Theorem~\ref{ta-thm:opt}(c)).

\paragraph{The two parts of $\so$.} The polymorphic reading splits $\so$ in two parts, only one
of which is specific to a library (Section~\ref{sec:poly}). Let me give the details for
\texttt{serve}; \texttt{start\_link} comes in the next paragraph.
\begin{itemize}
\item \emph{The protocol} is specific: the shape $\srv(c,r)$ (the tag, the return address, the
  reference) and the relation between a callback type and a server type. The author of the
  library states it once, with a parametric \texttt{\$type} and the signatures of
  \texttt{serve} and \texttt{call}, in the form of the model (I omit the bodies):
\begin{lstlisting}
(*@\N{\$type}@*) server(c, r) = pid({c, pid(r)})

(*@\N{\$}@*) ((c, pid(), s) -> {:reply, r, s}, s) -> server(c, r)
    when c: term(), r: term(), s: term()
def serve(h, s) do: ...

(*@\N{\$}@*) (server(c, r), c) -> r when c: term(), r: term()
def call(p, c) do: ...
\end{lstlisting}
  In a signature, \texttt{when c:\;term(),\;r:\;term()} quantifies the type variables
  \texttt{c} and \texttt{r} over all types, so that \texttt{\$\;$t$ when c:\;term(), r:\;term()}
  is the type $\forall c,r.\,t$, written as in \citet{Design24}. Appendix~\ref{ta-sec:otp}
  gives the OTP form of \texttt{server(c, r)}. The checker cannot verify these signatures, since
  OTP is untyped: they are trusted, like those of a foreign function. No operator appears in
  them: the checker needs only parametric types and the application rule with sets of
  substitutions. A new behaviour started from its callback is then a new signature, together
  with a proof that its instantiation at the regions is sound, which does not follow from the
  signature (for \texttt{serve}, it is Theorem~\ref{ta-thm:start}).
\item \emph{The regions} are not specific: they depend only on the arrows of the callback. For
  any protocol constructor $F(c,r)$ (here $\srv$), the type of the server started from the
  callback is $\bigwedge_P F(C_P,R_P)$, and $P\mapsto(C_P,R_P)$ is the normalization of an
  overloaded arrow into disjoint domains, a property of set-theoretic types alone. So the
  checker would offer this normalization once (or the inference of the set of substitutions would find
  it: here we only exhibit it), and $\so$ is the instance of a generic construction at
  $F=\srv$.
\end{itemize}

\paragraph{Through \texttt{start\_link} it fails.} \texttt{start\_link} does not receive
\texttt{handle\_call} but a module. Typed like \texttt{serve}, it would be
\[
  \begin{array}{l}
  \mathtt{start\_link} : \forall a,c,r.\\
  \quad(\inst{GenServer}{\prm{init\_arg}~$a$, \prm{call\_msg}~$c$, \prm{reply}~$r$},\ a)\to\{\at{ok},\ \srv(c,r)\}
  \end{array}
\]
In the module types of \citet{Bretagne26} and \citet{BC26}, parameters are, for the moment,
invariant: a module $m$ has the type
\inst{GenServer}{$\ldots$, \prm{call\_msg}~$c$, \prm{reply}~$r$} only when $c$ and $r$ are
equivalent to its declared parameters, that is, to $\bigvee_ic_i$ and $\bigvee_ir_i$ when the
declaration agrees with the arrows of \texttt{handle\_call}; for the first version of
\texttt{MyServer}, \texttt{:get\_int or \{:set\_int, integer()\}} and
\texttt{integer() or :ok}. The instance at a region would require the parameters $C_P$ and
$R_P$, which the module does not have unless they are equivalent to $\bigvee_ic_i$ and
$\bigvee_ir_i$. The only valid set of substitutions is then the singleton, and the result is $\srv(\bigvee_i c_i,\bigvee_i r_i)$
(Proposition~\ref{ta-prop:param}): sound, but it loses the correspondence between requests
and replies (Appendix~\ref{ta-sec:problem}). This is why the signature of \texttt{start\_link}
in Appendix~\ref{ta-sec:op} uses $\so(m)$, a type that depends on the module argument $m$:
\begin{lstlisting}
(*@\N{\$}@*) (m (*@\N{::}@*) GenServer(*@\N{[}@*)(*@\N{init\_arg:}@*) a(*@\N{]}@*), a) -> {:ok, (*@\N{server\_of}@*)(m)} when a: term()
def start_link(m, v) do: ...
\end{lstlisting}
This failure concerns the derivation of $\so(m)$ from the parameters, not $\so(m)$ itself: with
this signature, a \texttt{start\_link} that receives \texttt{\_\_MODULE\_\_} or the name of the
module gets the regions, since the type of the module shows the intersection type of
\texttt{handle\_call}. Only a function that receives a module typed by the behaviour loses them
(Appendix~\ref{ta-sec:conditions}).

So, for \texttt{start\_link}, polymorphism explains $\so$ but does not yet replace it, and a new
behaviour started from a module still needs an operator like $\so$ in its signature.
Quantifying over the types of the callbacks of the module, rather than over the parameters of
the behaviour, or changing the typing of applications, so that a module is accepted at every
instance of the behaviour that its declarations satisfy, would restore the instantiation at the
regions and remove $\so$ from the signature. I outline these two remedies in \parref{ta-par:no-so}, though at this stage they are just working hypotheses, since they require further investigation in the module calculus.

\subsection{Deferred replies}\label{ta-sec:deferred}

So far, the clauses of \texttt{handle\_call} reply at once, by returning
$\{\at{reply},r,s\}$. Section~\ref{sec:server-of} explains why a clause may instead return
$\{\at{noreply},s\}$, keep the return address \texttt{from}, and let some process reply later
through \texttt{GenServer.reply(from, v)}, and why the type of \texttt{from} must then carry
the type of the reply, with a contravariant index. Here I give the details.

\paragraph{The type of \texttt{GenServer.reply}.} In the model of Section~\ref{sec:servers},
the return address is a pid and \texttt{GenServer.reply(w, v)} is just \texttt{send(w, v)} (in
OTP the reply also carries a tag: Appendix~\ref{ta-sec:otp}). So we give
\texttt{GenServer.reply} the input type of a send:
\[
  \mathtt{GenServer.reply} : \forall r.\ (\pid(r),\ r)\to\at{ok}
\]
Let us see what this type allows. If \texttt{w} has type $\pid(t)$, the instance at $r$
applies if and only if $\pid(t)\le\pid(r)$, that is, by~\eqref{eq:contra}, contravariance, $r\le t$, and then
\texttt{v} must have type $r$: so the call type-checks exactly when \texttt{v} has type $t$
(take $r=t$). With $\pid()=\pid(\Tempty)$ the only instances are at $r=\Tempty$, and no value
has type $\Tempty$: a body that calls \texttt{GenServer.reply} on a \texttt{from} of type
$\pid()$ is rejected. A clause that uses \texttt{from} therefore declares $\pid(r_i)$ for it,
and \texttt{handle\_call} has type
\[
  \textstyle\bigwedge_{i\in I}\,((c_i,\ \pid(r_i),\ \state) \to
  \{\at{reply},\ r_i,\ \state\}\lor\{\at{noreply},\ \state\})
\]
so that \texttt{GenServer.reply(from, v)} type-checks only if \texttt{v} has type $r_i$,
wherever \texttt{from} has been stored or sent. I call this type the \emph{$\pid(r_i)$ form},
and the type of Appendix~\ref{ta-sec:problem} the \emph{$\pid()$ form}. The programmer may mix
the two, one arrow per kind of request, as in the example below.

\paragraph{The third version of the example.} This gives the second fix of the second version
of \texttt{MyServer} (Appendix~\ref{ta-sec:problem}), the one that keeps the API: \texttt{get\_int}
waits while the stored value is \texttt{nil}, instead of replying \texttt{nil}. The server keeps
the return addresses of the waiting clients in its state, and \texttt{set\_int} wakes them
when it receives an integer. Here are the parts that change (the \texttt{\$behaviour} line and
the API are those of the second version):
\begin{lstlisting}
  (*@\N{\$opaque}@*) state = %{current: integer() or nil, waiting: [pid(integer())]}

  def init(v), do: {:ok, %{current: v, waiting: []}}


  (*@\N{\$}@*) ((:get_int, pid(integer()), state)
                         -> {:reply, integer(), state} or {:noreply, state}) and
    (({:set_int, integer() or nil}, pid(), state)
                         -> {:reply, :ok, state})
  def handle_call(:get_int, from, %{current: nil} = s),
    do: {:noreply, %{s | waiting: [from | s.waiting]}}   # no value yet: keep "from"
  def handle_call(:get_int, _from, %{current: v} = s) when is_integer(v),
    do: {:reply, v, s}
  def handle_call({:set_int, v}, _from, s) do
    for w <- s.waiting, do: GenServer.reply(w, v)        # rejected
    {:reply, :ok, %{s | current: v, waiting: []}}
  end
\end{lstlisting}
The state now has a second field, \texttt{waiting}, the list of the return addresses of the
clients that wait for an integer, which has type \texttt{[pid(integer())]}. The first arrow is in
the $\pid(r_i)$ form, since the first clause stores \texttt{from} to reply later, and its
result is the union of an immediate and a deferred reply; the second arrow is in the $\pid()$
form, since the clauses for \texttt{set\_int} do not use \texttt{from}.

The last clause handles the requests of the second arrow, so in its body \texttt{v} has type
\texttt{integer() or nil} and \texttt{s} has type \texttt{state}. Then \texttt{s.waiting} has
type \texttt{[pid(integer())]}, and in the comprehension \texttt{w <- s.waiting} the variable
\texttt{w} has type \texttt{pid(integer())}. By the type of \texttt{GenServer.reply}, the call
\texttt{GenServer.reply(w, v)} type-checks only if \texttt{v} has type \texttt{integer()};
since \texttt{integer() or nil} is not a subtype of \texttt{integer()}, the clause is rejected.
It must be: with today's spec of \texttt{GenServer.reply}, which accepts any reply, it would be
accepted, and a request \texttt{\{:set\_int, nil\}}, which the server accepts, would wake the
clients waiting in \texttt{get\_int} with \texttt{nil}. The fix is a clause for \texttt{\{:set\_int, nil\}} that leaves them
waiting, before the last clause, which is then guarded by \texttt{is\_integer(v)}, so that in
its body \texttt{v} has type \texttt{integer()} and the call type-checks:
\begin{lstlisting}
  def handle_call({:set_int, nil}, _from, s),
    do: {:reply, :ok, %{s | current: nil}}               # clients keep waiting
  def handle_call({:set_int, v}, _from, s) when is_integer(v) do
    for w <- s.waiting, do: GenServer.reply(w, v)        # accepted
    {:reply, :ok, %{s | current: v, waiting: []}}
  end
\end{lstlisting}
The two requests are disjoint, so there is one region per arrow, and the server has the type
\[
  \so(\mathtt{MyServer})=\srv(\at{get\_int},\Int)\land\srv(\{\at{set\_int},\Int\lor\Nil\},\at{ok})
\]
The API of the second version, with \texttt{\$ (server) -> integer()} for \texttt{get\_int}
(here \texttt{server} is the type defined in the API of Appendix~\ref{ta-sec:problem}, not to
be confused with the parameterized type $\srv(c,r)$), is thus
accepted: deferring the reply lets the stored value be \texttt{nil} without changing the API,
and the checker verifies that no \texttt{nil} reaches a client of \texttt{get\_int}, neither
through the reply of the loop (the second clause replies \texttt{v} only when it is an
integer) nor through \texttt{GenServer.reply}.

\paragraph{The type of the server in the $\pid(r_i)$ form.} The $\pid()$ form is the more
precise one for $\so$ (Appendix~\ref{ta-sec:op}), the $\pid(r_i)$ form is the one that lets the
body use \texttt{from}. Let us see what the second one costs. Write $\Hr$ for the type of
\texttt{handle\_call} in the $\pid(r_i)$ form, and $K_r$ for
$\{\at{reply},r,\state\}\lor\{\at{noreply},\state\}$. Typed like \texttt{handle\_call},
\[
  \mathtt{serve} : \forall c,r,s.\ \big((c,\pid(r),s)\to\{\at{reply},r,s\}\lor\{\at{noreply},s\},\ s\big)
  \to\srv(c,r)
\]
and a substitution $[c:=c',\ r:=r',\ s:=\state]$ is valid for the callback
\texttt{handle\_call} if $\Hr\le(c',\pid(r'),\state)\to K_{r'}$. The substitutions at the
arrows, $[c:=c_i,\ r:=r_i,\ s:=\state]$, are valid whatever the domains, since each of these
arrows is a component of $\Hr$; so
$\mathtt{serve}(\mathtt{handle\_call},\mathtt{s}):\bigwedge_i\srv(c_i,r_i)$. The substitutions
at the regions are another matter.

\begin{restatable}{claim}{claimregion}\label{ta-claim:region}
Let $C_P\neq\Tempty$ and $\state\neq\Tempty$. Then $\Hr\le(C_P,\pid(R_P),\state)\to K_{R_P}$ if and only if $r_i=R_P$
for some $i\in P$.
\end{restatable}
The client of a request in $C_P$ sends an address of type $\pid(R_P)$, and only an arrow of $P$
whose address type is $\pid(R_P)$ itself can accept it. So the substitution at a region is
valid exactly when one of its arrows already has the reply of the region, and then it adds
nothing, since $\srv(c_i,r_i)\le\srv(C_P,R_P)$ when $C_P\le c_i$ and $r_i=R_P$. In the
$\pid(r_i)$ form, the substitutions at the arrows and at the regions thus give only the plain
translation $\Tm=\bigwedge_i\srv(c_i,r_i)$, which, by Claim~\ref{ta-claim:plain}, loses the nonempty
regions $C_P$ such that $R_P<r_i$ for every $i\in P$. The reason is that the arguments of the
arrows are independent: their intersection cannot say that on a request in $C_P$ only values of
type $R_P$ are sent to \texttt{from}. Take, for instance, the callback $h$
of~\eqref{eq:overlap} in the $\pid(r_i)$ form: on \texttt{:b}, the address of the client has
type $\pid(\Int)$, while the first arrow asks for $\pid(\Int\lor\Nil)$ and the second one for
$\pid(\Int\lor\at{ok})$. With pairwise disjoint domains nothing is lost: this is the case of
\texttt{MyServer}, and in general of one arrow per clause with pairwise disjoint request
patterns.

So, on a module whose \texttt{handle\_call} is in the $\pid(r_i)$ form, the operator
\texttt{server\_of} of the signature of \texttt{start\_link} computes $\Tm$, and not the
$\so(m)$ of Appendix~\ref{ta-sec:op}, which nothing justifies there; $\Tm$ is sound
(\parref{ta-par:formhr}). For such a module to implement the behaviour GenServer, the callback
that the behaviour declares for \texttt{handle\_call} must type the return address as
$\pid(\mathit{reply})$, where $\mathit{reply}$ is the parameter of the replies, and not as
$\pid()$ (\parref{ta-par:conform}).

\paragraph{Recovering a lost region.} In the $\pid()$ form the programmer never writes the
regions. In the $\pid(r_i)$ form a lost region $C_P$ is recovered by declaring its arrow:
adding to the type of \texttt{handle\_call} the arrow $(C_P,\pid(R_P),\state)\to K_{R_P}$,
which the checker verifies against the body like the others. This changes neither the regions
nor their replies (\parref{ta-par:lostregion}), and the substitution at the region becomes valid by
Claim~\ref{ta-claim:region}, since the new arrow has the reply $R_P$. For $h$, the
arrow $(\at{b},\pid(\Int),\state)\to K_{\Int}$ is accepted if the body type-checks on
\texttt{:b} with \texttt{from} of type $\pid(\Int)$, and it restores $\srv(\at{b},\Int)$.
Since deferred replies are common, this is the price of precision when domains overlap.

\subsection{Conditions}\label{ta-sec:conditions}

Let me gather the conditions under which the API of a server is checked against its
implementation.
\begin{itemize}
\item The API must type its pid with $\so(m)$, or with a type written by hand that
  \texttt{start\_link} checks to be a supertype of $\so(m)$ (of $\Tm$ in the $\pid(r_i)$ form,
  Appendix~\ref{ta-sec:deferred}); not with \texttt{pid()}.
\item \texttt{handle\_call} must be given an intersection type, whose arrows type
  \texttt{from} as $\pid()$ or as $\pid(r_i)$ (Appendix~\ref{ta-sec:deferred}). The checker
  verifies this signature against the clauses, taking into account that a request is handled by
  the first clause that matches it \citep{CD26}, so $\so$ always receives a valid type of
  \texttt{handle\_call}. The type of the module passed to \texttt{start\_link} must show this
  intersection, not only the callback type declared by \texttt{\$behaviour}. The usual pattern
  satisfies this condition:
  \texttt{start\_link} is defined in the module itself and called on \texttt{\_\_MODULE\_\_}, as in
  \texttt{def start\_link(arg), do:\ GenServer.start\_link(\_\_MODULE\_\_, arg)}. Otherwise, as
  in generic code that receives a module typed by the behaviour,
  $\so(m)=\srv(\mathit{call\_msg},\mathit{reply})$, which is sound but coarse
  (Appendix~\ref{ta-sec:open} sketches two remedies).
\item The clients must call the server through the pid that \texttt{start\_link} returns. In
  practice this is the main limitation: a server is commonly registered under a name
  (\texttt{name:\ \_\_MODULE\_\_}) and its client API calls
  \texttt{GenServer.call(\_\_MODULE\_\_, ...)}, so that the pid of type $\so(m)$ returned by
  \texttt{start\_link} never reaches the client functions. Such calls, and the calls on a pid
  obtained from \texttt{whereis}, are untyped; typing them requires giving the registered name
  the type $\so(m)$, for instance by a typed \texttt{name:} option or by a module attribute
  that declares the type of the server.
\item A hot code upgrade must preserve the type of the server: the server type of the new
  module must be below that of the old one, and \texttt{code\_change} must map the old state
  into the new state type. (Asking that the new \texttt{handle\_call} have a subtype of the
  type of the old one is too strong, since the two state types may differ.)
\end{itemize}

\subsection{A minimal gen\_server model and the statements it satisfies}\label{ta-app:formal}

This section fixes a minimal model of the gen\_server loop, of \texttt{start\_link} and of
\texttt{call}, on which the claims of the previous sections become theorems.

\subsubsection{The calculus}

A direct-style local $\pi$-calculus in the style of \citet{YHM26}: functional expressions
(higher-order, since the loop receives the callback), pairs and tagged tuples, and the
primitives \texttt{spawn}, \texttt{send}, \texttt{receive} (on the mailbox of the running
process) and \texttt{alias}. A mailbox is a multiset of values (the order of messages plays no
role below). The expression \texttt{alias()} returns a fresh name $w$ owned by the caller: a
message sent to $w$ is received by the caller with \texttt{receive\_on($w$)}, never by a
\texttt{receive}, and nobody else can receive on $w$. This is the locality of the BEAM. As in
Appendix~\ref{ta-sec:problem}, the tag \texttt{:"\$gen\_call"} and the references are omitted,
and so are the process links. Types are those
of this appendix. We assume:
\begin{description}
\item[(A1)] The base system is sound: types are preserved by reduction, and no process is
  stuck on a receive or a match that its type excludes. This is an assumption, not a result of
  \citet{YHM26}, whose system is monomorphic and has no aliases: the base system extends it
  with polymorphic functions, aliases and the rule (receive) of Appendix~\ref{ta-sec:base}.
\item[(A2)] The laws~\eqref{eq:contra}--\eqref{eq:union} of $\pid$, recalled in
  Appendix~\ref{ta-sec:problem}, union law included, and the decompositions of arrows and of
  products of \citet{FCB08}, recalled in Appendix~\ref{ta-sec:decomp}.
\item[(A3)] $m$ is a module with state type $\state\neq\Tempty$,
  $\mathtt{init}:(a)\to\{\at{ok},\state\}$ (so $a$ is the argument type of \texttt{init}, the
  instance of the parameter \texttt{init\_arg} in $m$), and \texttt{handle\_call} typed in one of the two
  forms of Appendix~\ref{ta-sec:deferred}, with $\bigvee_{i\in I}c_i\neq\Tempty$:
  \begin{align*}
    \Hpid&=\textstyle\bigwedge_{i\in I}((c_i,\pid(),\state)\to\{\at{reply},r_i,\state\})\\
    \Hr&=\textstyle\bigwedge_{i\in I}((c_i,\pid(r_i),\state)\to
      \{\at{reply},r_i,\state\}\lor\{\at{noreply},\state\})
  \end{align*}
  An intersection that mixes the two forms is below $\Hr$ (\parref{ta-par:mixed}), and is
  treated as $\Hr$. Moreover, the body of \texttt{handle\_call} performs no
  \texttt{receive} and does not use \texttt{self()}: the only receive on the mailbox of the
  server is the one of the loop. (The body may call other servers, since \texttt{call} reads
  only its own alias.)
\end{description}
\pagebreak
\subsubsection{The encoding}

\begin{lstlisting}
def loop(h, s) do
  receive do
    {c, w} ->
      case h.(c, w, s) do
        {:reply, r, s2}  -> send(w, r); loop(h, s2)
        {:noreply, s2}   -> loop(h, s2)
      end
  end
end

def start_link(m, v) do                   # GenServer.start_link
  {:ok, s} = m.init(v)
  {:ok, spawn(fn -> loop(&m.handle_call/3, s) end)}
end

def call(p, c) do                         # GenServer.call
  w = alias()                             # fresh, owned by the caller
  send(p, {c, w})
  receive_on(w)                           # the reply, whatever process sends it
end

def reply(w, r), do: (send(w, r); :ok)    # GenServer.reply
\end{lstlisting}
The loop is that of Section~\ref{sec:servers}, extended with the clauses that do not reply:
Milner's encoding of a function, with the replicated input replaced by a recursive one that
takes one request at a time and threads the state through. In the form $\Hpid$ the
callback types \texttt{from} as $\pid()$, on which nothing can be sent, and always returns
$\{\at{reply},y,s\}$, so that the loop sends the reply itself; the \texttt{:noreply} branch and
\texttt{reply} are used only in the form $\Hr$, where the callback may call \texttt{reply} on \texttt{from}, or hand it to another
process, and return $\{\at{noreply},s\}$.

\subsubsection{Results}

On this model the claims of the previous sections become theorems, which
Appendix~\ref{ta-sec:statements} states and proves. The function \texttt{call} has the
polymorphic type of Appendix~\ref{ta-sec:problem}, and a call typed with the reply type $r$
receives only values of type $r$ (Theorem~\ref{ta-thm:call}). In the form $\Hpid$,
\texttt{start\_link} returns a pid of type $\so(m)$ (Theorem~\ref{ta-thm:start}); in the form
$\Hr$, a pid of type $\Tm$ (\parref{ta-par:formhr}). In a program that uses the pids at these
types, \texttt{handle\_call} is never applied outside its domain
(Corollary~\ref{ta-cor:safe}), and, in the form $\Hpid$, a call returns the reply that
\texttt{handle\_call} computes for its request, of the type computed in
Appendix~\ref{ta-sec:op} (Theorem~\ref{ta-thm:adequacy}). Theorems~\ref{ta-thm:exact}
and~\ref{ta-thm:opt} restate on the model Claims~\ref{ta-claim:exact} and~\ref{ta-claim:least}
and the computation of Appendix~\ref{ta-sec:poly}, and Proposition~\ref{ta-prop:param} the
failure of the instantiation through \texttt{start\_link}.

\subsubsection{What the model leaves out}

\texttt{handle\_cast} and \texttt{handle\_info}; receives and \texttt{self()} in the body of
\texttt{handle\_call}, which (A3) excludes; the other results of \texttt{handle\_call}
(\texttt{\{:stop, ...\}}, and replies with a timeout, \texttt{:hibernate} or
\texttt{\{:continue, ...\}}) and of \texttt{init} (\texttt{\{:stop, ...\}},
\texttt{:ignore}); the process in which \texttt{init} runs (the caller here, the server in
OTP); the monitor that \texttt{GenServer.call} sets on the server; timeouts, crashes and links (so no statement
on liveness: Theorem~\ref{ta-thm:adequacy} is about replies that are sent, not about replies
that must be sent); message order; registered names and \texttt{whereis}; hot code upgrade;
\texttt{gen\_statem}.

\subsection{From the model to OTP: what changes in GenServer}\label{ta-sec:otp}

Let me stress that nothing in this appendix requires a change in the current code of GenServer.
What changes are the
signatures of \texttt{start\_link}, \texttt{call} and \texttt{reply}, which the checker trusts
since the library is untyped Erlang (Elixir's \texttt{GenServer} delegates to the Erlang module
\texttt{gen\_server}), and the checker, which computes $\so(m)$ (Appendix~\ref{ta-sec:op}) or
instantiates with sets of substitutions (Appendix~\ref{ta-sec:poly}). Let us compare the
model of Appendix~\ref{ta-app:formal} with the code of OTP~29.1.1 and Elixir~1.20.4.

\paragraph{The loop.} When a server starts, \texttt{gen\_server} extracts the function
\texttt{handle\_call} from the module, written \texttt{\&m.handle\_call/3} in Elixir, and its
loop applies this function to each request (functions \texttt{server\_data/4} and
\texttt{try\_handle\_call/4} of \texttt{gen\_server.erl}). The loop of OTP thus has the
structure of the \texttt{loop(h, s)} of Sections~\ref{sec:servers} and~\ref{ta-app:formal}: the
function \texttt{loop/4} of \texttt{gen\_server.erl} has among its arguments \texttt{h}, which is
a function and not a module, and the state \texttt{s} of the server, which it replaces with the
new state after each request. The function \texttt{serve} of Appendix~\ref{ta-sec:poly} starts
this loop from a given \texttt{h} and a given initial state \texttt{s}. The API of
\texttt{gen\_server} has no such function, since all its functions that start a server take a
module, but \texttt{gen\_server} performs this step internally, once it has extracted
\texttt{handle\_call} and computed the initial state with \texttt{init}.

\paragraph{The return address.} In the \texttt{call} function of the model the return address is a fresh name
$\mathtt{w}:\pid(r)$, on which the client waits with \texttt{receive\_on(w)}. In OTP it is a pair
\texttt{\{pid, tag\}} formed by the pid of the client and a fresh tag (built on a reference, a value
that is unique in the running system): the request is the message
\texttt{\{:"\$gen\_call", \{pid, tag\}, c\}}, the server sends the reply \texttt{v} to the
client as the message \texttt{\{tag, v\}}, and the client waits for a message that carries
\texttt{tag}. Since the tag is fresh, no other message can be taken for the reply: this is the
guarantee that the model gives with a fresh name, and that the checker trusts. The new
signatures therefore index the tag by the type of the reply:
\[
  \from(r)=\{\pid(),\ \rtag(r)\}
  \qquad\qquad
  \srv(c,r)=\pid(\{\at{"\$gen\_call"},\ \from(r),\ c\})
\]
(written \texttt{from(r)} and \texttt{server(c, r)} in Table~\ref{ta-tab:signatures}), where
$\rtag(r)$ is an abstract type with the three laws of $\pid(r)$
(Appendix~\ref{ta-sec:problem}): together with the pid, the tag is an address for the replies
of type $r$. Since products distribute over intersections and unions, $\from(r)$ has the same laws, and
the claims of Appendices~\ref{ta-sec:op}--\ref{ta-sec:deferred} (with $\MT(c\to r)=\srv(c,r)$),
which use only these laws and the decompositions of products and arrows, carry over to this
definition of $\srv(c,r)$. The arrows of \texttt{handle\_call}
type \texttt{from} as $\from()=\from(\Tempty)$, the largest type of return addresses, or as
$\from(r_i)$; the examples of this appendix write $\pid()$ and $\pid(r_i)$, as in the model.

\paragraph{Today's specs and the new signatures.} Table~\ref{ta-tab:signatures} compares the
specs of GenServer in Elixir~1.20.4 with the new signatures of the module \texttt{GenServer},
written with Elixir types. Only the types change: the functions are those of today. The
type variables are quantified with \texttt{when} as in Appendix~\ref{ta-sec:poly}
(\texttt{\$\,$t$ when a:\;term()} is the type $\forall a.\,t$), and $m\named T$ is the binder of
Appendix~\ref{ta-sec:op}.

\begin{table}[ht]
  \centering\footnotesize\frenchspacing
  \begin{tabular}{@{}>{\raggedright\arraybackslash}p{0.43\textwidth}
                    >{\raggedright\arraybackslash}p{0.55\textwidth}@{}}
    \toprule
    Elixir 1.20.4 & New signatures\\
    \midrule
    \texttt{@type server :: pid | name | \{atom, node\}}
      & \texttt{\$type server(c, r) = pid(\{:"\$gen\_call", from(r), c\})}; names keep today's type\\
    \addlinespace
    \texttt{@type from :: \{pid, tag :: term\}}
      & \texttt{\$opaque reply\_tag(r) = ...}\newline
        \texttt{\$type from(r) = \{pid(), reply\_tag(r)\}}; \newline \texttt{from()}, that is, \texttt{from(none())}, takes the place of today's type\\
    \addlinespace
    \texttt{@spec call(server, term, timeout) :: term}
      & \texttt{\$ (server(c, r), c, timeout()) -> r }
      \newline\hspace*{4mm}\texttt{when c: term(), r: term()}\\
    \addlinespace
    \texttt{@spec reply(from, term) :: :ok}
      & \texttt{\$ (from(r), r) -> :ok when r: term()}\\
    \addlinespace
    \texttt{@spec start\_link(module, term, options) :: on\_start}
      & \texttt{\$ (m }\N{\texttt{::}}\texttt{ }\inst{GenServer}{\prm{init\_arg}~a}\texttt{, a, options()) -> }
      \newline\hspace*{6mm}\texttt{\{:ok, GenServer.}\N{\texttt{server\_of}}\texttt{(m)\} or :ignore or }
      \newline\hspace*{6mm}\texttt{\{:error, term()\} }
      \newline\hspace*{4mm}\texttt{when a: term()}\\
    \addlinespace
    \texttt{@spec cast(server, term) :: :ok}
      & \texttt{\$ (pid(\{:"\$gen\_cast", k\}), k) -> :ok when k: term()}\\
    \addlinespace
    \texttt{@callback handle\_call(request :: term, from, state :: term) :: \{:reply, reply, new\_state\} | ...}
      & \texttt{\$callback handle\_call(call\_msg, from(reply), state)}
      \newline\hspace*{4mm}\texttt{= \{:reply, reply, state\} or}
      \newline\hspace*{4mm}\texttt{\ \ \{:noreply, state\} or ...}\\
    \bottomrule
  \end{tabular}
  \caption{The types of GenServer in Elixir~1.20.4 and the new signatures.}
  \label{ta-tab:signatures}
\end{table}

Let us see how these signatures fit together. The functions of OTP are long (monitors, timeouts,
registered names, the other callbacks), so Figure~\ref{ta-fig:genserver} gives the code of the
model of Appendix~\ref{ta-app:formal}, rewritten with the messages of OTP and the results of
\texttt{init}, and typed with the signatures of Table~\ref{ta-tab:signatures} (without the arguments
\texttt{options()} and \texttt{timeout()}, since the model has neither).

\begin{figure}[!t]
\begin{lstlisting}[stringstyle=\color{atc}]
defmodule GenServer do
  (*@\N{\$opaque}@*) reply_tag(r) = pid({term(), r})
  (*@\N{\$type}@*) from(r) = {pid(), reply_tag(r)}
  (*@\N{\$type}@*) server(c, r) = pid({:"$gen_call", from(r), c})
  # computed by the checker, from the regions (C_P, R_P) of m.handle_call:
  #   server_of(m)   = the intersection of the server(C_P, R_P)
  #   messages_of(m) = the type t such that server_of(m) = pid(t)

  # the behaviour
  (*@\N{\$param}@*) init_arg
  (*@\N{\$param}@*) call_msg
  (*@\N{\$param}@*) reply
  (*@\N{\$opaque}@*) state
  (*@\N{\$callback}@*) init(init_arg) = {:ok, state} or :ignore or {:stop, term()}
  (*@\N{\$callback}@*) handle_call(call_msg, from(reply), state) =
              {:reply, reply, state} or {:noreply, state}

  # the functions
  (*@\N{\$}@*) (m (*@\N{::}@*) GenServer(*@\N{[}@*)(*@\N{init\_arg:}@*) a(*@\N{]}@*), a) ->
      {:ok, (*@\N{server\_of}@*)(m)} or :ignore or {:error, term()} when a: term()
  def start_link(m, v) do
    case m.init(v) do
      {:ok, s} ->
        (*@\N{\$interface}@*) (*@\N{messages\_of}@*)(m)
        {:ok, spawn(fn -> loop(m, s) end)}
      :ignore -> :ignore
      {:stop, reason} -> {:error, reason}
    end
  end

  (*@\N{\$}@*) (m (*@\N{::}@*) GenServer, (*@\N{m.state}@*)) -> none()
  defp loop(m, s) do
    receive do
      {:"$gen_call", {_, tag} = from, c} ->
        case m.handle_call(c, from, s) do
          {:reply, v, s2} -> send(tag, {tag, v}); loop(m, s2)
          {:noreply, s2}  -> loop(m, s2)
        end
    end
  end

  (*@\N{\$}@*) (server(c, r), c) -> r when c: term(), r: term()
  def call(p, c) do
    tag = alias()                     # fresh, of type reply_tag(r)
    send(p, {:"$gen_call", {self(), tag}, c})
    {_, v} = receive_on(tag)          # the reply, whatever process sends it
    v
  end

  (*@\N{\$}@*) (from(r), r) -> :ok when r: term()
  def reply({_pid, tag}, v), do: (send(tag, {tag, v}); :ok)
end
\end{lstlisting}
\caption{The module \texttt{GenServer} of the model, with the messages of OTP and the signatures
of Table~\ref{ta-tab:signatures}.}
\label{ta-fig:genserver}
\end{figure}

The first three declarations of the figure are the types of the table. Today's \texttt{server} becomes the parametric
\texttt{server(c, r)}, written \texttt{GenServer.server(c, r)} outside the module
(Appendix~\ref{ta-sec:problem}). The pair \texttt{from(r)} is the one that \texttt{gen\_server.erl}
declares, a pid and an opaque tag, with the index \texttt{r} added. As for the \texttt{\$opaque} of
Appendix~\ref{ta-sec:problem}, inside \texttt{GenServer} the type \texttt{reply\_tag(r)} denotes its
definition, an alias that accepts the pairs of a tag, of type \texttt{term()} since nobody inspects
it, and a reply of type \texttt{r}, while outside it is abstract. This declaration goes beyond that
of Appendix~\ref{ta-sec:problem} in two ways: it has a parameter, and the abstract type keeps the
three laws of $\pid(r)$, which this definition satisfies (\parref{ta-par:otp-model}). The alias is
what OTP uses on its main path, where the tag contains an alias and the reply is sent to this alias
(functions \texttt{do\_call} and \texttt{reply} of \texttt{gen.erl}). On the other path, a local
call without timeout, the tag is a plain reference and the reply is sent to the pid of the pair:
this is why the table does not give the definition, and for OTP the checker just trusts the
signatures.

As in today's Elixir, a single \texttt{defmodule} defines both a behaviour and the functions of
the library. The behaviour is the part that each module instantiates, with three declarations
that I have not explained yet: \texttt{\$param $p$} declares a parameter, which a module fixes in its
line \texttt{\$behaviour} (e.g. the first three lines of the body of \texttt{MyServer} in Appendix~\ref{ta-sec:problem}); \texttt{\$opaque state}, without a definition, declares a type that each
module defines, as \texttt{MyServer} does right after its behaviour declaration; and
\texttt{\$callback $f$($t_1,\ldots,t_n$) = $t$} requires the function $f$ of each module to have a
type below $(t_1,\ldots,t_n)\to t$. (The dots of the table stand for the other results of OTP,
which the model leaves out.) So each module gives \texttt{handle\_call} its own
intersection, \texttt{((c, from(), state) -> \{:reply, r, state\}) and ...}, with \texttt{from(r)} in
the arrows of the clauses that use \texttt{from} (Appendix~\ref{ta-sec:deferred}), and this
intersection, not the type of the callback, is what $\so$ takes.

The signature of \texttt{start\_link} mentions \texttt{server\_of(m)}, which the checker computes
for each module, so the body cannot be checked against it once and for all:
Theorem~\ref{ta-thm:start} types it for each $m$ (\parref{ta-par:otp-model}). The spawn declares
the interface of the new process, with the syntax of \citet{YHM26}: \texttt{\$interface $t$}, before
a spawn, says that the process accepts the messages of type $t$, and the spawn returns a pid of
type \texttt{pid($t$)}. Here $t$ is \texttt{messages\_of(m)}, which the checker computes together
with \texttt{server\_of(m)}: it is the union of the messages
$\{\at{"\$gen\_call"},\from(R_P),C_P\}$ of the regions, so that \texttt{server\_of(m)} is
\texttt{pid(messages\_of(m))} by the second law of $\pid$. In the result of \texttt{start\_link} only
the \texttt{:ok} case changes. In the code the other two come from \texttt{init}; in OTP the
errors also include a name already taken, for which \texttt{start\_link} returns
\texttt{\{:error, \{:already\_started, pid\}\}}, where \texttt{pid} is the process registered under
that name, whose type is not known. The function
\texttt{start\_link} of an API such as that of \texttt{MyServer} declares this union as well.

In the signature of \texttt{loop}, the binder $m\named\texttt{GenServer}$ says that the first
argument is a module that implements the behaviour, whatever its parameters, \texttt{m.state} is
the type \texttt{state} of this module, and \texttt{none()} says that the function never returns.
By the convention of Appendix~\ref{ta-sec:op}, \texttt{GenServer} without brackets is the instance
at three type variables. This tells the callers which modules \texttt{loop} accepts; as for
\texttt{start\_link}, the body is typed for each $m$, at the type of \texttt{m.handle\_call} and not
at the callback type (\parref{ta-par:otp-model}).
Since \texttt{start\_link} applies \texttt{loop} to its own \texttt{m}, this is the module of the
declaration \texttt{\$interface messages\_of(m)}. This is why \texttt{loop} receives the module and
not, as in the model, the function \texttt{h}: the type of \texttt{h} depends on the module. The
difference does not matter at run time: \texttt{m.handle\_call(c, from, s)} calls the same function
as \texttt{\&m.handle\_call/3}, and OTP keeps both the module and this function in the data of its
loop (record \texttt{server\_data} of \texttt{gen\_server.erl}). In \citet{YHM26} an arrow also has decorations, which say how the function uses the mailbox
of its process; I leave them out, since the mailbox that \texttt{loop} reads is the one that the
\texttt{\$interface} of its spawn declares (\parref{ta-par:otp-model}). In the pattern of
\texttt{loop}, \texttt{\{\_, tag\} = from} binds \texttt{from} to the return address and \texttt{tag}
to the second component of this address.

The signature of \texttt{call} is where \texttt{server(c, r)} is used, and this type is what relates
\texttt{call} to \texttt{start\_link}: the pid that \texttt{start\_link} returns has the type
\texttt{server\_of(m)}, an intersection of such types, and \texttt{call} can be applied to this pid
at \texttt{c} and \texttt{r} if and only if \texttt{server\_of(m)} is below \texttt{server(c, r)},
the condition computed in Appendix~\ref{ta-sec:op}. In the body of \texttt{call}, the alias is created at
the type \texttt{reply\_tag(r)}, where \texttt{r} is the reply type of the instance, and the client
sends it together with its own pid, \texttt{self()}.
This pid travels at the type \texttt{pid()}, to which nobody can send, so the type of the mailbox of
the client need not include the replies. For the same reason \texttt{reply} sends on the tag: with
\texttt{send(pid, \{tag, v\})} its body would be rejected. Thus the bodies of \texttt{call} and
\texttt{reply} are well typed under their signatures (\parref{ta-par:otp-model}). The client
discards the tag, since \texttt{receive\_on} reads only what is sent to the alias; in OTP the reply
arrives in the mailbox of the client, which selects it by the tag (this difference shows only in the
programs at the end of this section).

The code leaves out casts, the requests without reply, as the model does. They are sent as
\texttt{\{:"\$gen\_cast", c\}}: the pid returned by \texttt{start\_link} also has the type
$\pid(\{\at{"\$gen\_cast"},k_m\})$, where $k_m$ is the domain of \texttt{handle\_cast}, and by the
second law of $\pid$ the intersection of this type and $\so(m)$ is the type of a process that
accepts both kinds of messages. With casts, then, \texttt{messages\_of(m)} has this further summand,
and \texttt{server\_of(m)} in the code denotes this intersection, and not only the type of the regions of Appendix~\ref{ta-sec:op} (or $\Tm$ in the
$\pid(r_i)$ form), since otherwise \texttt{cast} could not be applied to the pid
(\parref{ta-par:otp-model}); it is also the type that \texttt{start\_link} compares with a type
written by hand (Appendix~\ref{ta-sec:conditions}). For a module without \texttt{handle\_cast}, such as
\texttt{MyServer}, $k_m$ is $\Tempty$ and the intersection is $\so(m)$. Since a cast has no reply,
the overloading of \texttt{handle\_cast} plays no role. The asynchronous calls of the
Erlang module \texttt{gen\_server} (\texttt{send\_request} and \texttt{receive\_response}),
which Elixir~1.20.4 does not wrap, are typed in the same way, by indexing the request
identifier by the type of the reply. A server also accepts system messages and the messages
for \texttt{handle\_info}, which $\srv(c,r)$ does not list; by contravariance this is harmless,
since a process that accepts more messages has a smaller pid type.

\paragraph{What the signatures do not type.} Calls on a registered name keep today's type
(Appendix~\ref{ta-sec:conditions}). Debugging functions such as \texttt{sys:replace\_state},
which replace the state of a running server, escape the typing, since $\srv(c,r)$ does not
mention the type of the state: they are sound only if the new state has the type of the old
one.

Finally, on OTP two programs that the signatures accept can put a message where a type excludes
it, which cannot happen in the model. A clause in the $\pid(r_i)$ form that replies through
\texttt{from} and also returns \texttt{\{:reply, v, s\}} sends two replies: in the model the second
one stays on the alias, while in OTP it remains in the mailbox of the client after the call has
returned, where a later \texttt{receive} of the client can take it, whatever the type of this
mailbox. And since \texttt{from(r)} is a pair, \texttt{\{\_, tag\} = from; GenServer.reply(\{p, tag\}, v)} is well typed for every pid \texttt{p}: on the path without alias, OTP sends the reply to
\texttt{p}, on which the type \texttt{pid()} forbids sending, and not to the caller.

We are currently working on a layer of behavioral types over the interfaces of \citet{YHM26}: such a system
ensures that for every request there is at most one reply and that the obligation to reply cannot
be discarded, which excludes the first of these programs.

\subsection{Open questions}\label{ta-sec:open}

I close with the questions that this appendix leaves open.

\paragraph{A \texttt{start\_link} without $\so$.}\label{ta-par:no-so} What blocks the generic reading of
Appendix~\ref{ta-sec:poly} for \texttt{start\_link} is not the operator but the way
\texttt{start\_link} receives the module: module parameters are invariant, and the type of
the module shows only the callback type declared by the behaviour. Making the parameters
variant, contravariant in \texttt{call\_msg} and covariant in \texttt{reply} as for an arrow,
would not suffice: the instance at a region would need $\bigvee_ir_i\le R_P$, that is, the
type of $m$ would have to show the intersection type of \texttt{handle\_call}, and not only the
callback type declared by the behaviour (the second condition of
Appendix~\ref{ta-sec:conditions}). A naive solution would be to expose the callbacks, that is, to
quantify over their types rather than over the parameters of the behaviour. A module is then typed
by a record of its exports, open since a module exports more than its callbacks, and covariant in
the type of each function. The signature of \texttt{start\_link} becomes
\[
  \begin{array}{l}
  \mathtt{start\_link} : \forall a,c,r,s.\\
  \quad\big(\{\mathtt{init}{:}\ a\to\{\at{ok},s\},\
  \mathtt{handle\_call}{:}\ (c,\pid(r),s)\to\{\at{reply},r,s\}\lor\{\at{noreply},s\}\},\ a\big)\\
  \qquad\to\{\at{ok},\srv(c,r)\}
  \end{array}
\]
We instantiate it as we did for \texttt{serve}. Let $a_m$ be the argument type of \texttt{init} in
$m$. In the form $\Hpid$, take the set of substitutions $[a:=a_m,\ c:=C_P,\ r:=R_P,\ s:=\state]_{P:C_P\not=\Tempty}$, one
for each nonempty region. The module $m$ has the type of the first argument of every instance: for
\texttt{handle\_call} by~\eqref{ta-eq:regions-h}, since $\pid(R_P)\le\pid()$ and
$\{\at{reply},R_P,\state\}$ is below the codomain of the instance, and for \texttt{init} by condition (A3).
Therefore, applied to $m$ and to an argument of type $a_m$, \texttt{start\_link} returns the type
$\bigwedge_P\{\at{ok},\srv(C_P,R_P)\}=\{\at{ok},\so(m)\}$. In the form $\Hr$, take the substitutions at
the arrows $[a:=a_m,\ c:=c_i,\ r:=r_i,\ s:=\state]$: every arrow of $\Hr$ is the type of the field
\texttt{handle\_call} at one of them, and the result is $\{\at{ok},\Tm\}$, where
$\Tm=\bigwedge_i\srv(c_i,r_i)$ is the plain translation of Claim~\ref{ta-claim:plain}. The behaviour
instantiation \inst{GenServer}{\prm{init\_arg}~$a$} can be kept on top of it, to check the argument
against \texttt{init}. Then \texttt{start\_link} would not need $\so$ either. This is a sketch: I have
not checked it against the module calculus, in particular the instantiation of $s$ with the
opaque \texttt{state} of $m$, and its soundness would rest, as for \texttt{serve}, on
Theorem~\ref{ta-thm:start} (in the form $\Hr$, on \parref{ta-par:formhr}). The solution also has a
weakness: the variable $s$ allows instances that give the state different types,
which is unsound (\parref{ta-par:spawn}). Moreover, the models of \citet{BC26} and \citet{Bretagne26} can
already type a module as the open record of its exports, but do not expose these types to
programmers, so this solution requires changing this design choice. 

A better but more disruptive
solution would be to change the typing of applications. The body of \texttt{start\_link} uses the module only through its
callbacks, whose types it takes from the declarations of the behaviour, with the parameters
instantiated. This is also what the module calculus compares when a module declares a behaviour. In
\citet{Bretagne26}, let $D_{\mathcal{X}}$ be the declarations of a behaviour $\mathcal{X}$ that has type parameters $\overline{x}$ (which thus appear free in $D_{\mathcal{X}}$); a
module with declarations $D$ that declares $\mathcal{X}$ at the types $\overline{t_0}$ has the type
$\{D\}_{\mathcal{X}[\overline{x{=}t_0}]}$, which is well formed only if
$\{D\}\preceq\{D_{\mathcal{X}}\}[\overline{x:=t_0}]$, where $\preceq$ is the subtyping relation on module types.
Nothing in this comparison is tied to the types $\overline{t_0}$. The idea, then, is to keep the
signature of Appendix~\ref{ta-sec:poly} that quantifies over the parameters, that is
\[
  \begin{array}{l}
  \mathtt{start\_link} : \forall a,c,r.\\
  \quad(\inst{GenServer}{\prm{init\_arg}~$a$, \prm{call\_msg}~$c$, \prm{reply}~$r$},\ a)\to\{\at{ok},\ \srv(c,r)\}
  \end{array}
\]
and to type its
applications by the following rule, which I give for a function with one argument (the other
arguments are typed as usual, at each instance), and where $\Gamma$ is the typing environment and the
$\sigma_k$ are substitutions for the variables $\overline{\alpha}$:
\[
  \frac{\begin{array}{c}
    \Gamma\vdash E:\forall\overline{\alpha}.\,(N'{:}\,\mathcal{X}[\overline{x{=}t}])\to t'
    \qquad
    \Gamma\vdash N:\{D\}_{\mathcal{X}[\overline{x{=}t_0}]}\\[1mm]
    \Gamma\vdash\{D\}\preceq\{D_{\mathcal{X}}\}[\overline{x:=t\sigma_k}]\quad\text{for every }k\in K
  \end{array}}
  {\Gamma\vdash E(N):\bigwedge_{k\in K}t'[N/N']\sigma_k}
\]
In words, if $E$ is a polymorphic function that expects a module of behaviour $\mathcal{X}$ instantiated at
$\overline{t}$, and $N$ is a module that declares $\mathcal{X}$ at $\overline{t_0}$, then $E$ can be applied
to $N$ at any set of instances $\sigma_k$ such that the declarations of $N$ are a subtype of those
of $\mathcal{X}$ at the types $\overline{t\sigma_k}$, and the application has the intersection of the results
of these instances.
The module must still declare the behaviour, but the types $\overline{t\sigma_k}$ are no longer
compared with the types $\overline{t_0}$ that it declares. For \texttt{start\_link}, $\mathcal{X}$ is \texttt{GenServer},
$\overline{t}$ are the variables $a,c,r$ of the signature, and $t'$ is $\{\at{ok},\srv(c,r)\}$. In the
form $\Hpid$ the substitutions at the regions $[a:=a_m,\ c:=C_P,\ r:=R_P]$ are the $\sigma_k$ in the rule, they satisfy the premise
by~\eqref{ta-eq:regions-h}, and thus the rule types the result as
$\bigwedge_P\{\at{ok},\srv(C_P,R_P)\}=\{\at{ok},\so(m)\}$, with no $\so$ in the signature. In the form
$\Hr$ the substitutions at the arrows $[a:=a_m,\ c:=c_i,\ r:=r_i]$ satisfy it, and the result is
$\{\at{ok},\Tm\}$. The programmer
would keep writing behaviours, and the state would need no type variable, since $D$ and $D_{\mathcal{X}}$ both
declare \texttt{state} and are compared at the same one. This is a rule for applications, not a rule of
subtyping. It is only an idea, which we will study in the module calculus, together with its soundness.

\paragraph{Sets of substitutions and processes.}\label{ta-par:subst} Since instantiating with
a set of substitutions a function that spawns a process is unsound in general
(\parref{ta-par:spawn}), a new behaviour needs, besides its signature, its own proof that the
instantiation at the regions is sound, as Theorem~\ref{ta-thm:start} is for \texttt{serve}
(Appendix~\ref{ta-sec:poly}). A criterion on the signature, for instance that the instances
agree on everything that the function shares between requests (the state, for
\texttt{serve}), would spare these proofs; I have not looked for one. The same question arises
for \texttt{call}. Take the callback $h$ of~\eqref{eq:overlap}, a server of type $\MT(h)$, and
one instance of \texttt{call} for each of the two arrows of $h$. The arguments of a call on
\texttt{:b} are in the domains of both instances, so \texttt{call} would return the
intersection of their replies, $\Int$: the axiom~\eqref{eq:bcd} would come back, applied to the
arrows of \texttt{call}, and the regions would not be needed to type calls. But the instances
share the single return address of the call, to which they give the types
$\pid(\Int\lor\Nil)$ and $\pid(\Int\lor\at{ok})$. In the $\pi$-calculus of \citet{CDV08},
where a process can test the type of a channel, this is unsound: the server that replies
\texttt{nil} when the return address accepts $\Int\lor\Nil$, and $1$ otherwise, has type
$\MT(h)$, and makes \texttt{call} return \texttt{nil} where the intersection promises $\Int$.
This is the output-driven overloading of Section~\ref{sec:lost} again. Whether the
instantiation of \texttt{call} is sound on the BEAM, which cannot test the type of a pid, is
open.

\subsection{Proofs and technical details}\label{ta-sec:proofs}

This section gives the proofs and the technical details that the previous sections leave
out. Appendix~\ref{ta-sec:decomp} recalls the decompositions of products and of arrows, and
derives from them the two lemmas on which the proofs of the claims rest.
Appendices~\ref{ta-sec:claims} and~\ref{ta-sec:claims-deferred} prove
the claims of Appendices~\ref{ta-sec:op} and~\ref{ta-sec:deferred}, restated with their numbers,
and justify the facts that these sections use without proof. Appendices~\ref{ta-sec:base}
and~\ref{ta-sec:statements} concern the model of Appendix~\ref{ta-app:formal}: the first lists
the rules of the base system that (A1) assumes, the second states and proves the theorems
announced there. Appendix~\ref{ta-sec:cdv-details} compares $\so$ with the translation with
$\lambda$-channels of \citet{CDV06}.

\subsubsection{Preliminaries}\label{ta-sec:decomp}

Besides the laws~\eqref{eq:contra}--\eqref{eq:union} of $\pid$, recalled in
Appendix~\ref{ta-sec:problem}, the proofs use the decompositions of products and of arrows of
\citet{FCB08}. For a finite set $J$:
\begin{align}
  \textstyle t\times s\le\bigvee_{j\in J}(t_j\times s_j)
  &\iff\textstyle\forall Q\subseteq J.\ \ t\le\bigvee_{j\in Q}t_j\ \text{ or }\
  s\le\bigvee_{j\in J\setminus Q}s_j \label{ta-eq:dprod}\\
  \textstyle\bigwedge_{j\in J}(t_j\to s_j)\le t\to s
  &\iff\textstyle\forall Q\subseteq J.\ \ t\le\bigvee_{j\in Q}t_j\ \text{ or }\
  \big(Q\neq J \text{ and } \bigwedge_{j\in J\setminus Q}s_j\le s\big) \label{ta-eq:darrow}
\end{align}
where the second equivalence requires $J\neq\emptyset$, and a union indexed by the empty set is
$\Tempty$. Triples are treated as pairs, $(t,p,s)$ as $t\times(p\times s)$. The laws of $\pid$ and the two decompositions form the
assumption~(A2) of Appendix~\ref{ta-app:formal}.

The first lemma drops the components that are the same in all the tuples of an inequality,
such as the return address $\pid()$ and the state in the type of \texttt{handle\_call}.

\begin{lemma}[common components]\label{ta-lem:common}
Let $J$ be finite and $u\neq\Tempty$.
\begin{enumerate}
\item[(a)] $t\times u\le\bigvee_{j\in J}(t_j\times u)$ if and only if $t\le\bigvee_{j\in J}t_j$.
  The same holds when $u$ is the first component of the pairs, and when it is one of the
  components of triples, the pair of the other two components playing the role of $t$.
\item[(b)] Let $J\neq\emptyset$ and $p,s\neq\Tempty$. Then
  $\bigwedge_{j\in J}((c_j,p,s)\to\{\at{reply},r_j,s\})\le(c,p,s)\to\{\at{reply},r,s\}$ if and
  only if $\bigwedge_{j\in J}(c_j\to r_j)\le c\to r$.
\end{enumerate}
\end{lemma}
\begin{proof}
\emph{(a).} By~\eqref{ta-eq:dprod}, the inequality holds if and only if, for every
$Q\subseteq J$, either $t\le\bigvee_{j\in Q}t_j$ or $u\le\bigvee_{j\in J\setminus Q}u$. Since
$u\neq\Tempty$, the second disjunct holds if and only if $Q\neq J$, so only the condition for
$Q=J$ remains. When $u$ is the first component, exchange the roles of $Q$ and $J\setminus Q$.
For triples, apply~\eqref{ta-eq:dprod} to the outer pair and the case of pairs to its second
disjunct.

\emph{(b).} By~\eqref{ta-eq:darrow}, the first inequality holds if and only if, for every
$Q\subseteq J$, either $(c,p,s)\le\bigvee_{j\in Q}(c_j,p,s)$, or $Q\neq J$ and
$\bigwedge_{j\in J\setminus Q}\{\at{reply},r_j,s\}\le\{\at{reply},r,s\}$. By~(a), the first
disjunct is $c\le\bigvee_{j\in Q}c_j$. Since
$\bigwedge_{j\in J\setminus Q}\{\at{reply},r_j,s\}=\{\at{reply},\bigwedge_{j\in J\setminus Q}r_j,s\}$,
by~(a) the second one is $Q\neq J$ and $\bigwedge_{j\in J\setminus Q}r_j\le r$. These are the
conditions that~\eqref{ta-eq:darrow} gives for the second inequality.
\end{proof}

\paragraph{Regions.} Let $C_P$ and $R_P$ be as in Appendix~\ref{ta-sec:op}. By definition, the
regions are pairwise disjoint, $C_P\le c_i$ if $i\in P$ and $C_P\land c_i=\Tempty$ otherwise,
$c_i=\bigvee_{P\ni i}C_P$, and $R_P\le r_i$ if $i\in P$ (Section~\ref{sec:server-of}).
In~\eqref{ta-eq:regions} the empty regions can be dropped, since $\Tempty\to t$ is the largest
arrow type. By Lemma~\ref{ta-lem:common}(b), applied to each arrow of either side with
$p=\pid()$ and $s=\state$, which is not empty by (A3), \eqref{ta-eq:regions}
gives~\eqref{ta-eq:regions-h} (the lemma applies since $I$ is not empty and, as
$\bigvee_ic_i\neq\Tempty$, neither is the set of the nonempty regions), whose domains $(C_P,\pid(),\state)$ are pairwise disjoint, and
nonempty when $C_P$ is.

The second lemma is the pattern of all the proofs on server types: an intersection of server
types is below $\srv(c,r)$ if and only if the components whose reply is in $r$ cover $c$.

\begin{lemma}[server types]\label{ta-lem:srv}
Let $J$ be finite, and let $J_r=\{j\in J\mid r_j\le r\}$. Then
\[
  \textstyle c\times\pid(r)\le\bigvee_{j\in J}(c_j\times\pid(r_j))
  \iff c\le\bigvee_{j\in J_r}c_j .
\]
Hence, for $J\neq\emptyset$,
$\bigwedge_{j\in J}\srv(c_j,r_j)\le\srv(c,r)$ if and only if $c\le\bigvee_{j\in J_r}c_j$.
\end{lemma}
\begin{proof}
By~\eqref{ta-eq:dprod}, the first inequality holds if and only if, for every $Q\subseteq J$,
either $c\le\bigvee_{j\in Q}c_j$ or $\pid(r)\le\bigvee_{j\in J\setminus Q}\pid(r_j)$. By the
union law~\eqref{eq:union}, the second disjunct holds if and only if $r_j\le r$ for some
$j\in J\setminus Q$, that is, if and only if $J_r\not\subseteq Q$. So the inequality holds if
and only if $c\le\bigvee_{j\in Q}c_j$ for every $Q\supseteq J_r$, that is, for $Q=J_r$, which
implies the other cases. The second equivalence follows, since
$\bigwedge_{j\in J}\srv(c_j,r_j)=\pid(\bigvee_{j\in J}(c_j\times\pid(r_j)))$
by~\eqref{eq:pidand}, and $\pid$ is contravariant~\eqref{eq:contra}.
\end{proof}
For a single component and $c\neq\Tempty$, this is the variance of $\srv$ stated in
Section~\ref{sec:servers}.

\subsubsection{Proofs of the claims on \texorpdfstring{$\so$}{server\_of}}\label{ta-sec:claims}

This section proves Claims~\ref{ta-claim:exact}--\ref{ta-claim:plain} of
Appendix~\ref{ta-sec:op}, justifies the type that \texttt{call} returns, which
Appendix~\ref{ta-sec:op} uses without proof, and characterizes the solutions that the plain
translation gives to \texttt{call}.

\claimexact*
\begin{proof}
\emph{(ii)$\iff$(iii).} By Lemma~\ref{ta-lem:srv}, (ii) holds if and only if
$c\le\bigvee_{j\in J_r}C_j$, where $J_r=\{j\in J\mid R_j\le r\}$. Since the $C_j$ are pairwise
disjoint, this holds if and only if $c\le\bigvee_{j\in J}C_j$ and $c\land C_j=\Tempty$ for
every $j\notin J_r$, which is (iii).

\emph{(i)$\Rightarrow$(iii).} By~\eqref{ta-eq:darrow}, which applies since $J\neq\emptyset$,
(i) holds if and only if, for every $Q\subseteq J$, either $c\le\bigvee_{j\in Q}C_j$, or
$Q\neq J$ and $\bigwedge_{j\in J\setminus Q}R_j\le r$. For $Q=J$ this gives
$c\le\bigvee_{j\in J}C_j$. For $Q=J\setminus\{j\}$ with $c\land C_j\neq\Tempty$, the first
disjunct fails, since $C_j$ is disjoint from the other $C_k$; hence $R_j\le r$.

\emph{(iii)$\Rightarrow$(i).} Let $Q\subseteq J$ be such that $c\not\le\bigvee_{j\in Q}C_j$.
Since $c\le\bigvee_{j\in J}C_j$, there is $j\notin Q$ such that $c\land C_j\neq\Tempty$; hence
$Q\neq J$ and $\bigwedge_{k\in J\setminus Q}R_k\le R_j\le r$.
\end{proof}
The hypothesis $J\neq\emptyset$ excludes a degenerate case: for $J=\emptyset$, (i) and (ii) are
false, since the empty intersection $\Tany$ is neither an arrow type nor a pid type, while
(iii) holds for $c=\Tempty$.

\paragraph{What \texttt{call} returns.} In Appendix~\ref{ta-sec:op},
$\mathtt{call}(\mathtt{p},\mathtt{c})$, with $\mathtt{p}:\so(m)$ and $\mathtt{c}:c$, can be
given the type $r$ if and only if $\so(m)\le\srv(c,r)$: instantiating the request variable of
\texttt{call} at a type $c'$ requires $c\le c'$ and $\so(m)\le\srv(c',r)$, and $c'=c$ is the
best choice, since $\srv$ is contravariant in its first argument. Claim~\ref{ta-claim:exact}
applies to the nonempty regions, of which there is at least one since
$\bigvee_ic_i\neq\Tempty$.

\claimleast*
\begin{proof}
Let $j\in J$. Since the two intersections are equivalent,
$\bigwedge_i(c_i\to r_i)\le d_j\to s_j$, hence, by~\eqref{ta-eq:regions},
$\bigwedge_P(C_P\to R_P)\le d_j\to s_j$, where $P$ ranges over the nonempty regions, of
which there is at least one since $\bigvee_ic_i\neq\Tempty$. By
(i)$\Rightarrow$(ii) of Claim~\ref{ta-claim:exact},
$\so(m)=\bigwedge_P\srv(C_P,R_P)\le\srv(d_j,s_j)$. Hence
\[
  \textstyle\so(m)\le\bigwedge_{j\in J}\srv(d_j,s_j)=\MT\big(\bigwedge_{j\in J}(d_j\to s_j)\big)
\]
since $\MT$ commutes with intersections.
\end{proof}
The proof uses only one half of the equivalence: $\so(m)$ is below the translation of every
intersection of arrows that is a supertype of the type of \texttt{handle\_call}.

\claimplain*
\begin{proof}
Below, $P$ and $P'$ range over the nonempty regions, of which $\so(m)$ is the intersection.

\emph{$\so(m)\le\Tm$.} Let $i\in I$. By Lemma~\ref{ta-lem:srv}, $\so(m)\le\srv(c_i,r_i)$ if and
only if $c_i\le\bigvee_{P'\,:\,R_{P'}\le r_i}C_{P'}$, which holds since
$c_i=\bigvee_{P'\ni i}C_{P'}$ and $R_{P'}\le r_i$ when $i\in P'$.

\emph{$\Tm\le\srv(C_P,R_P)$.} By Lemma~\ref{ta-lem:srv}, this holds if and only if
$C_P\le\bigvee_{i\,:\,r_i\le R_P}c_i$. Since $C_P$ is nonempty, below $c_i$ for $i\in P$ and
disjoint from $c_i$ for $i\notin P$, this holds if and only if $r_i\le R_P$ for some $i\in P$,
that is, since $R_P\le r_i$, if and only if $r_i=R_P$.

\emph{Equivalence.} By the first part, $\so(m)$ and $\Tm$ are equivalent if and only if
$\Tm\le\srv(C_P,R_P)$ for every $P$; conclude by the second part.
\end{proof}

\paragraph{The solutions with the plain translation.} By Claim~\ref{ta-claim:plain}, every
solution for $r$ that the plain translation gives to \texttt{call} is sound. By
Lemma~\ref{ta-lem:srv}, $\Tm\le\srv(c,r)$ if and only if $c\le\bigvee_{i\,:\,r_i\le r}c_i$,
that is, since $c_i=\bigvee_{P\ni i}C_P$ and the regions are pairwise disjoint, if and only if
$c\le\bigvee_ic_i$ and every region $C_P$ that $c$ meets has an arrow $i\in P$ with
$r_i\le r$. So the solutions are the supertypes of the unions
$\bigvee_{P\,:\,c\land C_P\neq\Tempty}r_{\varphi(P)}$, for the functions $\varphi$ that choose an
arrow $\varphi(P)\in P$ in each region met by $c$, while the exact result is
$\bigvee_{P\,:\,c\land C_P\neq\Tempty}R_P$. The two coincide when $c$ meets only regions with
$|P|=1$; otherwise a minimal solution replaces $R_P$, on each region with $|P|\ge2$ that $c$
meets, by the reply $r_{\varphi(P)}$ of a single arrow of $P$.

\subsubsection{Proofs of the claims on deferred replies}\label{ta-sec:claims-deferred}

This section proves Claim~\ref{ta-claim:region} of Appendix~\ref{ta-sec:deferred}, and
justifies three facts that Appendices~\ref{ta-sec:deferred} and~\ref{ta-app:formal} use: an
intersection that mixes the two forms is below the $\pid(r_i)$ form; declaring the arrow of a
lost region changes neither the regions nor their replies; and a module in the $\pid(r_i)$
form conforms to GenServer only if the behaviour types the return address as
$\pid(\mathit{reply})$.

\paragraph{Mixed forms.}\label{ta-par:mixed} By contravariance of the domain and covariance of
the codomain,
\[
  (c,\pid(),s)\to\{\at{reply},r,s\}\ \le\ (c,\pid(r),s)\to\{\at{reply},r,s\}\lor\{\at{noreply},s\}
\]
since $\pid(r)\le\pid()$ by~\eqref{eq:contra}: an arrow in the $\pid()$ form is below the
corresponding arrow in the $\pid(r_i)$ form, so an intersection that mixes the two forms, as in
the third version of \texttt{MyServer} (Appendix~\ref{ta-sec:deferred}), is below the
$\pid(r_i)$ form.

\claimregion*
\begin{proof}
Let $D=(C_P,\pid(R_P),\state)$ and $E_i=(c_i,\pid(r_i),\state)$. For every $Q\subseteq I$,
\[
  \textstyle D\le\bigvee_{i\in Q}E_i\iff r_i\le R_P\text{ for some }i\in Q\cap P.
  \quad(\dagger)
\]
Indeed, by Lemma~\ref{ta-lem:common}(a), which drops the component $\state\neq\Tempty$, and by
Lemma~\ref{ta-lem:srv}, the left-hand side holds if and only if
$C_P\le\bigvee_{i\in Q\,:\,r_i\le R_P}c_i$, that is, if and only if some $i\in Q\cap P$ has
$r_i\le R_P$, since $C_P$ is nonempty, below $c_i$ for $i\in P$ and disjoint from $c_i$ for
$i\notin P$. By~\eqref{ta-eq:darrow}, with $I\neq\emptyset$, $\Hr\le D\to K_{R_P}$ if and only
if, for every $Q\subseteq I$, either $D\le\bigvee_{i\in Q}E_i$, or $Q\neq I$ and
$\bigwedge_{i\in I\setminus Q}K_{r_i}\le K_{R_P}$.

\emph{Only if.} For $Q=I$ the second disjunct is excluded, so $(\dagger)$ gives an $i\in P$
with $r_i\le R_P$, hence $r_i=R_P$, since $R_P\le r_i$.

\emph{If.} Let $i_0\in P$ with $r_{i_0}=R_P$, and let $Q\subseteq I$. If $i_0\in Q$, the first
disjunct holds by $(\dagger)$. Otherwise $Q\neq I$ and
$\bigwedge_{i\in I\setminus Q}K_{r_i}\le K_{r_{i_0}}=K_{R_P}$.
\end{proof}
In particular, when the claim fails for a nonempty region, $\Hr$ is not below the arrow of
that region: the equation~\eqref{ta-eq:regions-h} has no counterpart in the $\pid(r_i)$ form.

\paragraph{Recovering a lost region.}\label{ta-par:lostregion} Adding to $\Hr$ the arrow
$(C_P,\pid(R_P),\state)\to K_{R_P}$ (Appendix~\ref{ta-sec:deferred}) changes neither the regions
nor their replies, since the new domain $C_P$ is included in every $c_i$ with $i\in P$ and
disjoint from the others: the region $C_P$ is now in the domains of the arrows of $P$ and of
the new one, and its reply is $R_P\land R_P=R_P$.

\paragraph{Conformance to the behaviour.}\label{ta-par:conform} A module implements a
behaviour when its type is a subtype of the type of the behaviour. By contravariance of the
domain, an arrow that asks for $\pid(r_i)$ is not below an arrow that offers only $\pid()$:
for a module whose \texttt{handle\_call} is in the $\pid(r_i)$ form to implement GenServer, the
callback that the behaviour declares for \texttt{handle\_call} must therefore type the return
address as $\pid(\mathit{reply})$, where $\mathit{reply}$ is the parameter of the replies. With
the callback type
\[
  (\mathit{call\_msg},\pid(\mathit{reply}),\state)\to\{\at{reply},\mathit{reply},\state\}\lor\{\at{noreply},\state\}
\]
both forms conform, since $r_i\le\mathit{reply}$ gives
$\pid(\mathit{reply})\le\pid(r_i)\le\pid()$; and with only this callback type the server gets
$\srv(\mathit{call\_msg},\mathit{reply})$, as in Appendix~\ref{ta-sec:conditions}.

\subsubsection{The base system}\label{ta-sec:base}

The proofs of Appendix~\ref{ta-sec:statements} use the following rules and properties of the
base system of~(A1): rules for the primitives on processes, (send) to (receive); rules for the
functional part, (case) to (progress); and two properties of names, (names) and (opacity);
the second is used only in the proof of Theorem~\ref{ta-thm:adequacy} and in the paragraph
``\nameref{ta-par:otp-model}''.
\begin{description}
\item[(send)] \texttt{send(p, v)} is well typed if $p:\pid(t)$ and $v:t$ for some $t$.
\item[(spawn)] \texttt{spawn(f)} has type $\pid(M)$ if the body of \texttt{f} is well typed
  in a process whose mailbox has type $M$, that is, whose receives expect messages of
  type $M$.
\item[(alias)] \texttt{alias()} creates a name of type $\pid(t)$, for a $t$ chosen by the
  caller at creation, and then \texttt{receive\_on} on this name has type $t$, the type of
  creation, in the caller, the only process that can receive on it. (A polymorphic
  \texttt{receive\_on} of type $\forall t.\,\pid(t)\to t$ would be unsound, by
  contravariance.)
\item[(receive)] A receive on a mailbox of type $M=\bigvee_kt_k$, a union of products, types
  its branch once for each summand $t_k$, with the variables of the pattern at the
  components of $t_k$. This is sound since a message is a value, which belongs to one of
  the summands, provided that a value can be given every type that contains it; for names,
  this means that a name created at $\pid(t)$ has the type $\pid(s)$ for every $s\le t$ and
  the type $\neg\pid(s)$ for every $s\not\le t$, as in the semantic reading of pid types
  (Section~\ref{sec:servers}). This rule is not the one of \citet{YHM26}, which types the
  variables of the pattern by projection.
\item[(case)] A branch of a \texttt{case} whose pattern is disjoint from the type of the
  scrutinee is never taken, and is not typed.
\item[(app)] An application of a function whose type is an intersection of arrows to an
  argument in the domain of one of the arrows has the type of the codomain of this arrow
  (by subsumption, then application of one arrow). When the domains $D_k$ of the arrows are
  pairwise disjoint and the argument has a type $D\le\bigvee_kD_k$, the application has the
  type $\bigvee_{k\,:\,D\land D_k\neq\Tempty}T_k$, where $T_k$ is the codomain of the $k$-th
  arrow (a special case of the type of applications of \cite[\S4.4.3]{Cas15}).
\item[(inst)] A polymorphic function is instantiated by one substitution at a time, and an
  occurrence is instantiated by the same substitution in every typing of the enclosing body
  (when the body is typed several times, against the arrows of an intersection or the
  summands of (receive)). Sets of substitutions appear only in
  Theorem~\ref{ta-thm:opt}(b) and~(c), which compute types with them but do not claim them
  sound.
\item[(progress)] An application of a function defined by clauses, whose declared type is
  checked against the clauses, to an argument in the domain of this type matches one of the
  clauses.
\item[(names)] If a name (a pid created by \texttt{spawn}, an alias created by
  \texttt{alias}) created at type $\pid(t)$ is used at a type $\pid(s)$, in any process and
  at any step of any run, then $\pid(t)\le\pid(s)$, that is, $s\le t$: this is how subject
  reduction types names.
\item[(opacity)] Code in which a name $n$ is bound to a variable of type $\pid()$, and
  whose every route to $n$ (variables, closure environments, received messages, results of
  functions, state) carries $n$ only at type $\pid()$, never sends on $n$, and neither do the
  processes to which it passes $n$. The reason is that no run-time test tells which messages
  a pid accepts (\texttt{is\_pid} only gives $\pid()$), so the code must also work with a pid
  that accepts no message, on which nothing can be sent. This is a parametricity property,
  not a consequence of soundness.
\end{description}

\subsubsection{Statements on the model}\label{ta-sec:statements}

This section states and proves the results announced in Appendix~\ref{ta-app:formal}, with the
rules of Appendix~\ref{ta-sec:base}; its last two paragraphs treat the form $\Hr$ and the code
of Appendix~\ref{ta-sec:otp}. Let $C_P$, $R_P$ and $\so(m)$ be as in Appendix~\ref{ta-sec:op}, and let $D_P$ denote the
triple $(C_P,\pid(),\state)$, so that, by~\eqref{ta-eq:regions-h},
$\Hpid=\bigwedge_{P\,:\,C_P\neq\Tempty}(D_P\to\{\at{reply},R_P,\state\})$ with pairwise disjoint
domains $D_P$. In the proofs, $P$ ranges over the nonempty regions.

\begin{theorem}[\texttt{call} is well typed]\label{ta-thm:call}
$\vdash\mathtt{call}:\forall c,r.\ (\srv(c,r),\,c)\to r$. In particular, if $\mathtt{call}(\mathtt{p},\mathtt{c})$, with $\mathtt{c}:c$,
is typed with the instance $(\srv(c,r),c)\to r$ of this type, then every value sent to its
alias, hence every value that it returns, in any run of any well-typed program, has type
$r$.
\end{theorem}
\begin{proof}[Sketch]
\emph{Typing.} Let $\mathtt{p}:\srv(c,r)$ and $\mathtt{c}:c$. By (alias), \texttt{w = alias()}
creates a name of type $\pid(r)$, and $\mathtt{receive\_on(w)}:r$; by (inst), each instance of
\texttt{call} creates its alias at the $r$ of the instance. Then
$\mathtt{\{c, w\}}:c\times\pid(r)$, and \texttt{send(p, \{c, w\})} is well typed by (send),
since $\srv(c,r)=\pid(c\times\pid(r))$. So the body has type $r$, for all $c$ and $r$.

\emph{Runs.} By (A1), the processes of a run of a well-typed program stay well typed. A process
that sends a value $v$ on \texttt{w} does it by (send), at a type $\pid(t)$ of \texttt{w}, with
$v:t$; by (names), $\pid(r)\le\pid(t)$, so $t\le r$ by~\eqref{eq:contra}, and $v:r$. The call
returns the value of $\mathtt{receive\_on(w)}$, which was sent on \texttt{w}.
\end{proof}

\begin{theorem}[\texttt{start\_link} is well typed]\label{ta-thm:start}
Under (A3), in the form $\Hpid$, for $v:a$ we have
$\vdash\mathtt{start\_link}(m,v):\{\at{ok},\so(m)\}$.
\end{theorem}
\begin{proof}[Sketch]
Let $M=\bigvee_P(C_P\times\pid(R_P))$, so that $\pid(M)=\bigwedge_P\srv(C_P,R_P)=\so(m)$
by~\eqref{eq:pidand}. We type the body of \texttt{start\_link}, and then the body of
\texttt{loop} by the rule for recursive functions, that is, under the hypothesis that
\texttt{loop(h, s')} is well typed, for $\mathtt{h}:\Hpid$ and every $\mathtt{s'}:\state$, in a
process whose mailbox has type $M$.
\begin{description}
\item[\texttt{\{:ok, s\} = m.init(v)}] By (A3), $\mathtt{m.init(v)}:\{\at{ok},\state\}$, so
  the match binds $\mathtt{s}:\state$.
\item[\texttt{spawn(fn -> loop(\&m.handle\_call/3, s) end)}] The capture has the declared
  type $\Hpid$ of \texttt{handle\_call}. By (spawn), the expression has type $\pid(M)$ if
  \texttt{loop(h, s)}, with $\mathtt{h}:\Hpid$ and $\mathtt{s}:\state$, is well typed in a
  process whose mailbox has type $M$. Hence the result
  $\{\at{ok},\pid(M)\}=\{\at{ok},\so(m)\}$.
\item[\texttt{receive do \{c, w\} -> ...}] By (receive), the branch is typed once for each
  $P$, with $\mathtt{c}:C_P$ and $\mathtt{w}:\pid(R_P)$.
\item[\texttt{h.(c, w, s)}] $\mathtt{(c, w, s)}:(C_P,\pid(R_P),\state)\le D_P$, since
  $\pid(R_P)\le\pid()$ by~\eqref{eq:contra}, and this triple is disjoint from the other
  domains $D_{P'}$. By (app), $\mathtt{h.(c, w, s)}:\{\at{reply},R_P,\state\}$.
\item[\texttt{case}] The pattern \texttt{\{:noreply, s2\}} is disjoint from this type, so its
  branch is not typed (case). The pattern \texttt{\{:reply, r, s2\}} binds $\mathtt{r}:R_P$ and
  $\mathtt{s2}:\state$.
\item[\texttt{send(w, r); loop(h, s2)}] The send is well typed by (send), since
  $\mathtt{w}:\pid(R_P)$ and $\mathtt{r}:R_P$; the recursive call by the hypothesis on
  \texttt{loop}, since $\mathtt{s2}:\state$.
\end{description}
\end{proof}
The receive of the loop is the only receive on the mailbox of the server, since the body of
\texttt{handle\_call} performs none (A3). Without this hypothesis a clause such as
\texttt{handle\_call(:a, \_, s) -> receive n -> \{:reply, n, s\} end} could take another
request for its reply.

\begin{corollary}[no run-time type error]\label{ta-cor:safe}
Consider a program, well typed under (A1), whose servers are started by \texttt{start\_link}
and whose pids are used only at the types given by Theorem~\ref{ta-thm:start} (or by
Proposition~\ref{ta-prop:formhr}, in the form $\Hr$), with \texttt{call} used at the types of
Theorem~\ref{ta-thm:call}. In every run of this program: (i) \texttt{handle\_call} is applied
only to requests in its domain $\bigvee_i c_i$ (no \texttt{function\_clause} error, the error
raised when no clause matches); (ii) every message sent to the alias of a call used at type
$r$ has type $r$.
\end{corollary}
\begin{proof}[Sketch]
\emph{(i).} The pid $q$ of the server is created at type $\pid(M)$, with $M$ as in the proof of
Theorem~\ref{ta-thm:start}. Every message that the loop receives was sent by (send) at some
type $\pid(t)$ of $q$ with a value of type $t$; by (names), $\pid(M)\le\pid(t)$, so $t\le M$
by~\eqref{eq:contra}: every message received has type $M$. By induction on the number of
consumed requests, the state passed to the loop has type $\state$: the initial one by (A3), and
the next one, \texttt{s2}, by the proof of Theorem~\ref{ta-thm:start}. So each application of
\texttt{handle\_call} has its arguments in some $(C_P,\pid(R_P),\state)\le D_P$, below the
domain of $\Hpid$, and by (progress) some clause matches.

\emph{(ii).} Theorem~\ref{ta-thm:call}.
\end{proof}

\begin{theorem}[adequacy of the loop, form $\Hpid$]\label{ta-thm:adequacy}
Let $q$ be the pid of the server. In every run of every program that satisfies the
hypotheses of Corollary~\ref{ta-cor:safe}, let $(x_1,w_1),(x_2,w_2),\ldots$ be the requests
consumed by the loop, in the order of consumption, and let $\{\at{reply},y_k,s_k\}$ be the
value returned by the $k$-th application of \texttt{handle\_call}, if it returns. If the
$k$-th request was sent by a call $\mathtt{call}(q,x_k)$, then the loop sends $y_k$ to $w_k$
after consuming the request, and nothing else is sent to $w_k$. Hence a call
$\mathtt{call}(q,\mathtt{c})$ on a request of type $c\le\bigvee_i c_i$ returns a value of type
$\bigvee_{P\,:\,c\land C_P\neq\Tempty}R_P$: the result of applying \texttt{handle\_call} to
$c$ (Appendix~\ref{ta-sec:op}).
\end{theorem}
\begin{proof}[Sketch]
\emph{Step 1: the loop sends $y_k$ to $w_k$, and nothing else.} In the form $\Hpid$ the
callback returns only \texttt{:reply} triples, so the loop, after consuming the $k$-th
request, takes the branch \texttt{\{:reply, r, s2\}} with $\mathtt{r}=y_k$ and
$\mathtt{s2}=s_k$, and sends $y_k$ to $w_k$; this is its only \texttt{send}.

\emph{Step 2: no other process sends on $w_j$}, for a request sent by $\mathtt{call}(q,x_j)$.
The name $w_j$ is held by the client, which creates it in \texttt{call}, sends it only to $q$,
and only receives on it; by the loop, which takes it from its mailbox (no other code of the
server receives, by (A3)) and passes it only to \texttt{handle\_call}; and by the body of
\texttt{handle\_call}, together with the processes to which this body passes it, directly or
through further messages. The body is typed with its declared type $\Hpid$, so in it
\texttt{from} has type $\pid()$, whatever the loop passes, and every route from the body to
$w_j$ carries it at type $\pid()$: by (opacity), neither the body nor these processes send on
it. If the body sends $w_j$ back to a server, inside a request, then this request, whose
address has type $\pid()$, is in a region with $R_P=\Tempty$, on which \texttt{handle\_call}
returns no reply, so the loop does not send on it either. Hence the only value sent to $w_j$
is $y_j$, and the call that created $w_j$ returns $y_j$.

\emph{Step 3: the type of $y_j$.} Let $x_j$ have type $c\le\bigvee_ic_i=\bigvee_PC_P$. Then
$(c,\pid(),\state)\le\bigvee_PD_P$, and this triple meets $D_P$ if and only if $c$ meets $C_P$,
since $\pid()$ and $\state$ are not empty. By (app), applied to
$\Hpid=\bigwedge_P(D_P\to\{\at{reply},R_P,\state\})$, the result of \texttt{handle\_call} has
type $\bigvee_{P\,:\,c\land C_P\neq\Tempty}\{\at{reply},R_P,\state\}$, and its second component,
$y_j$, has the stated type.
\end{proof}
The restriction to the calls on $q$ is necessary: a process $p'$ of type $\srv(c,r)$ may
forward the request $\{x,w\}$ of a client to $q$ and also send on $w$ a value of its own, so
that $w$ receives two values.

\begin{theorem}[exactness of $\so$, Claim~\ref{ta-claim:exact}]\label{ta-thm:exact}
Under (A2) and (A3), form $\Hpid$, for all closed types $c,r$:
\[
  \so(m)\le\srv(c,r)\iff\Hpid\le(c,\pid(),\state)\to\{\at{reply},r,\state\}.
\]
Hence, for a type $T=\bigwedge_{j\in J}\srv(d_j,s_j)$ written by hand, $\so(m)\le T$ if and
only if every arrow $(d_j,\pid(),\state)\to\{\at{reply},s_j,\state\}$ is a supertype of
$\Hpid$: this is what \texttt{start\_link} checks against the API.
\end{theorem}
\begin{proof}
By~\eqref{ta-eq:regions-h}, $\Hpid=\bigwedge_P(D_P\to\{\at{reply},R_P,\state\})$, where $P$
ranges over the nonempty regions, of which there is at least one since
$\bigvee_ic_i\neq\Tempty$ (A3). By Lemma~\ref{ta-lem:common}(b), with $p=\pid()$ and
$s=\state$, which are not empty (A3), the right-hand side is equivalent to $\bigwedge_P(C_P\to R_P)\le c\to r$, hence, the
$C_P$ being pairwise disjoint, by (i)$\iff$(ii) of Claim~\ref{ta-claim:exact}, to $\bigwedge_P\srv(C_P,R_P)\le\srv(c,r)$, the
left-hand side. The second part follows, since a type is below an intersection if and only if
it is below each of its components.
\end{proof}

\begin{theorem}[optimality of $\so$, Claim~\ref{ta-claim:least}]\label{ta-thm:opt}
Under (A2) and (A3), form $\Hpid$, call a set of substitutions for \texttt{serve}
\emph{valid} when \texttt{handle\_call} and the initial state, of type $\state$, are in the
domain of every instance. Then:
\begin{enumerate}
\item[(a)] if $\bigwedge_{j\in J}(d_j\to s_j)\equiv\bigwedge_{i\in I}(c_i\to r_i)$ then
  $\so(m)\le\MT\big(\bigwedge_{j\in J}(d_j\to s_j)\big)$;
\item[(b)] let $\mathtt{serve}:\forall c,r,s.\ \big((c,\pid(),s)\to\{\at{reply},r,s\},\,s\big)\to\srv(c,r)$
  be the function of Appendix~\ref{ta-sec:poly}, which spawns the loop on a callback and an
  initial state. Instantiating it with the set of substitutions
  $[c:=C_P,\ r:=R_P,\ s:=\state]_{P\,:\,C_P\neq\Tempty}$ \citep{CNXL14} is valid when the
  callback is \texttt{handle\_call}, and gives $\so(m)$;
\item[(c)] if $[c:=c_k,\ r:=r_k,\ s:=S_k]_{k\in K}$ is any valid set of substitutions, then
  $\so(m)\le\bigwedge_{k\in K}\srv(c_k,r_k)$.
\end{enumerate}
\end{theorem}
\begin{proof}[Sketch]
\emph{(a).} Claim~\ref{ta-claim:least}.

\emph{(b).} The rule of \citet{CNXL14} gives a polymorphic function instantiated with a set
of substitutions the intersection of its instances, and then applies the rule for
intersections of arrows. The instance of $P$ is
$\big(D_P\to\{\at{reply},R_P,\state\},\ \state\big)\to\srv(C_P,R_P)$. By~\eqref{ta-eq:regions-h},
$\Hpid\le D_P\to\{\at{reply},R_P,\state\}$ for every $P$, and the initial state has type
$\state$: the arguments are in the domain of every instance. Since
$\bigwedge_k(\sigma_k\to\tau_k)\le(\bigwedge_k\sigma_k)\to\bigwedge_k\tau_k$ (by contravariance
of the domains, then by~\eqref{eq:bcd}), the application has type
$\bigwedge_P\srv(C_P,R_P)=\so(m)$.

\emph{(c).} The validity of the $k$-th substitution is $\state\le S_k$ and
$\Hpid\le(c_k,\pid(),S_k)\to\{\at{reply},r_k,S_k\}$. If $c_k=\Tempty$, then
$\srv(c_k,r_k)=\pid(\Tempty)$, which is above $\so(m)$. Otherwise~\eqref{ta-eq:darrow}, for $Q$
the set of all the regions, gives
$(c_k,\pid(),S_k)\le\bigvee_PD_P=(\bigvee_PC_P,\pid(),\state)$, hence $S_k\le\state$, since
$c_k$ and $\pid()$ are not empty. So $S_k\simeq\state$, the validity is the right-hand side of
Theorem~\ref{ta-thm:exact} for $c_k$ and $r_k$, and $\so(m)\le\srv(c_k,r_k)$.
\end{proof}
Item (b) computes the type, but does not prove it sound (\parref{ta-par:spawn}). Without the
initial state in the definition of validity, (c) would be false: the substitution
$[c:=\at{foo},\ r:=\Tempty,\ s:=\Tempty]$ has the empty domain $(\at{foo},\pid(),\Tempty)$, so
\texttt{handle\_call} is in it, but $\srv(\at{foo},\Tempty)$ is not above $\so(m)$ when
\texttt{:foo} is not a request of \texttt{handle\_call}.

\paragraph{Sets of substitutions and spawning.}\label{ta-par:spawn} Instantiating with a set of
substitutions a function that spawns a process is unsound in general: the process has one
mailbox for all the instances. For example, the function that spawns a process of interface
$\alpha$ that receives two messages and sends their pair to its argument $p$ has type
$\forall\alpha.\ \pid(\alpha\times\alpha)\to\pid(\alpha)$, but with one instance for each spawn,
not the intersection of several. Instantiated at both $\Int$ and $\Bool$, and applied to
$p:\pid((\Int\times\Int)\lor(\Bool\times\Bool))$, which is below both
$\pid(\Int\times\Int)$ and $\pid(\Bool\times\Bool)$ by~\eqref{eq:contra}, it would give a process of
type $\pid(\Int)\land\pid(\Bool)=\pid(\Int\lor\Bool)$, which may receive $1$ and
\texttt{true} and send to $p$ the pair $\{1,\mathtt{true}\}$, which $p$ does not accept. For
\texttt{serve} the type is sound, because the substitutions agree on $s$ and the loop handles
each request independently, given the state: this is what the proof of
Theorem~\ref{ta-thm:start} shows, by typing the loop directly. With two types for $s$, the
state produced by a request of one instance could reach a clause of the other.

\begin{proposition}[module parameters]\label{ta-prop:param}
Let
\[
  \begin{array}{l}
  \mathtt{start\_link}:\forall a,c,r.\\
  \quad(\inst{GenServer}{\prm{init\_arg}~$a$, \prm{call\_msg}~$c$, \prm{reply}~$r$},\,a)\to\{{\at{ok}},\srv(c,r)\}
  \end{array}
\]
with invariant parameters. A module $m$ whose declared parameters are $\bigvee_i c_i$ and
$\bigvee_i r_i$ is accepted only at $c\simeq\bigvee_ic_i$ and $r\simeq\bigvee_ir_i$, so the
result is $\srv(\bigvee_ic_i,\bigvee_ir_i)$. In the form $\Hpid$,
$\so(m)\le\srv(\bigvee_ic_i,\bigvee_ir_i)$, and the inequality is strict for the first version
of \texttt{MyServer}.
\end{proposition}
\begin{proof}
\emph{Step 1: $\mathsf{eq}(t)\le\mathsf{eq}(t')$ if and only if $t\simeq t'$.} In the encoding
of \citet{BC26}, the type of a module records each parameter instantiated at $t$ as
$\mathsf{eq}(t)=(t\lor\{\mho\})\times(\neg t\lor\{\mho\})$, where $\mho$ is a distinguished
constant that, like the element $\Omega$ of \citet{FCB08}, does not belong to the domain of
values: no type of the language contains it, not even \texttt{term()}. The ``if'' part is
immediate. For the ``only if'' part, both products contain the pair $(\mho,\mho)$, so they are
nonempty, and~\eqref{ta-eq:dprod} reduces the inequality to its components:
$t\lor\{\mho\}\le t'\lor\{\mho\}$ and $\neg t\lor\{\mho\}\le\neg t'\lor\{\mho\}$, that is,
$t\setminus t'\le\{\mho\}$ and $t'\setminus t\le\{\mho\}$ (since
$\neg t\setminus\neg t'=t'\setminus t$). Since $\mho$ belongs to no type of the language,
$t\setminus t'=\Tempty$ and $t'\setminus t=\Tempty$.

\emph{Step 2: the instance.} The module $m$ has the type of the first argument only if
$\mathsf{eq}(\bigvee_ic_i)\le\mathsf{eq}(c)$ and $\mathsf{eq}(\bigvee_ir_i)\le\mathsf{eq}(r)$,
that is, by Step~1, $c\simeq\bigvee_ic_i$ and $r\simeq\bigvee_ir_i$.

\emph{Step 3: the inequality.} By covariance of the codomains and the law~\eqref{eq:bddl},
\begin{align*}
  \Hpid&\textstyle\le\bigwedge_i((c_i,\pid(),\state)\to\{\at{reply},\bigvee_jr_j,\state\})\\
  &\textstyle=(\bigvee_ic_i,\pid(),\state)\to\{\at{reply},\bigvee_jr_j,\state\}
\end{align*}
so that $\so(m)\le\srv(\bigvee_ic_i,\bigvee_ir_i)$ by Theorem~\ref{ta-thm:exact}.

\emph{Step 4: strictness.} For the first version of \texttt{MyServer},
$\so(m)\le\srv(\at{get\_int},\Int)$ by Claim~\ref{ta-claim:exact}, since \texttt{:get\_int}
meets only the region $C_1=\at{get\_int}$, whose reply is $R_1=\Int$. Instead,
$\srv(\at{get\_int}\lor\{\at{set\_int},\Int\},\Int\lor\at{ok})$ is not below
$\srv(\at{get\_int},\Int)$, by Lemma~\ref{ta-lem:srv}, since $\Int\lor\at{ok}\not\le\Int$ and
$\at{get\_int}\neq\Tempty$. This is the type with the union of the requests and the union of
the replies of Appendix~\ref{ta-sec:problem}.
\end{proof}

\paragraph{The form $\Hr$.}\label{ta-par:formhr} Let $\Tm=\bigwedge_{i\in I}\srv(c_i,r_i)$.

\begin{proposition}[form $\Hr$]\label{ta-prop:formhr}
Under (A3), in the form $\Hr$:
\begin{enumerate}
\item[(a)] $\vdash\mathtt{start\_link}(m,v):\{\at{ok},\Tm\}$ for $v:a$;
\item[(b)] Corollary~\ref{ta-cor:safe} holds.
\end{enumerate}
\end{proposition}
\begin{proof}[Sketch]
\emph{(a).} As for Theorem~\ref{ta-thm:start}, with $M=\bigvee_i(c_i\times\pid(r_i))$, so that
$\pid(M)=\Tm$ by~\eqref{eq:pidand}, and with the following changes. By (receive), the branch
is typed once for each $i$, with $\mathtt{c}:c_i$ and $\mathtt{w}:\pid(r_i)$. Then
$\mathtt{(c, w, s)}$ is in the domain of the $i$-th arrow of $\Hr$, so, by (app),
$\mathtt{h.(c, w, s)}:\{\at{reply},r_i,\state\}\lor\{\at{noreply},\state\}$. Both branches of
the \texttt{case} are typed: \texttt{\{:reply, r, s2\}} binds $\mathtt{r}:r_i$ and
$\mathtt{s2}:\state$, and \texttt{send(w, r)} is well typed since $\mathtt{w}:\pid(r_i)$;
\texttt{\{:noreply, s2\}} binds $\mathtt{s2}:\state$.

\emph{(b).} As for Corollary~\ref{ta-cor:safe}, with this $M$: a message of type $M$ is in one
summand $c_i\times\pid(r_i)$, so the arguments of \texttt{handle\_call} are in the domain of
its $i$-th arrow, and both branches of the \texttt{case} give $\mathtt{s2}:\state$.
\end{proof}
When the $c_i$ are pairwise disjoint, the nonempty regions are the nonempty $c_i$ and
$\Tm=\so(m)$ (the other components of $\Tm$ are $\srv(\Tempty,r_i)=\pid()$); otherwise
Claim~\ref{ta-claim:plain} shows where $\Tm$ is coarser than $\so(m)$, and
Appendix~\ref{ta-sec:deferred} how the programmer recovers the lost regions. In (b), that a
message lies in one summand is the justification of (receive); projections would not suffice,
since $(\pi_1M,\pi_2M,\state)$ contains triples that are in no domain
$(c_i,\pid(r_i),\state)$. Theorem~\ref{ta-thm:adequacy} holds only in part: a call may never
return, since a clause may return $\{\at{noreply},\state\}$ and never reply, and it may return
a value sent through \texttt{from} instead of $y_k$, since a clause may reply through
\texttt{from} and also return $\{\at{reply},y_k,s_k\}$. What remains is
Corollary~\ref{ta-cor:safe}(ii): if $w_k$ was created by a call (on $q$, or on a process that
forwards the request to $q$), every value sent to $w_k$ has the type $r$ of that call; in
general, it has the type of creation of $w_k$, by~(names).

\paragraph{The model with the messages of OTP.}\label{ta-par:otp-model} The code of
Figure~\ref{ta-fig:genserver} differs from the model of Appendix~\ref{ta-app:formal} in the
messages and in the results of \texttt{init}, and \texttt{loop} receives the module instead of
the function \texttt{handle\_call}. The tag plays the role of the return address, so that the
proofs of Theorems~\ref{ta-thm:call} and~\ref{ta-thm:start} carry over. Let
$\rtag(r)=\pid(\Tany\times r)$, $\from(r)=\{\pid(),\rtag(r)\}$ and
$\srv(c,r)=\pid(\{\at{"\$gen\_call"},\from(r),c\})$.

\begin{lemma}[laws of $\rtag$ and $\from$]\label{ta-lem:rtag}
The laws~\eqref{eq:contra}--\eqref{eq:union} hold with $\rtag$, and with $\from$, in place of
$\pid$. In particular, $\from(r)\le\from()$ and $\from()\neq\Tempty$.
\end{lemma}
\begin{proof}
By Lemma~\ref{ta-lem:common}(a), $\Tany\times s\le\Tany\times t$ if and only if $s\le t$.
Hence, by~\eqref{eq:contra}, $\rtag(t)\le\rtag(s)$ if and only if $s\le t$;
by~\eqref{eq:pidand}, and since products distribute over unions,
$\rtag(t_1)\land\rtag(t_2)=\pid((\Tany\times t_1)\lor(\Tany\times t_2))=\rtag(t_1\lor t_2)$;
and by~\eqref{eq:union}, $\rtag(t)\le\bigvee_{n\in N}\rtag(s_n)$ if and only if
$\Tany\times s_n\le\Tany\times t$, that is, $s_n\le t$, for some $n\in N$. For $\from$:
products distribute over intersections and unions, and, by Lemma~\ref{ta-lem:common}(a) with
$u=\pid()$, $\from(t)\le\bigvee_{n\in N}\from(s_n)$ if and only if
$\rtag(t)\le\bigvee_{n\in N}\rtag(s_n)$.
\end{proof}
Outside \texttt{GenServer} the checker knows of $\rtag(r)$ only these laws, so a program that
type-checks with the abstract type also type-checks with this definition, which the next
proposition uses.

\begin{proposition}[the code with the messages of OTP]\label{ta-prop:otp}
Let (A3) hold with $\from()$ and $\from(r_i)$ in place of $\pid()$ and $\pid(r_i)$ in the arrows
of \texttt{handle\_call}, and with
$\mathtt{init}:(a)\to\{\at{ok},\state\}\lor\at{ignore}\lor\{\at{stop},\Tany\}$. Then the bodies of
\texttt{call}, \texttt{reply} and \texttt{start\_link} in Figure~\ref{ta-fig:genserver} are well
typed under their signatures, where \texttt{server\_of(m)} is $\so(m)$ in the form $\Hpid$ and
$\Tm$ in the form $\Hr$, and the second part of Theorem~\ref{ta-thm:call} holds for this
\texttt{call}.
\end{proposition}
\begin{proof}[Sketch]
\emph{\texttt{call}.} Let $\mathtt{p}:\srv(c,r)$ and $\mathtt{c}:c$. By (alias) at the type
$\Tany\times r$, \texttt{tag = alias()} creates a name of type $\rtag(r)$, and
$\mathtt{receive\_on(tag)}:\Tany\times r$, so that the match binds $\mathtt{v}:r$. The value of
\texttt{self()} is the pid of the running process, whose type is below $\pid()$
by~\eqref{eq:contra}, whatever the type of its mailbox, so that
$\mathtt{\{self(), tag\}}:\from(r)$, and the send is well typed by (send). The second part of
Theorem~\ref{ta-thm:call} follows as there: by (names), a value sent on \texttt{tag} has type
$\Tany\times r$, so the second component of the value that \texttt{receive\_on(tag)} returns
has type $r$.

\emph{\texttt{reply}.} Let $\mathtt{tag}:\rtag(r)$ and $\mathtt{v}:r$. Then
$\mathtt{\{tag, v\}}:\Tany\times r$, and the send is well typed by (send).

\emph{\texttt{start\_link} and \texttt{loop}.} The region form~\eqref{ta-eq:regions-h} holds as
before, since $\from()\neq\Tempty$. In the form $\Hpid$, the proof of
Theorem~\ref{ta-thm:start} applies with
$M=\bigvee_{P}\{\at{"\$gen\_call"},\from(R_P),C_P\}$, the type \texttt{messages\_of(m)}, so that
$\pid(M)=\so(m)$ by~\eqref{eq:pidand}. By (spawn), the body of the spawn is typed in a process
whose mailbox has type $M$, as the \texttt{\$interface} declares; there \texttt{loop} is applied
to the module $m$ of \texttt{start\_link}, so that \texttt{m.handle\_call} has the type $\Hpid$
and $\mathtt{s}:\state$. By (receive), the branch is typed once for each $P$, with the
variables of the pattern, the nested ones included, at the components of the summand:
$\mathtt{from}:\from(R_P)$, $\mathtt{tag}:\rtag(R_P)$ and $\mathtt{c}:C_P$. Since
$\from(R_P)\le\from()$, the arguments of \texttt{m.handle\_call} are in the domain of the arrow
of the region, so that, by (app),
$\mathtt{m.handle\_call(c, from, s)}:\{\at{reply},R_P,\state\}$, and
\texttt{send(tag, \{tag, v\})} is well typed by (send), with $\mathtt{v}:R_P$. The other two
branches of the \texttt{case} of \texttt{start\_link} return \texttt{:ignore} and
\texttt{\{:error, reason\}}, with $\mathtt{reason}:\Tany$, by the type of \texttt{init}, and
both are in the declared result. The form $\Hr$ is treated in the same way, with
$M=\bigvee_{i\in I}\{\at{"\$gen\_call"},\from(r_i),c_i\}$, as in
Proposition~\ref{ta-prop:formhr}.
\end{proof}
Three remarks on this code. In \texttt{call}, the route to the pid of the caller through
\texttt{self()} carries it only at the type $\pid()$, so that, as in (opacity), it adds no
sender: a body of \texttt{handle\_call} may still call other servers, as (A3) says, although
\texttt{call} uses \texttt{self()}. In \texttt{reply}, the first component of \texttt{from}
cannot take the place of the tag: it has type $\pid()=\pid(\Tempty)$, and no value has type
$\Tempty$. The signature of \texttt{loop} has no decorations. In \citet{YHM26} its arrow would
have those that say that it sends nothing to the mailbox of its process and that it reads
messages of type $M$. The body alone does not determine them (it is typed, for instance, at
any mailbox type disjoint from its pattern) and (receive) needs the summands of $M$, so they
come from the \texttt{\$interface} of the spawn. The loop sends directly, as in the model, and
does not call the polymorphic \texttt{reply}: by (inst), that occurrence would have one
instance $r_0$ for all the summands of (receive), which would require $r_0\le R_P$ (for
$\from(R_P)\le\from(r_0)$) and $R_P\le r_0$ (for $\mathtt{v}:R_P$) for every $P$, hence one
reply type for all the regions.

Finally, casts, which the code leaves out. The type $\so(m)$ alone is below
$\pid(\{\at{"\$gen\_cast"},k\})$ only for $k=\Tempty$, by~\eqref{eq:contra}, since the pairs tagged
by \at{"\$gen\_cast"} are disjoint from the triples of $M$: this is why the operator
\texttt{server\_of} must include the type of the casts. Let $K=\{\at{"\$gen\_cast"},k_m\}$; then
\texttt{messages\_of(m)} is $M\lor K$. On the side of the server, \texttt{loop} needs a branch for these pairs, which applies
\texttt{handle\_cast} (a function that, like \texttt{handle\_call} in (A3), performs no receive) and
loops on the new state: (receive) types this branch on the summand $K$, and the branch of the calls
on the summands of $M$, since each pattern is disjoint from the summands of the other. On the side
of the client, this changes no result: let $T=\{\at{"\$gen\_call"},\from(r),c\}$;
by~\eqref{eq:pidand} and~\eqref{eq:contra}, $\so(m)\land\pid(K)\le\srv(c,r)$ if and only if
$T\le M\lor K$, and since $T$ and $K$ are disjoint, if and only if $T\le M$, that is, if and only if
$\so(m)\le\srv(c,r)$. In the form $\Hr$ the same holds with $\Tm$ in place of $\so(m)$.

\subsubsection{The translation with \texorpdfstring{$\lambda$}{lambda}-channels}\label{ta-sec:cdv-details}

\citet{CDV06} encodes the $\lambda$-calculus with semantic subtyping (the functional core of
CDuce) into the $\pi$-calculus with semantic subtyping. This encoding explains why $\so$
normalizes into regions, and why the extra precision of its $\lambda$-channels is out of reach
on the BEAM. Section~\ref{sec:lost} tells the story; here I give the details.

\paragraph{Two translations of arrows.} \citet{CDV06} considers two translations of function
types into channel types. Both commute with $\lor$, $\land$, $\neg$ and $\times$, and translate
the types under the arrows as well; here I apply them only to the arrows of the type of
\texttt{handle\_call}, and leave the request and reply types $c$, $r$ unchanged. They differ
on arrows:
\[
  \MT(c\to r) = \pid(c\times\pid(r)) = \srv(c,r)
  \qquad\qquad
  \LC(c\to r) = \pid(c\times\chl(r))
\]
$\MT$ is the Milner--Turner translation \citep[\S4.1]{CDV06}: the request travels with a
return channel on which the reply can be written. The \emph{$\lambda$-channel translation}
$\LC$ \citep[\S5.1--5.2]{CDV06} gives the return channel the type of a
\emph{$\lambda$-channel} $\chl(t)=(\neg\pid(\neg t)\lor\pid(\Tany))\land\pid(\Tempty)$: it
contains $\pid(t)$, plus ``noise'', the channels that accept neither all of $t$ nor all of
$\neg t$ \citep[\S5.2, property (c)]{CDV06}. The property that matters is
\begin{equation}\label{ta-eq:chl}
  \chl(t_1\land t_2)=\chl(t_1)\lor\chl(t_2)
  \qquad\text{whereas}\qquad
  \pid(t_1)\lor\pid(t_2)\le\pid(t_1\land t_2)
\end{equation}
The equation is the condition that defines $\lambda$-channels in \citet[\S5.1]{CDV06}, and it
follows from~\eqref{eq:pidand} and De Morgan's laws:
$\neg\pid(\neg(t_1\land t_2))=\neg\pid(\neg t_1\lor\neg t_2)=\neg(\pid(\neg t_1)\land\pid(\neg t_2))
=\neg\pid(\neg t_1)\lor\neg\pid(\neg t_2)$; then $\lor\,\pid(\Tany)$ and
$\land\,\pid(\Tempty)$ distribute over the union. The inequality holds by
contravariance~\eqref{eq:contra}, and it is strict exactly when $t_1$ and $t_2$ are
incomparable: by the union law, $\pid(t_1\land t_2)\le\pid(t_1)\lor\pid(t_2)$ holds if and only
if $t_1\le t_1\land t_2$ or $t_2\le t_1\land t_2$, that is, $t_1\le t_2$ or $t_2\le t_1$.
Applied to the type of \texttt{handle\_call}, by~\eqref{eq:pidand}:
\begin{align*}
  \textstyle\MT\big(\bigwedge_i(c_i\to r_i)\big)&\textstyle=\pid\big(\bigvee_i(c_i\times\pid(r_i))\big)=\bigwedge_i\srv(c_i,r_i)\\
  \textstyle\LC\big(\bigwedge_i(c_i\to r_i)\big)&\textstyle=\pid\big(\bigvee_i(c_i\times\chl(r_i))\big)
\end{align*}

\paragraph{$\LC$ needs no regions; $\MT$ does.} Since $c_i=\bigvee_{P\ni i}C_P$,
distributing $\times$ over $\lor$ gives, for any types $X_i$,
\[
  \textstyle\bigvee_{i\in I}(c_i\times X_i)=\bigvee_P(C_P\times\bigvee_{i\in P}X_i) .
\]
With $X_i=\chl(r_i)$, \eqref{ta-eq:chl} turns the inner union into $\chl(R_P)$, so that
\[
  \textstyle\LC(\bigwedge_i(c_i\to r_i))=\LC(\bigwedge_P(C_P\to R_P))
\]
consistently with the fact that the full $\LC$ translation respects type equality
\citep[\S5.1]{CDV06}. With $X_i=\pid(r_i)$, the inner union is only below $\pid(R_P)$ (by
contravariance, since $R_P\le r_i$), so that $\MT$ maps the two equivalent function types
$\bigwedge_i(c_i\to r_i)$ and $\bigwedge_P(C_P\to R_P)$ to $\bigwedge_i\srv(c_i,r_i)$ and to
$\so(m)$, which Claim~\ref{ta-claim:plain} compares. The normalization into regions is
therefore required by $\MT$, not by $\LC$.

The rewriting of Section~\ref{sec:lost}, which computes the regions, appears in
\citet[\S3]{CDV06}, in the discussion of the rule (abstr), whose premise requires the domains of
the interface of a $\lambda$-abstraction to be pairwise disjoint. It serves the encoding of
terms, not of types: the body of an encoded function is a choice with one branch per arrow,
each the input of a pair $(x:\sigma_i,\ w:\pid(\tau_i))$; if
$\sigma_h\land\sigma_k\neq\Tempty$, an argument in $\sigma_h\land\sigma_k$ would select branch
$h$ or $k$ according to the type of the return channel $w$.

\paragraph{Why not use $\LC$ then?} $\chl(r)$ is a type of our system, since it uses only
Boolean connectives over pid types, and $\LC$ is more precise than $\so$; but where it is more
precise, a server typed with it cannot reply.

\begin{lemma}[$\lambda$-channels]\label{ta-lem:chl}
For all types $t$ and $s$:
\begin{enumerate}
\item[(a)] $\pid(t)\le\chl(t)$, and $\chl(t)\le\pid(t)$ if and only if $t=\Tempty$ or
  $t=\Tany$;
\item[(b)] if $t\neq\Tany$, then $\chl(t)\le\pid(s)$ if and only if $s=\Tempty$.
\end{enumerate}
\end{lemma}
\begin{proof}
\emph{(a).} $\pid(t)\le\pid(\Tempty)$ by~\eqref{eq:contra}, and
$\pid(t)\le\neg\pid(\neg t)\lor\pid(\Tany)$ because
$\pid(t)\land\pid(\neg t)=\pid(t\lor\neg t)=\pid(\Tany)$, by~\eqref{eq:pidand}. For the
equivalence, $\chl(\Tempty)=\pid(\Tempty)$, and $\chl(\Tany)=\pid(\Tany)$ since
$\neg\pid(\Tempty)\land\pid(\Tempty)=\Tempty$ and $\pid(\Tany)\le\pid(\Tempty)$. Conversely,
$\neg\pid(\neg t)\land\pid(\Tempty)\le\chl(t)$, so $\chl(t)\le\pid(t)$ gives
$\pid(\Tempty)\le\pid(t)\lor\pid(\neg t)$, hence $t=\Tempty$ or $t=\Tany$ by the union
law~\eqref{eq:union}.

\emph{(b).} $\chl(t)\le\pid(\Tempty)$ by definition. Conversely, the pid that accepts no
message is in $\chl(t)$: it is in $\pid(\Tempty)$, and it is in $\neg\pid(\neg t)$, since it
would be in $\pid(\neg t)$ only if $\neg t\le\Tempty$, that is, $t=\Tany$. And it is in
$\pid(s)$ only for $s=\Tempty$.
\end{proof}
By contravariance~\eqref{eq:contra} and Lemma~\ref{ta-lem:chl}(a),
\begin{align*}
  \textstyle\LC(\bigwedge_i(c_i\to r_i))&\textstyle=\pid(\bigvee_P(C_P\times\chl(R_P)))\\
  &\textstyle\le\pid(\bigvee_P(C_P\times\pid(R_P)))=\so(m)
\end{align*}
and, since the regions are pairwise disjoint, by~\eqref{ta-eq:dprod} the converse inequality
holds if and only if $\chl(R_P)\le\pid(R_P)$ for every nonempty region: the inequality is
strict exactly when some $R_P$ with $C_P\neq\Tempty$ is neither $\Tempty$ nor $\Tany$. But on
such a region a server of this type cannot reply: by Lemma~\ref{ta-lem:chl}(b), to reply on a
return channel $w:\chl(r)$, a server must first check that $w:\pid(r)$, and
Section~\ref{sec:lost} explains why this check cannot be programmed on the BEAM. So the
gen\_server loop, which sends every reply that \texttt{handle\_call} returns, can be typed by
$\LC$ only when $\LC$ adds nothing to $\so$.

\paragraph{The BEAM does not make $\MT$ faithful.} Section~\ref{sec:lost} explains why the
absence of type tests on pids does not make $\MT$ respect type equality: the gap between
$\bigwedge_i\srv(c_i,r_i)$ and $\so(m)$ comes from the union law, and closing it would require
the law $\pid(t_1\land t_2)\le\pid(t_1)\lor\pid(t_2)$, which is false for pids. This law would
also identify too many types: it gives
$\pid(\Tempty)\le\pid(\Int\times\pid(\Bool))\lor\pid(\Bool\times\pid(\Int))$
\citep[\S4.3]{CDV06}, that is, every pid would be a server of $\Int\to\Bool$ or of
$\Bool\to\Int$. In short, $\so$ is justified by Claims~\ref{ta-claim:least}
and~\ref{ta-claim:exact}; the absence of type tests on pids only explains why the extra
precision of $\LC$ is out of reach.

\fi

\end{document}